\documentclass[prx,aps,reprint,floatfix]{revtex4-2} 
\usepackage{graphicx} 
\usepackage{amsmath}
\usepackage{amsfonts}
\usepackage{float}
\usepackage[english]{babel}
\usepackage[T1]{fontenc}

\usepackage[bookmarks=false]{hyperref}
\usepackage{xcolor}
\hypersetup{
    colorlinks,
    linkcolor={red!50!black},
    citecolor={blue!50!black},
    urlcolor={blue!80!black}
}

\newcommand{\dd}{\mathrm{d}}

\newcommand{\av}[1]{\left\langle #1 \right\rangle}
\newcommand{\smallav}[1]{\langle #1 \rangle}

\newcommand{\Fig}[1]{Fig.~\ref{#1}}
\newcommand{\Figs}[1]{Figs.~\ref{#1}} 
\newcommand{\App}[1]{Appendix~\ref{#1}}
\newcommand{\Sec}[1]{Sec.~\ref{#1}}

\newcommand{\Eq}[1]{Eq.~\ref{#1}}

\newcommand{\Sdot}{\dot{S}}
\newcommand{\SI}[1]{(see Appendix~\ref{#1})}
\newcommand{\dP}{P(\mathbf{x}) \dd \mathbf{x}}
\renewcommand{\vec}[1]{\mathbf{#1}}
\newcommand{\heat}{\Pi}

\usepackage{hyperref}

\begin{document}

\title{ Learning force fields from stochastic trajectories}
  
\author{Anna Frishman}
\email{frishman@technion.ac.il}
\affiliation{Department of Physics, Technion Israel Institute of Technology, 32000 Haifa, Israel}

\author{Pierre Ronceray}
\email{ronceray@princeton.edu}
\affiliation{Center for the Physics of Biological Function, Princeton University, Princeton, NJ 08544, USA}


\begin{abstract}
  When monitoring the dynamics of stochastic systems, such as
  interacting particles agitated by thermal noise, disentangling
  deterministic forces from Brownian motion is challenging. Indeed, we
  show that there is an information-theoretic bound, the
  \emph{capacity} of the system when viewed as a communication
  channel, that limits the rate at which information about the force
  field can be extracted from a Brownian trajectory. This capacity
  provides an upper bound to the system's entropy production rate, and
  quantifies the rate at which the trajectory becomes distinguishable
  from pure Brownian motion. We propose a practical and principled
  method, Stochastic Force Inference, that uses this information to
  approximate force fields and spatially variable diffusion
  coefficients. It is data efficient, including in high dimensions,
  robust to experimental noise, and provides a self-consistent
  estimate of the inference error.  In addition to forces, this
  technique readily permits the evaluation of out-of-equilibrium
  currents and the corresponding entropy production with a limited
  amount of data.
\end{abstract}

\maketitle

\begin{figure}[b]
  \centering
  {\includegraphics[width=1.\columnwidth]{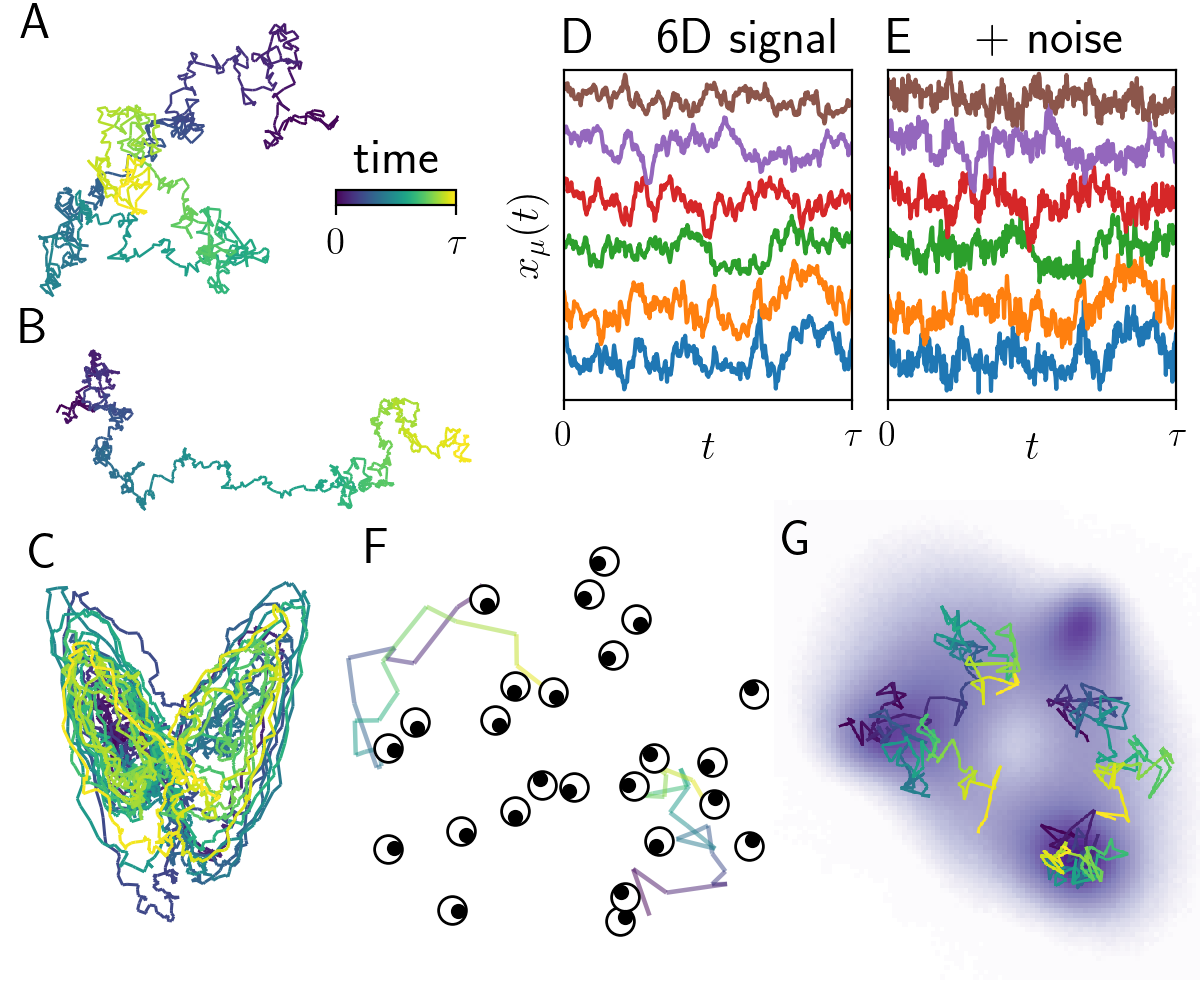}}
  \caption{Typical trajectories of example Brownian systems studied in
    this article. \textbf{A.} Pure Brownian motion in 2D, without
    forces. \textbf{B.} A drifted Brownian motion
    trajectory. \textbf{C.}  The stochastic Lorenz process (see
    \Fig{fig:nonlinear}). \textbf{D.}  Time series of a 6D
    out-of-equilibrium Ornstein-Uhlenbeck process (see
    \Fig{fig:6D}). \textbf{E.}  The same trajectories as in {D}, with
    additional time-uncorrelated measurement noise. \textbf{F.}
    Self-propelled active Brownian particles with soft repulsion and
    harmonic confinement (see \Fig{fig:particles}).  \textbf{G}
    Simulated single-molecule trajectories in a complex environment
    with space-dependent diffusion (see \Fig{fig:comparison}).}
  \label{fig:demo}
\end{figure}

From nanometer-scale proteins to micron-scale colloids, particles in
biological and soft matter systems undergo Brownian
dynamics~\cite{brown_brief_1828,perrin_mouvement_1909}: their
deterministic motion due to the forces competes with the random
diffusion due to thermal noise from the solvent. At a larger scale,
the overdamped Langevin equation describing Brownian dynamics is
commonly used as an effective model for the stochastic evolution of
complex systems such as motile cells~\cite{li_dicty_2011}, financial
markets~\cite{oksendal_stochastic_2003} or climate
dynamics~\cite{hasselmann_stochastic_1976}, where the noise
corresponds to the random influence of fast, unresolved degrees of
freedom, while force fields model persistent, deterministic trends. In
the absence of forces, all trajectories would thus look alike
(\Fig{fig:demo}A): the force field simultaneously shapes a system's
trajectory (\Fig{fig:demo}B-C) and encompasses most physical
information about the system.  The inference of such force fields from
experimental data is therefore crucial to problems as varied as
understanding the dynamics of single molecules in complex cellular
environments~\cite{knight_dynamics_2015,sungkaworn_single-molecule_2017},
quantifying the interactions between self-propelled colloidal
particles~\cite{palacci_living_2013}, calibrating devices to optically
trap particles~\cite{gavrilov_real-time_2014}, or identifying the laws
governing the motion of cells~\cite{bruckner_stochastic_2019}. This
problem is particularly relevant in the context of living or driven
out-of-equilibrium systems, where active forces induce dissipative
currents at the mesoscale~\cite{gnesotto_broken_2018}. The knowledge
of the force field in such cases would permit to measure the mean
entropy production rate and thus quantify the irreversibility of the
dynamics, a question which gained attention
recently~\cite{roichman_influence_2008,battle_broken_2016,gladrow_broken_2016,li_quantifying_2019,seara_entropy_2018,gonzalez_experimental_2019}. Moreover,
it would also enable one to measure the fluctuations of heat, work and
entropy production -- the subject of stochastic thermodynamics
\cite{seifert_stochastic_2012} -- which is so far only possible in
highly controlled systems~\cite{ciliberto_experiments_2017}.

Numerous previous studies have proposed methods to reconstruct force
fields, motivated by applications in soft
matter~\cite{garcia_high-performance_2018}, cell
biology~\cite{hoze_heterogeneity_2012,turkcan_bayesian_2012,beheiry_inferencemap:_2015},
climate
dynamics~\cite{gottwald_stochastic_2017,Bottcher_reconstruction_2006}
finance~\cite{gobet_nonparametric_2004,comte_penalized_2007,hoffmann_adaptive_1999,kutoyants_statistical_2004,papaspiliopoulos_nonparametric_2012}
and other complex systems~\cite{friedrich_approaching_2011}. However,
force inference in Brownian systems remains a hard problem, and a
general method is still missing, in particular one addressing the many
challenges associated with experimental data in soft matter and
biological systems.  First, there needs to be enough information about
the force available in the trajectory: short trajectories are
dominated by noise (\Fig{fig:demo}A), and only after a long enough
observation time does the effect of the force field become apparent
(\Fig{fig:demo}B). Second, one needs a practical method to extract
that information and reconstruct the force field, which is challenging
for out-of-equilibrium systems with a complex spatial structure
(\Fig{fig:demo}C), in particular for high-dimensional processes
(\Fig{fig:demo}D-F) and in the presence of measurement error
(\Fig{fig:demo}E) and multiplicative noise (\Fig{fig:demo}G). 

Here we address these challenges for steady-state Brownian
trajectories. We first use communication-theory tools to quantify the
maximal rate at which information about a force field can be inferred
from a trajectory (\Sec{sec:capacity_main}). We relate this rate, that
we term channel capacity of the system, to the entropy production
rate, thus providing a novel link between stochastic thermodynamics
and information theory. We then propose a practical procedure,
Stochastic Force Inference (SFI), to use the information in a
trajectory and reconstruct the force field by projecting it onto a
finite-dimensional functional space (\Sec{sec:SFI}). By inferring the
information contained in a trajectory, we propose a practical
criterion to control overfitting, an aspect generally overlooked by
previous approaches. We ensure that this method is robust to the
presence of experimental noise. Finally, the diffusion coefficient can
depend on the state of the system, which significantly complicates
force inference: in such cases, we adapt our method to infer the
space-dependent diffusion and force field (\Sec{sec:inhomogeneous}).
Using simple model stochastic processes, we demonstrate that our
method permits a quantitative evaluation of phase space forces,
currents and diffusion coefficients, and estimate the entropy
production with a minimal amount of data.

\begin{figure*}[t]
  \centering
  \includegraphics[width=0.8\textwidth]{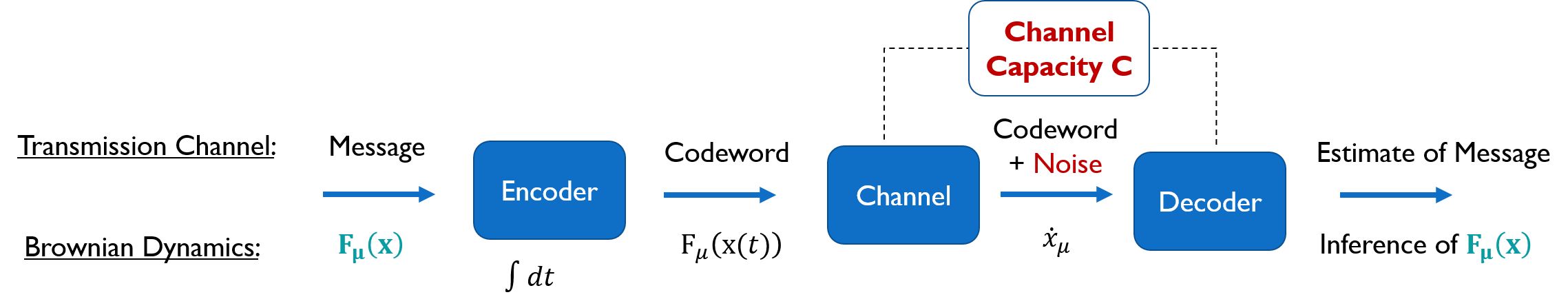}
  \caption{ \label{fig:schematic} The dynamics of an overdamped system
    can be seen as a noisy data transmission channel, encoding
    information about the force field, with a rate bounded by the
    channel capacity $C$ as defined in \Eq{eq:capacity}. Note that
    this definition does not include the information loss stemming
    from the measurement device. This analogy is further discussed in
    \App{sec:capacity}. }
\end{figure*}

We focus in this article on stochastic systems governed by the
overdamped Langevin equation, where friction dominates over inertia,
as is typically the case in sub-cellular biological systems for
instance.  We thus consider a system where the phase space coordinates
$x_\mu$ obey Brownian dynamics,
\begin{equation}
  \dot{x}_\mu = F_\mu(\mathbf{x}) + \sqrt{2D}_{\mu\nu} \xi_\nu
  \label{eq:Langevin}
\end{equation}
where $F_\mu(\mathbf{x})$ is the force field (we absorb the mobility
matrix in its definition), $D_{\mu\nu}$ is the diffusion tensor, and
$\xi_\mu$ is a Gaussian white noise,
$\av{\xi_\mu(t) \xi_\nu(t')} = \delta(t-t')$. In the first two
sections of this article, we assume that $D_{\mu\nu}$ is
space-independent and
known~\cite{qian_single_1991,vestergaard_optimal_2014}; in the third
section we address the case of inhomogeneous diffusion, which modifies
\Eq{eq:Langevin}.

\section{The information content of Brownian trajectories}
\label{sec:capacity_main}

We propose to interpret Brownian dynamics (\Eq{eq:Langevin}) as a
noisy transmission channel, where the force is the encoded signal and
$\sqrt{2\mathbf{D}} \xi$ is the noise (\Fig{fig:schematic}). Information can be read out
from such a channel at a maximal rate $C$, called the channel
capacity, which relates to the signal-to-noise ratio of the
input~\cite{cover_elements_2006}. This fundamentally limits the
ability to infer forces by monitoring the dynamics. To build up
intuition, consider the simplest case of a spatially constant force
with isotropic diffusion, corresponding to drifted Brownian
motion (\Fig{fig:demo}B). The capacity is then given by $C=F^2/4D$ (expressed in natural
information units, or nats, per time unit --- 1 nat = $1/\log 2$
bits).  The force to infer is here equal to the persistent velocity,
which can be estimated as $\hat{F}_\mu = \Delta x_\mu / \tau$, where
$\Delta \mathbf{x}$ is the end-to-end vector along the trajectory of
duration $\tau$. The relative error on this estimator due to
random diffusion is
$\av{||\hat{\mathbf{F}} - \mathbf{F}||^2/F^2} = 2dD/\tau F^{2} =
d/2I$, where $d$ is the space dimension. We have identified here
$I = C\tau$, defining it as the information in the trajectory.
Persistent motion thus starts to emerge from the noise if the
trajectory duration $\tau$ is longer than $d/C$, corresponding to the
diffusive-to-persistent transition for the mean-squared
displacement. Equivalently, the force starts to be resolved if
$I > d$, \emph{i.e.} if more than one bit of information is available
for each degree of freedom $\hat{F}_\mu$ to infer.

We now give a precise meaning to the notion of capacity for general
Brownian systems, where inter-particle interactions and external
fields lead to a force that depends on the position $\mathbf{x}$ in
phase space.  We recognize that within communication theory, the
dynamics of a Brownian system (\Eq{eq:Langevin}) corresponds to an
infinite-bandwidth Gaussian channel~\cite{cover_elements_2006}. The
signal transmitted is the force, with signal power equal to its
time-averaged square. The corresponding channel's capacity, which
we refer to as the system's capacity, is thus~\SI{sec:capacity}
\begin{equation}
  C = \frac{1}{4} \int F_\mu(\mathbf{x}) D^{-1}_{\mu\nu} F_\nu(\mathbf{x}) \dP
  \label{eq:capacity}
\end{equation}
where $P(\mathbf{x})$ is the steady-state probability distribution
function of the process, and we use the Einstein convention of
summation over repeated indices throughout. This quantity was
previously considered as a penalty term to regularize force
inference~\cite{batz_variational_2016}. 

The steady-state Fokker-Planck equation allows to decompose the force
into a sum of two terms,
\begin{equation}
  F_\mu = v_\mu + D_{\mu\nu} \partial_\nu \log P
  \label{eq:FP}
\end{equation}
where $v_\mu$ is the average phase space velocity, quantifying the
presence of irreversible currents, and
$D_{\mu\nu} \partial_\nu \log P$ quantifies reversible, diffusive
currents. Interestingly, this implies that the capacity defined in
\Eq{eq:capacity} decomposes into two non-negative parts, one related
to dissipation and the other to spatial structure, as
\begin{equation}
4C = \Sdot + G
\label{eq:C_decomposition}
\end{equation}
Here $\Sdot$ is the steady-state entropy production of the
process~\cite{seifert_stochastic_2012},
$\Sdot = \int v_\mu D^{-1}_{\mu\nu} v_\nu \dP$ (we set the Boltzmann
constant $k_B=1$ throughout).  In the case of thermal systems
satisfying the Einstein relation, $\Sdot$ corresponds to the rate at
which the system dissipates heat into the bath, divided by the
temperature; in other cases, $\Sdot$ quantifies the irreversibility of
the dynamics. The second term, named \emph{inflow rate}
$G = \int g_\mu D_{\mu\nu} g_\nu \dP$ with
$g_\mu = \partial_\mu \log P$, was previously introduced and studied
in Ref.~\cite{baiesi_inflow_2015}. It reflects the amount of
information that the force field injects into the system in order to
maintain probability gradients against diffusion, and is positive even
at equilibrium. Indeed, in a thought experiment where the force field
would be suddenly switched off, $G$ would correspond to the
instantaneous entropy production rate due to the relaxation of
probability gradients~\SI{sec:inflow}. The inflow rate quantifies the
fact that in steady state, the system dwells in convergent regions of
the force field: an equivalent expression for it is
indeed~\cite{baiesi_inflow_2015}
$G = - \int \partial_\mu F_\mu(\mathbf{x}) \dP$. In a deterministic
system, it would thus correspond to the average phase space
contraction rate. The connection between the inflow rate and the
previously introduced notions of traffic and
frenesy~\cite{maes_steady_2008,chetrite_fluctuation_2008} is explored
in \App{sec:traffic}. As $G\geq 0$, \Eq{eq:C_decomposition} provides a
generic upper bound to the entropy production in Brownian systems,
$\Sdot \leq 4C$.

The decomposition of the information into dissipative and structural
contributions introduced in \Eq{eq:C_decomposition} can be expressed
at the level of individual trajectories in phase space. Indeed, the
entropy production rate corresponds to the rate at which trajectories, $\mathcal{C} = \{\mathbf{x}(t)\}_{t=0..\tau}$,
become distinguishable from their time-reversed version, $-\mathcal{C} = \{\mathbf{x}(\tau-t)\}_{t=0..\tau}$, as quantified
by the Kullback-Leibler divergence rate~\cite{seifert_stochastic_2012}:
$\Sdot = \lim_{\tau\to\infty} \frac{1}{\tau} \av{ \log
  \mathcal{P}(\mathcal{C}|F)/ \mathcal{P}(\mathcal{-C}|F)}_F$.
Here $\mathcal{P}(\mathcal{C}|F)$ is the probability that the system
follows a trajectory $\mathcal{C} $
under Brownian dynamics (\Eq{eq:Langevin}) in the force field $F$,
 and $\av{\ \cdot\ }_F$ corresponds to
averaging over all possible trajectories $\mathcal{C}$ with weight
$\mathcal{P}(\mathcal{C}|F)$. Time reversal
$(\mathcal{C},F)\mapsto (-\mathcal{C},F)$ changes the sign of the heat
produced along the trajectory, and thus connects dissipation and
irreversibility of the dynamics.  Interestingly, a similar expression
can be derived for the inflow rate~\cite{baiesi_inflow_2015}:
$G = \lim_{\tau\to\infty} \frac{1}{\tau} \av{ \log
  \mathcal{P}(\mathcal{C}|F)/ \mathcal{P}(\mathcal{-C}|-F)}_F$,
where $-F$ corresponds to the reversed force field. Indeed, the
operation $(\mathcal{C},F)\mapsto (-\mathcal{C},-F)$ now leaves the
heat unchanged, but reverses the sign of the divergence of the
force. At equilibrium, this corresponds to inverting the energy
landscape: for a typical trajectory that dwells in potential wells,
the reverse trajectory is atypical in the force field $-F$, as it
spends time around unstable maxima of energy. Finally, the capacity
can be expressed as
$4C = \lim_{\tau\to\infty} \frac{1}{\tau} \av{ \log
  \mathcal{P}(\mathcal{C}|F)/ \mathcal{P}(\mathcal{C}|-F)}_F$:
this operation reverses both heat and force divergence. Intuitively,
there is information about the force in a trajectory if it allows to
distinguish the force field from its reverse. More naturally, the
capacity quantifies the rate at which a trajectory becomes
distinguishable from force-free Brownian motion: indeed, it can be
written as
$C = \lim_{\tau\to\infty} \frac{1}{\tau} \av{ I(\mathcal{C})}_F$, where we define 
\begin{equation}
   I(\mathcal{C}) = \log \frac{ \mathcal{P}(\mathcal{C}|F)}{ \mathcal{P}(\mathcal{C}|0)}
  \label{eq:information}
\end{equation}
as the trajectory-wise information gain about the force field.

\section{Stochastic force inference}
\label{sec:SFI}

A trajectory of finite duration contains finite information,
quantified by \Eq{eq:information}. We now show how to use this
information in practice and reconstruct the force field through
Stochastic Force Inference (SFI).  In contrast with the drifted
Brownian motion, a spatially variable force field is in principle
characterized by an infinite number of degrees of freedom: the force
value at each point in space. With a finite trajectory, only a finite
number of combinations of degrees of freedom can be estimated. It is
therefore natural to approximate the force field as a linear
combination of a finite basis of $n_b$ known functions
$b = \{b_\alpha(\mathbf{x})\}_{\alpha = 1..n_b}$. The force can, in
principle, be approximated arbitrarily well by using a large enough
set of functions from a complete basis, such as polynomials or Fourier
modes. Alternatively, a limited number of functions might suffice if
an educated guess for the functional form of the force field can be
made. We propose to perform this approximation by projecting the force
field onto the space spanned by $b_\alpha(\mathbf{x})$ using the
steady-state probability distribution function $P$ as a measure. This
corresponds to a least-squares fit of the force field by linear
combinations of the $b_\alpha$'s. To this aim, we
define the projector
$c_\alpha(\mathbf{x}) = B^{-1/2}_{\alpha\beta} b_\beta(\mathbf{x})$,
where $B_{\alpha\beta}$ is an orthonormalization matrix such that
$\int c_\alpha c_\beta \dP = \delta_{\alpha\beta}$.  Our approximation
of the force field is then
$F_\mu(\mathbf{x}) \approx F_{\mu\alpha} c_\alpha(\mathbf{x})$ with
the projection coefficient
\begin{equation}
  \label{eq:F_moments}
  F_{\mu\alpha} = \int F_\mu(\mathbf{x}) c_\alpha(\mathbf{x}) \dP.
\end{equation}
This is akin to projecting the dynamics onto a finite-dimensional
sub-channel of capacity
$C_b = \frac{1}{4}D^{-1}_{\mu\nu} F_{\mu\alpha} F_{\nu\alpha} < C $.
Similarly, we can define the projection $v_{\mu\alpha}$ of the phase
space velocity. The corresponding entropy production
$\Sdot_b = D^{-1}_{\mu\nu} v_{\mu\alpha} v_{\nu\alpha}$ is then a
lower bound to the total entropy production. Interestingly, for a
system obeying Brownian dynamics (\Eq{eq:Langevin}) but where only a
subset of degrees of freedom can be observed, our framework gives the
force averaged over hidden variables, and provides a lower bound on
the entropy production limited to the observable
currents~\SI{sec:time}.

The projected force field has a finite number of degrees of freedom
$N_b = d n_b$, one per element of the $d \times n_b$ tensor
$F_{\mu\alpha}$, and corresponds to a finite capacity $C_b$. Inferring
the approximate force with a finite trajectory is thus in principle
possible if the information $I_b = \tau C_b > N_b$. However, the force
coefficients introduced in \Eq{eq:F_moments} are not directly
accessible from experimental data. Indeed, neither the force nor the
probability distribution function $P$ are known, the latter being also
required in the definition of the orthonormal projectors
$c_\alpha$. Instead, the available data is typically a discrete time
series $\mathbf{x}(t_i)$ of phase space positions, at sampling times
$t_i = i \Delta t$. We thus propose to estimate phase space averages
by discrete time integrals along the trajectory. The empirical
projectors are defined as
$\hat{c}_\alpha = \hat{B}^{-1/2}_{\alpha\beta} b_\beta$, with
$\hat{B}_{\alpha\beta} = \sum_i b_\alpha(\mathbf{x}(t_i))
b_\beta(\mathbf{x}(t_i)) \frac{\Delta t}{\tau}$.  Furthermore, the
force can be expressed in terms of a local It\^o average of
$\dot{\mathbf{x}}$~\cite{risken_fokker-planck_1996}: a local estimator
for the force at $\mathbf{x}(t_i)$ is thus
$\Delta \mathbf{x}(t_i)/\Delta t$, with
$\Delta \mathbf{x}(t_i) = \mathbf{x}(t_{i+1}) - \mathbf{x}(t_i)$.
Combining these two insights yields an operational definition for the
estimator of \Eq{eq:F_moments} in terms of a discrete It\^o
integral~\SI{sec:trajectory},
\begin{equation}
  \label{eq:F_moments_empirical}
  \hat{F}_{\mu\alpha} = \frac{1}{\tau} \sum_i \Delta x_\mu(t_i) \hat{c}_\alpha(\mathbf{x}(t_i)) 
\end{equation}
which is the discretized version of the It\^o integral
$\frac{1}{\tau}\int_0^\tau \hat{c}_\alpha(\mathbf{x}(t)) \dd
x_\mu(t)$.  Indeed, discretizing \Eq{eq:Langevin} yields
$\Delta \mathbf{x}(t_i) = \mathbf{F}(\mathbf{x}(t_i)) \Delta t +
\sqrt{2\mathbf{D}}\Delta \xi_i$, where $\Delta \xi_i$ is independent
of $\mathbf{x}(t_i)$: in the long trajectory limit, the main
contribution comes from the force, while the noise averages to
zero. Equation~\ref{eq:F_moments_empirical} corresponds to a linear
regression of the local force estimator, previously suggested for
one-dimensional systems~\cite{comte_penalized_2007}, and coincides
with the maximum-likelihood estimator of the force projection
coefficients.  The typical squared relative error on the inferred
coefficients due to the diffusive noise can be estimated in practice
as
$\delta \hat{F}^2 / \hat{F}^2 \sim
N_b/2\hat{I}_b$~\SI{sec:trajectory}, where
$\hat{I_b} = \frac{\tau}{4}D^{-1}_{\mu\nu} \hat{F}_{\mu\alpha}
\hat{F}_{\nu\alpha}$ is the empirical estimate of information
contained in the trajectory. This formula indicates that again, in
order to resolve the force coefficients, the information in the
trajectory should exceed the number of inferred parameters.  Another
source of error stems from the fact that the force varies over a
finite time step $\Delta t$; we provide an estimator for the magnitude
of the resulting bias in~\App{sec:noise}.

\begin{figure}[b]
  \centering
  { \includegraphics[width=1.\columnwidth]{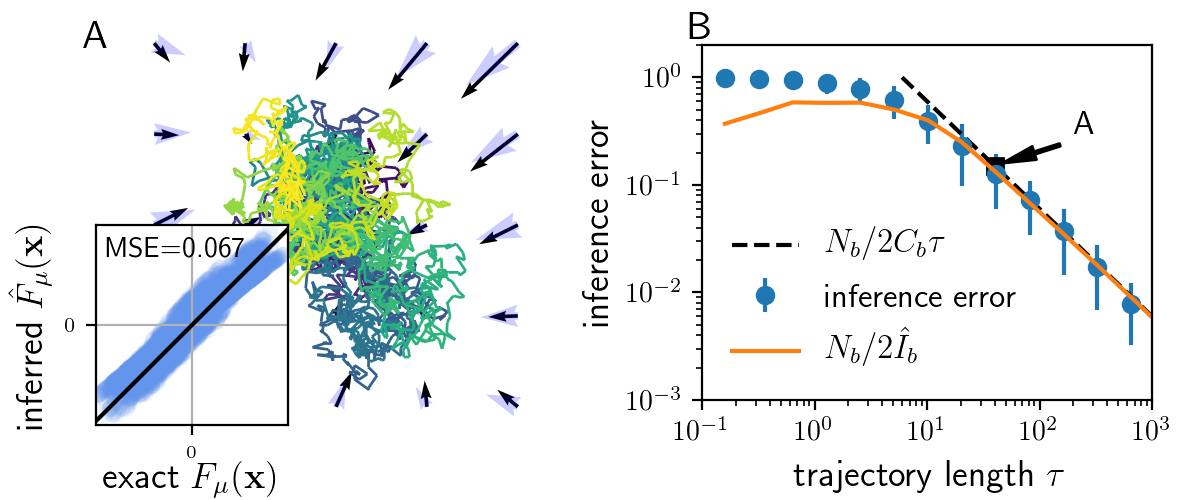}}
  \caption{ Stochastic force inference for a 2D Ornstein-Uhlenbeck
    process, with force field
    $F_\mu(\mathbf{x}) = -\Omega_{\mu\nu} x_\nu$ and isotropic
    diffusion. \textbf{A.} An example trajectory. The inferred force
    field for this trajectory, using SFI with functions
    $b=\{1,x_\mu\}$ (blue arrows), is compared to the exact force
    field (black arrows). \emph{Inset:} the inferred force components
    along the trajectory \emph{versus} the exact force components,
    with normalized mean-squared error (MSE).  \textbf{B.} The average of the
    relative error
    $[(\hat{F}_{\mu\alpha}-F^{\tau}_{\mu\alpha})D^{-1}_{\mu\nu}(\hat{F}_{\nu\alpha}-F^{\tau}_{\nu\alpha})]/[\hat{F}_{\mu\alpha}D^{-1}_{\mu\nu}\hat{F}_{\nu\alpha}]$
    on the inferred projection coefficients $\hat{F}_{\mu\alpha}$ and
    its self-consistent estimate $N_b/2\hat{I}_b$ both converge to
    $N_b/2I_b$, as expected from theory~\SI{sec:trajectory}. Here
    $F^{\tau}_{\mu\alpha} = \int F_\mu(\mathbf{x}(t))
    \hat{c}_\alpha(\mathbf{x}(t)) \frac{\dd t}{\tau}$ is the
    projection of the exact force on the empirical projectors. }
  \label{fig:2dOU}
\end{figure}

We now demonstrate the utility of our method using simulated data of
simple models. The simplest spatially varying force field is a
harmonic trap, \emph{i.e.} an Ornstein-Uhlenbeck process (\Fig{fig:2dOU}). We
benchmark our method by using a first-order polynomial basis,
$b=\{1,x_\mu\}$, which can capture the exact force field. The 2D
trajectory displayed in \Fig{fig:2dOU}A has an information content of $I=27.6$
bits, while this linear channel has $N_b=6$ degrees of freedom,
allowing precise inference of the projected force field (\Fig{fig:2dOU}A). Indeed, the squared relative error on the force coefficients is
$0.15$; this is consistent with the operational estimate of this
error, $N_b/2\hat{I_b} = 0.16$. The force along the trajectory is thus
inferred to a good approximation (\Fig{fig:2dOU}A, \emph{inset}). Furthermore,
the projected force field
$\hat{F}_{\mu\alpha} \hat{c}_\alpha(\mathbf{x})$ provides an ansatz
that can be extrapolated beyond the trajectory (\Fig{fig:2dOU}A), which works
equally well here as the functional form of the force field is fully
captured by our choice of basis. More quantitatively, we confirm the
predicted behavior for the squared relative error by studying an
ensemble of trajectories (\Fig{fig:2dOU}B).

In the case of out-of-equilibrium Brownian systems, our method also
permits the approximation of phase space currents and entropy
production. Indeed, the phase space velocity $\mathbf{v}$ can be
expressed in terms of a local Stratonovich average of
$\dot{\mathbf{x}}$, reflecting the fact that it is odd under time
reversal~\cite{chetrite_eulerian_2009}. Our estimator for the
projection coefficients of the phase space velocity is thus~\SI{sec:entropy_production}
\begin{equation}
  \hat{v}_{\mu\alpha} = \frac{1}{\tau} \sum_i \Delta x_\mu(t_i)\hat{c}_\alpha\left(\frac{\mathbf{x}(t_{i+1}) +
    \mathbf{x}(t_i)}{2}\right) \label{eq:v}
\end{equation}
which is the discretized version of the Stratonovich integral
$\frac{1}{\tau}\int_0^\tau \hat{c}_\alpha(\mathbf{x}(t)) \circ \dd
x_\mu(t) $. This allows the inference of the entropy production rate
$\hat{\Sdot}_b = D^{-1}_{\mu\nu} \hat{v}_{\mu\alpha}
\hat{v}_{\nu\alpha}$ associated to the observed currents.  This is a
biased estimator of the entropy production, with an error that can be
self-consistently controlled as
$\hat{\Sdot}_b = \Sdot_b + 2N_b/\tau +
O((2\hat{\Sdot}_b/\tau+(2N_b/\tau)^2)^{1/2})$: the entropy production
rate in the channel can thus be inferred using a single trajectory
provided that several $k_B$'s per degree of freedom have been
dissipated.

\begin{figure}[t]
  \centering
{ \includegraphics[width=1.\columnwidth]{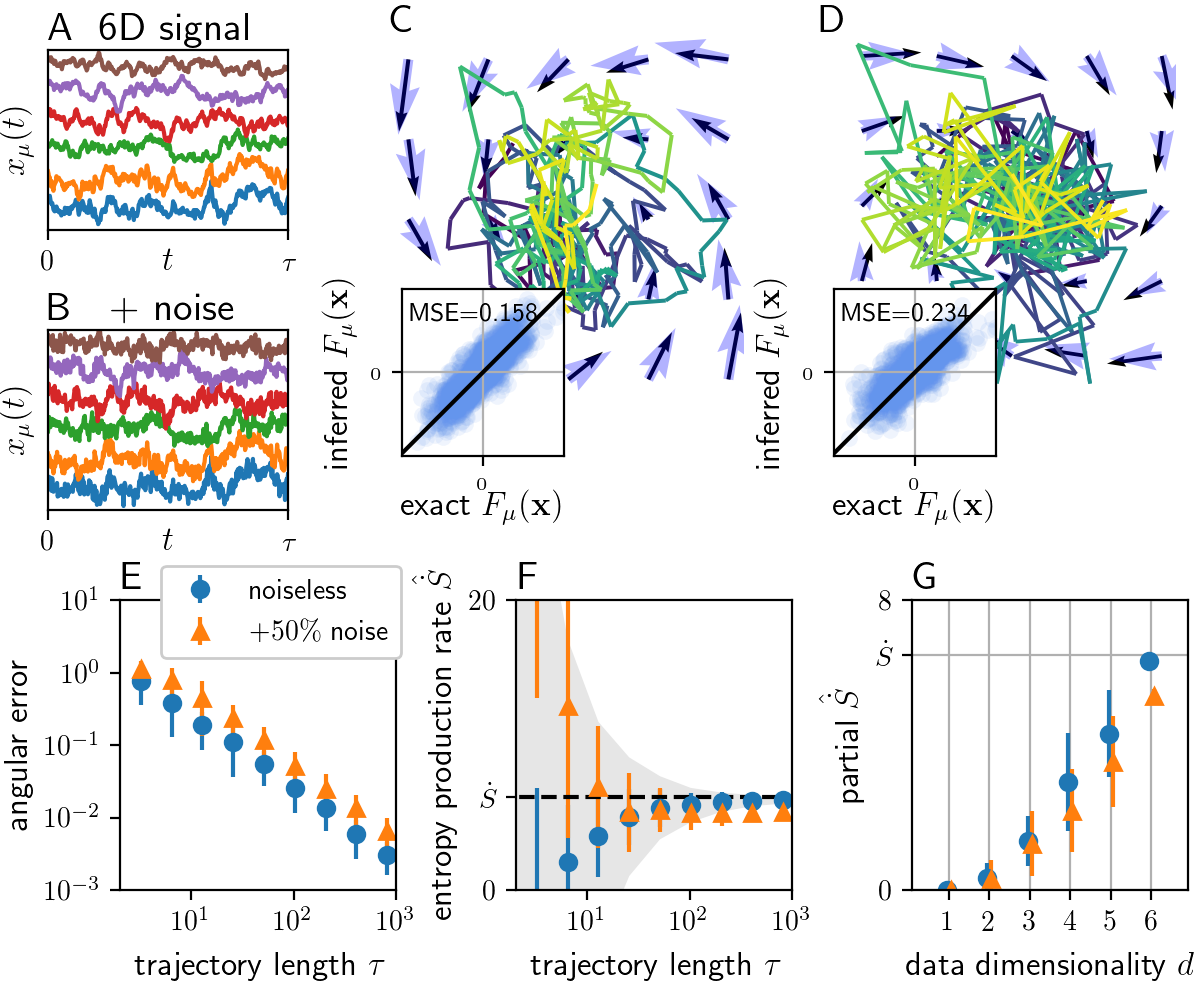}}
\caption{\textbf{A.}  Time series of a 6D out-of-equilibrium
  Ornstein-Uhlenbeck process, with anisotropic harmonic confinement
  and diffusion tensor, and circulation. The force field is
  $F_\mu(\mathbf{x}) = -\Omega_{\mu\nu} x_\nu$. The matrix $\Omega$
  and the diffusion matrix are chosen from a random ensemble. The
  antisymmetric part of $D^{-1}\Omega$ has rank 2, thus inducing
  circulation in a randomly chosen plane.  \textbf{B.}  The same
  trajectories as in {D}, with additional time-uncorrelated
  measurement noise.  \textbf{C.} SFI for the trajectory in {A} allows
  precise identification of the plane of circulation and
  reconstruction of the force along the trajectory. \textbf{D.} SFI
  applied to the trajectory in {B}, with measurement noise. It can
  still detect forces accurately. \textbf{E.}  Convergence of the
  angular error for cycle detection with increasing trajectory length,
  for the process shown in {D-E}. \textbf{F.}  Inferred entropy
  production rate for this process, with and without measurement noise
  (we subtracted here the systematic bias $2N_b/\tau$). The shadowed
  area indicates the self-consistent confidence interval for the
  inferred entropy production. The dotted line shows the exact value
  of the entropy produced; for the noisy process SFI underestimates
  this value due to blurring of the currents. \textbf{G.} Entropy
  production captured when observing a $d$-dimensional projection of
  the trajectory, averaged over direction of observation, for long
  trajectories.  In plots {E,F,G}, error bars indicate standard
  deviation over an ensemble of $32$ trajectories. Parameters of the
  simulations are presented in Appendix~\ref{sec:simus}. }
  \label{fig:6D}
\end{figure}

The simplest structure for phase space currents corresponds to cyclic
circulation around a point. The detection of such features in active biological
systems has been the focus of a number of recent studies, which employ
phase space
coarse-graining~\cite{battle_broken_2016,gnesotto_broken_2018,seara_entropy_2018}. This
method is however limited to low-dimensional systems, and even then
requires large amounts of data: indeed, the capacity per degree of
freedom is low, as each grid cell is visited infrequently. In
contrast, our method provides a way to detect circulation in any
dimension with minimal data. Using the centered linear basis
$b_\alpha(\mathbf{x})=\bar{x}_\alpha=x_\alpha-\int x_\alpha\frac{\dd
  t}{\tau}$, we can infer the velocity coefficients
$\hat{v}_{\mu\alpha}$, which have a matrix structure. This matrix
reads $\hat{v}_{\mu\alpha}=C^{-1/2}_{\alpha\beta} A_{\beta\mu}$, where
$C_{\mu\nu} = \int \bar{x}_\mu \bar{x}_\nu\frac{\dd t}{\tau}$ is the
covariance matrix, and the antisymmetric part of $A_{\mu\nu} $ is
$A_{\{\mu\nu\}} = \frac{1}{2\tau}\int \bar{x}_\mu\dd x_\nu -
\bar{x}_\nu\dd x_\mu$, which is the rate at which the process
encircles area in the $(\mu,\nu)$
plane~\cite{ghanta_fluctuation_2017,gonzalez_experimental_2019}. This
rate, sometimes called probability angular
momentum~\cite{shkarayev_exact_2014,zia_manifest_2016}, intuitively
quantifies circulation and closely connects to cycling
frequencies~\cite{gladrow_broken_2016,mura_nonequilibrium_2018}. Indeed,
the eigenvectors of $A_{\{\mu\nu\}}$ can be used to define cycling
planes~\SI{sec:simus}. The entropy production rate due to cycling reads
$\hat{\Sdot}_b = D^{-1}_{\mu\nu} A_{\nu\rho} C^{-1}_{\rho\sigma}
A_{\sigma\mu}$.

We demonstrate the potency of our cycle-detection method on a
challenging dataset: a short trajectory of an out-of-equilibrium
Ornstein-Uhlenbeck process in dimension $d=6$ (\Fig{fig:6D}A), which
is equivalent to popularly used bead-spring
models~\cite{battle_broken_2016,mura_nonequilibrium_2018,li_quantifying_2019}. Our
method identifies the principal circulation plane accurately, together
with the force field (\Fig{fig:6D}C). Quantitatively, we demonstrate
that the angular error in the identification of this plane vanishes
with increasing trajectory length (\Fig{fig:6D}E), concomitant with
the convergence of $\hat{\Sdot}_b$ to the exact value
(\Fig{fig:6D}F). The entropy production inferred is associated to the
observable currents: if only a fraction of the degrees of freedom can
be observed, $\hat{\Sdot}_b$ is a lower bound to the total entropy
production of the system (\Fig{fig:6D}G), as some currents are not
observable. In particular, if only one degree of freedom can be
measured, this technique will yield $\hat{\Sdot}_b = 0$; alternative
techniques based on the non-Markovianity of the dynamics are better
suited to inferring entropy production in this
case~\cite{roldan_arrow_2018}.

A major challenge in the inference of dynamical properties of
stochastic systems from real data is time-uncorrelated measurement
noise, which dominates time derivatives of the signal. Indeed, in our
inference scheme, \Eq{eq:F_moments_empirical} is highly sensitive to
such noise. In contrast, the time-reversal antisymmetry of the
velocity coefficients $\hat{v}_{\mu\alpha}$ makes them robust against
measurement noise~\SI{sec:noise}. Exploiting this symmetry, we obtain
an unbiased estimator for the force by using the relation between
It\^o and Stratonovich integration,
\begin{equation}
  \hat{F}_{\mu\alpha} = \hat{v}_{\mu\alpha} + D_{\mu\nu}\hat{g}_{\nu\alpha}
\label{eq:F_corrected_estimator}
\end{equation}
where
$\hat{g}_{\mu\alpha} = - \sum_i \frac{\Delta t}{\tau}\partial_\mu
\hat{c}_\alpha(\mathbf{x}(t_i))$ is an estimator for the projection of
$g_\mu=\partial_\mu \log P$ onto the basis (note that while
$\hat{g}_\mu(\mathbf{x}) \equiv
\hat{g}_{\mu\alpha}\hat{c}_\alpha(\mathbf{x})$ is an estimate of
$\partial_\mu\log P(\mathbf{x})$, it is not a gradient, and thus
cannot be integrated to estimate $P(\mathbf{x})$). The modified
estimator proposed in \Eq{eq:F_corrected_estimator} can only be computed if the
projection basis is smooth, and would not apply to grid
coarse-graining, for instance. It requires knowledge
of the diffusion tensor $D_{\mu\nu}$, as discussed in
\Sec{sec:inhomogeneous}. Using this modified force estimator
allows precise reconstruction of the force field, circulation and
entropy production even in the presence of large measurement noise
(\Fig{fig:6D}B,D-G). The limiting factor on force inference due to
measurement noise then becomes the blurring of the spatial structure
of the process. For observations with a finite time step $\Delta t$,
the currents are also blurred by time discretization, introducing an
additional bias in the force estimator~\SI{sec:noise}, and resulting
in an underestimate of the entropy production. Note however that this
finite $\Delta t$ effect only induces a bias on $\hat{v}_{\mu\alpha}$:
for an equilibrium, time-reversible process,
$\hat{v}_{\mu\alpha}\to 0$ and the force estimator reduces to
$D_{\mu\nu}\hat{g}_{\nu\alpha}$, which is independent of the
time-ordering of the data.

\begin{figure}[t!]
  \centering
  \includegraphics[width=\columnwidth]{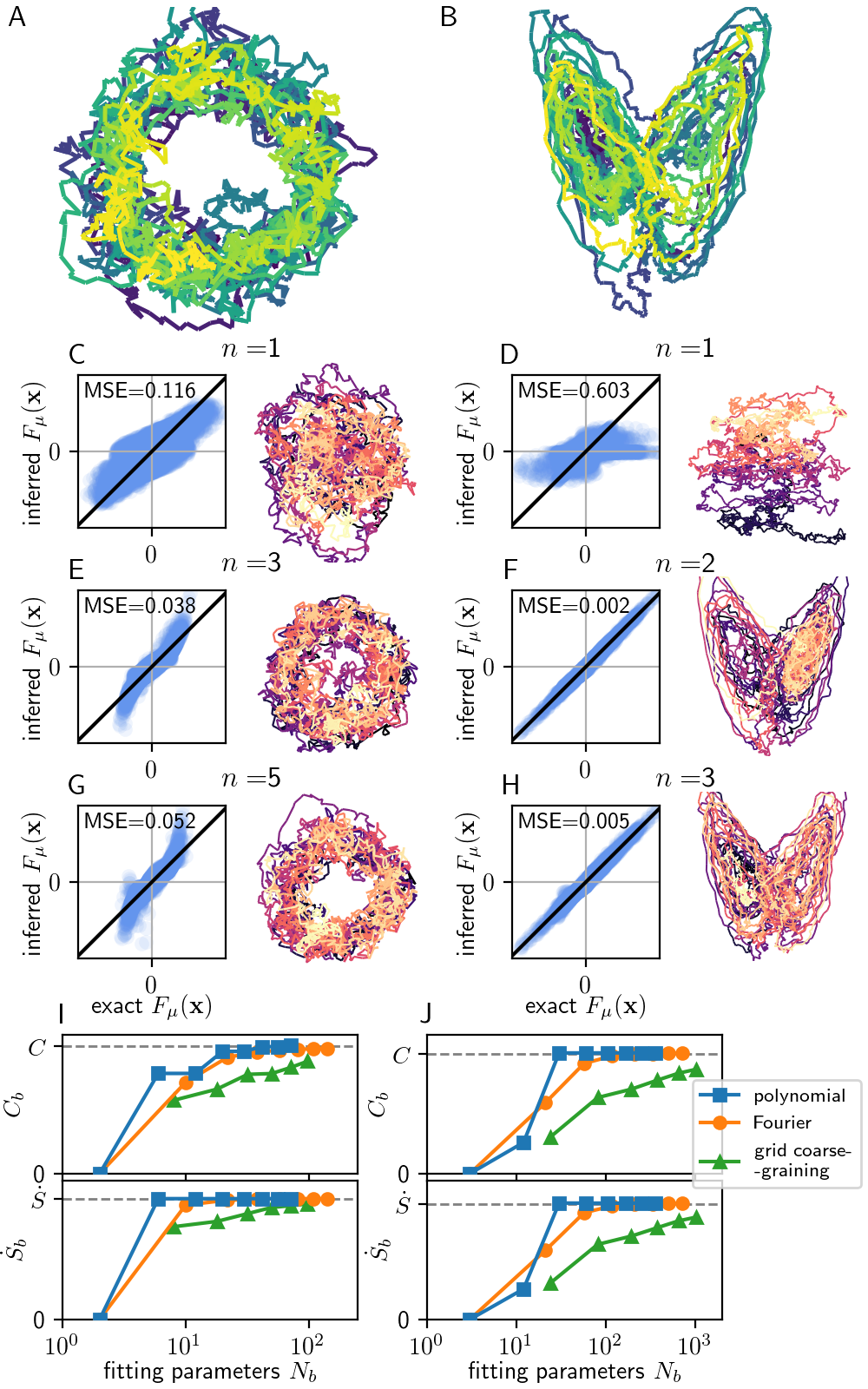}
  \caption{Stochastic force inference with non-linear force
    fields. \textbf{A.} Trajectory of an out-of-equilibrium process
    with harmonic trapping and circulation, and a Gaussian repulsive
    obstacle in the center. The force field is given by
    $F_\mu(\mathbf{x}) = -\Omega_{\mu\nu} x_\nu + \alpha
    e^{-x^2/2\sigma^2} x_\mu$ where $\Omega$ has both a symmetric and
    antisymmetric part. \textbf{B.} Trajectory of the stochastic
    Lorenz process, a 3D process with a chaotic attractor. The force
    field is $F_x = s(y-x),\ F_y = rx-y-zx,\ F_z = xy - bz$, where we
    choose $r=10$, $s=3$, and $b=1$. \textbf{C-H.} SFI for these two
    trajectories, respectively with polynomials of order $n=1,3,5$ and
    $n=1,2,3$: inferred force versus exact force (\emph{left}) and
    bootstrapped trajectory using the inferred force field
    (\emph{right}). \textbf{I-J.} Capacity (top) and entropy
    production (bottom) of each process projected on different bases
    for an asymptotically long trajectory, as a function of the number
    of degrees of freedom $N_b$ in the basis. These bases are
    polynomial and Fourier functions with order $n=0\dots 7$, and a
    coarse-grained approximation with a variable number of grid cells
    $n=2 \dots 7$ in each dimension. Parameters and details of the
    simulations are presented in Appendix~\ref{sec:simus}. }
  \label{fig:nonlinear}
\end{figure}

We have so far considered only the case of linear systems projected
onto linear functions. In general, force fields are nonlinear, which
can result in a complex spatial structure. We illustrate this in \Figs{fig:nonlinear}A-B for processes with, respectively, non-polynomial forces and a
complex attractor~\cite{allawala_statistics_2016}.  For such
processes, SFI with a linear basis captures the covariance of the data and
the circulation of its current. However, it fails to reproduce finer
features, as evident by inspecting bootstrapped trajectories generated
using the inferred force field (\Fig{fig:nonlinear}C-D).  A better approximation of
the force can be obtained by expanding the projection basis, for
instance by including higher-order polynomials
$\{x_\mu x_\nu\},\{x_\mu x_\nu x_\rho\}\dots$ (\Fig{fig:nonlinear}E-H) or Fourier
modes. The captured fraction of the capacity and entropy production
increases monotonically when expanding the basis (\Fig{fig:nonlinear}I-J),
corresponding to finer geometrical details: the force field is well
resolved if the measured capacity does not increase upon further
expansion of the basis.  However, expanding the basis also results in
an increase in the number of parameters to infer, which eventually
leads to overfitting. 

\begin{figure}[t!]
  \centering 
  \includegraphics[width=\columnwidth]{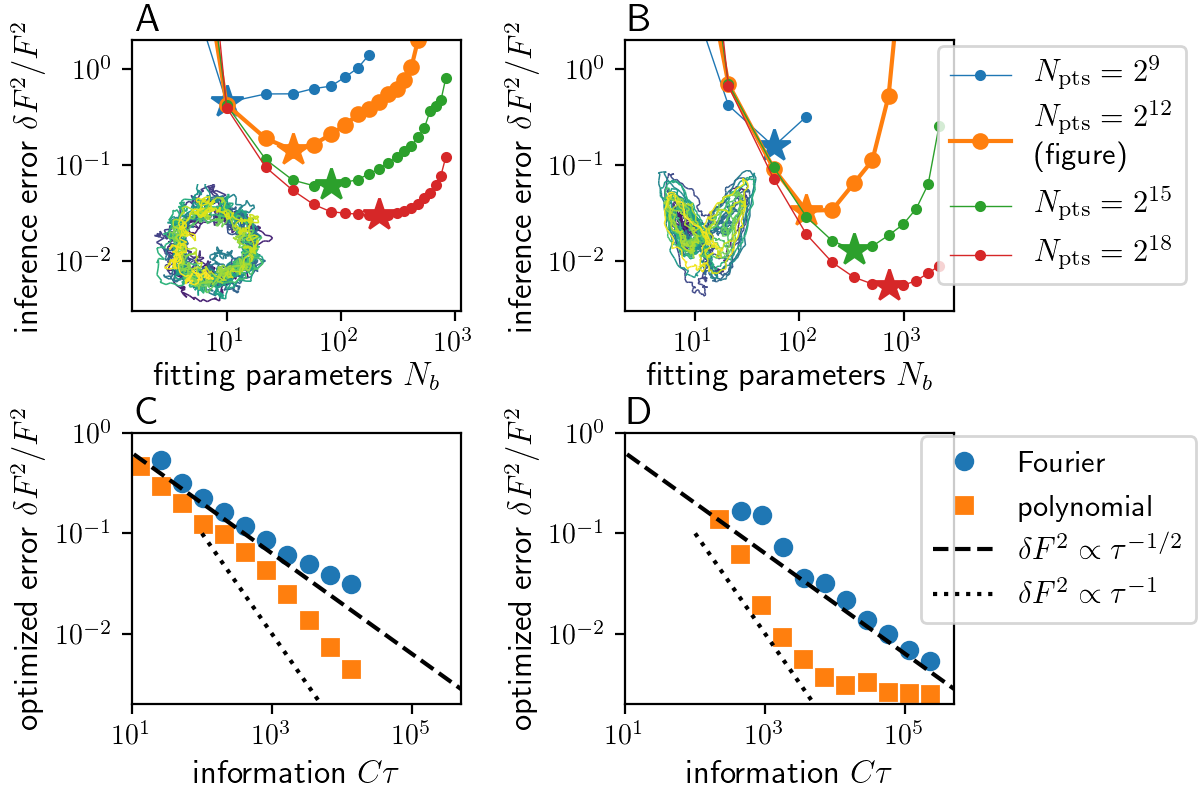}
  \caption{ Influence of the size of the basis on the precision of
    SFI. \textbf{A-B.} SFI error as a function of the number of fit
    parameters, respectively for the models presented in
    \Fig{fig:nonlinear}A-B, with a Fourier basis, and for different
    numbers of time steps in the trajectory.  Specifically, the
    $y$-axis is the mean squared relative error on the inferred force
    along the trajectory,
    $\smallav{(\hat{F}_\mu-F_\mu) D^{-1}_{\mu\nu} (\hat{F}_\nu-F_\nu)}
    / \smallav{\hat{F}_\mu D^{-1}_{\mu\nu} \hat{F}_\nu}$. The
    crossover from under- to overfitting is apparent, and takes place
    at larger $N_b$ and lower error with longer trajectories. The star
    symbols indicate the optimal basis size predicted by our
    self-consistent criterion of maximizing
    $\hat{I}_b - \delta \hat{I}_b$. \textbf{C-D}. The squared error as
    a function of the amount of information $C\tau$ in a trajectory of
    duration $\tau$, for the optimal basis, averaged over $n=3$
    trajectories. For the Lorenz process with a polynomial basis (D,
    orange squares), the convergence is fast as the basis is adapted
    to the exact force field, and the saturation of the error to a
    lower plateau is due to the finite time step (see \App{sec:noise}).  }
  \label{fig:overfitting}
\end{figure}

For a finite trajectory, there is therefore a trade-off between the
precision of the inferred force and the completeness of the force
field representation. This is demonstrated in \Fig{fig:overfitting}A-B
by plotting the force inference error along the trajectory as a
function of the number $N_b$ of degrees of freedom in the basis. At
small $N_b$, this error decreases, as it mostly originates from
underfitting. At large $N_b$, the error increases, as all
statistically significant information is already captured and adding
new functions primarily fits the noise.  This is reflected in the
inferred information $\hat{I}_b$ which steadily increases with the
number of fitting parameters $N_b$: the increase is initially mainly
due to the increase in the captured information $I_b$, but as $N_b$
grows, so does the typical error on $\hat{I}_b$,
$\delta \hat{I}_b \approx \sqrt{2\hat{I}_b + N_b^2/4 }$ (see
\App{app:information_error}), and this error eventually overwhelms the
gain in $I_b$.  As a practical criterion to optimize between under-
and overfitting and best estimate the force along the trajectory, we
thus propose to use the basis $b$ which maximizes the information
$I_b$ that can be statistically resolved.  In practice, we find that
choosing the basis size that maximizes $\hat{I}_b - \delta \hat{I}_b$
(\emph{i.e.} the inferred information minus one standard deviation)
robustly selects the optimal basis size for a given trajectory (star
symbols on \Fig{fig:overfitting}A-B).  An alternative optimization
procedure, based on a similar balance, was suggested
in~\cite{comte_penalized_2007} for one-dimensional processes. We
empirically observe that when using this criterion to adapt the basis
to the trajectory, the typical squared error on force inference scales
as $\tau^{-1/2}$ with the trajectory duration $\tau$
(\Fig{fig:overfitting}C-D). There is an exception to this scaling:
when the force field can be exactly represented by a finite number of
functions of the basis, such as the Lorenz process with order 2
polynomials, this same criterion selects the smallest adapted basis:
further adding functions does not resolve more information. This
results in a faster convergence of the force field as $\tau^{-1}$
(\Fig{fig:overfitting}D), which is the rate of convergence of the
force projections for a given basis size.

\begin{figure}[t]
  \centering
  \includegraphics[width=\columnwidth]{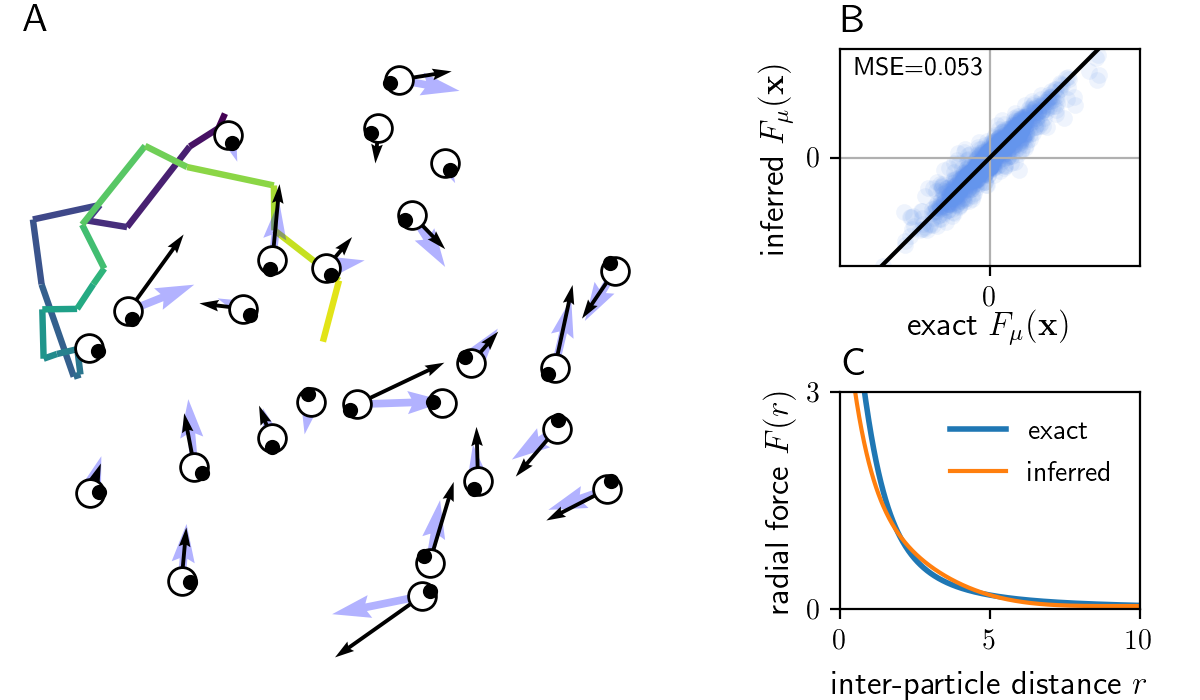}
  \caption{Stochastic force inference for harmonically trapped
    active Brownian particles with soft repulsive interactions
    $F(r) = 1/(1+r^2)$ between particles at distance $r$. \textbf{A.}
    Snapshot of a configuration for $25$ active particles. The black
    dots indicate the direction of self-propulsion. We perform SFI on
    a trajectory of only $25$ frames, blurred to mimic measurement
    noise. Background shows the trajectory of one particle, and force
    on each particle, inferred (blue arrows) and exact (black
    arrows). The fitting basis for SFI consists in a combination of
    harmonic trapping, constant velocity self-propulsion and radial
    interactions between particles with the form $r^k e^{-r/r_0}$ with
    $k = 0...5$ and $r_0$ a typical nearest-neighbour distance between
    particles. \textbf{B.} Inferred \emph{versus} exact components of
    the force on all particles along the trajectory. \textbf{C.}
    Inferred radial force between interacting particles, compared to
    the exact force.}
  \label{fig:particles}
\end{figure}

\begin{figure}[ht]
  \centering
  \includegraphics[width=\columnwidth]{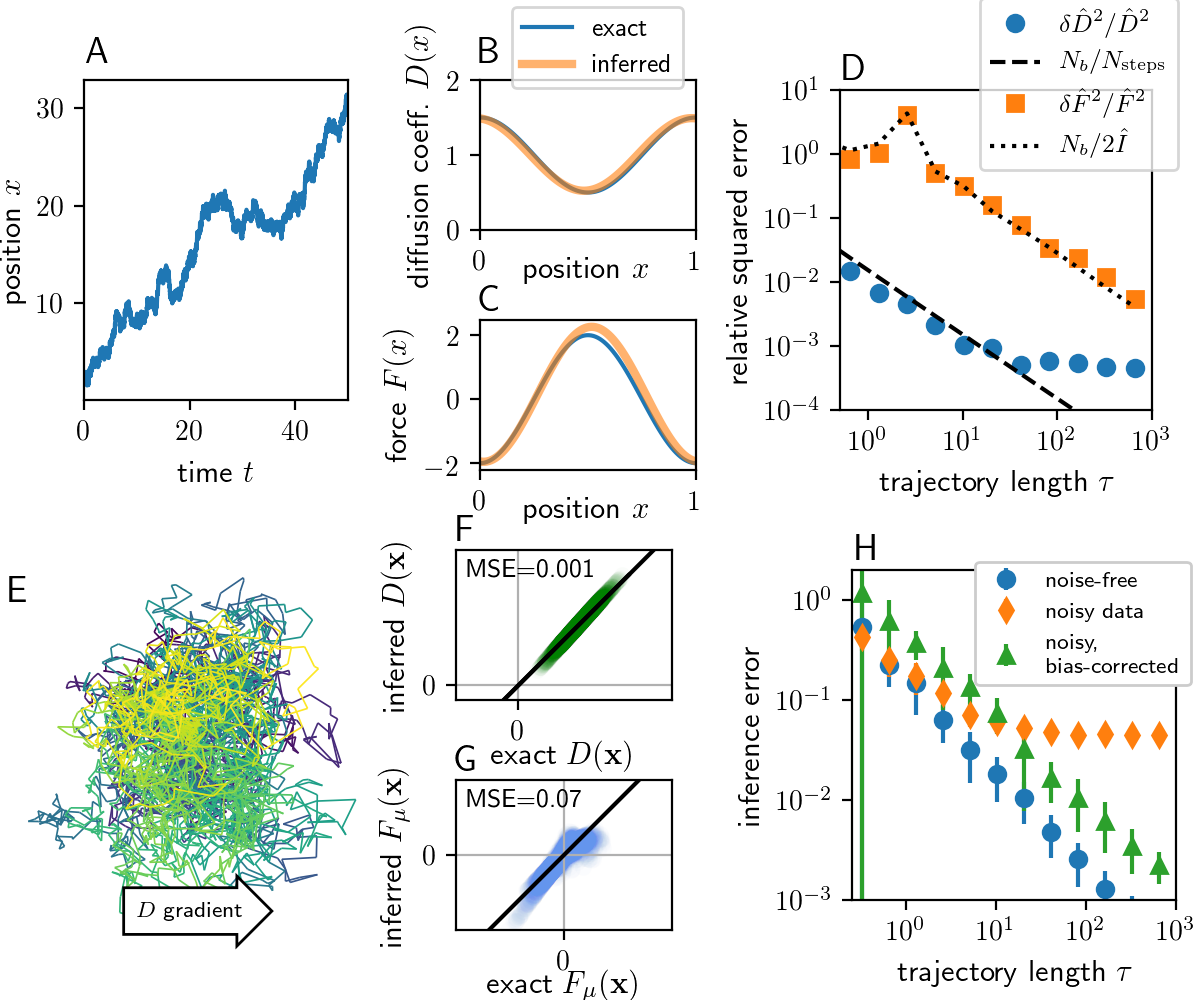}
  \caption{Stochastic inference of inhomogeneous diffusion and
    forces. \textbf{A.} A trajectory of a 1D ratchet model with
    $F(x)= F_0\cos(2\pi x)$ and $D(x) = 1 + a \cos(2\pi x)$, with
    periodic boundary conditions. \textbf{B-C} For the trajectory
    presented in A, inferred and exact diffusion coefficient (using
    \Eq{eq:D}) and force field (using \Eq{eq:F_D}) as a function of
    position. We use a $1$st order Fourier basis to infer both force
    and diffusion.  \textbf{D.} Analysis of the convergence of the
    diffusion (blue) and force (orange) estimators, as a function of
    trajectory duration, for the process presented in A. The dotted
    and dashed black lines are the self-consistent estimates for the
    squared error, respectively for the diffusion and the force. The
    plateau for the diffusion inference is due to the finite time
    step. \textbf{E.} A trajectory of a minimal 2D model, an isotropic
    harmonic trap at equilibrium,
    $F_\mu (\mathbf{x}) = - D_{\mu\nu}(\mathbf{x}) x_\nu$, in a
    constant gradient of isotropic diffusion,
    $D_{\mu\nu}(\mathbf{x}) = (1 + a_\rho x_\rho) \delta_{\mu\nu}$.
    \textbf{F-G.} Inferred \emph{versus} exact diffusion coefficient
    (using \Eq{eq:D}) and force components (using \Eq{eq:F_D}) along
    trajectory A. A linear polynomial basis was used to fit the
    diffusion coefficient, and a quadratic basis to fit $F_{\mu}$.
    \textbf{D.} Convergence of the diffusion projection estimator
    (normalized by the average diffusion tensor) to its exact value
    for the process shown in A. Circles: using \Eq{eq:D}, diamonds:
    using \Eq{eq:D} in the presence of time-uncorrelated measurement
    noise; triangles: using the bias-corrected local estimator. Error
    bars represent the standard deviation over $64$
    samples.  Details and parameters in Appendix~\ref{sec:simus}. }
  \label{fig:diffusion}
\end{figure}

Systems with many degrees of freedom, such as active interacting
particles (\Fig{fig:particles}A), are challenging to treat. Indeed,
with limited data, the criterion $\hat{I}_b \gg N_b$ precludes even
the inference of gross features of the force field. In such cases
however, the use of symmetries can make the problem tractable. For
instance, treating particles as identical implies that forces are
invariant under particle exchange, which greatly reduces the number of
parameters to infer. Forces can then be expanded as one-particle
terms, pair interactions, and higher orders, by choosing an
appropriate basis (see \App{sec:ABP}). With this scheme, a large
number of particles actually results in enhanced statistics, allowing
accurate inference of the force components (\Fig{fig:particles}A-B)
and reconstruction of the pair interactions (\Fig{fig:particles}C)
with a limited amount of data. This method could be straightforwardly
extended to include, \emph{e.g.}, alignment interactions between
particles. In contrast to standard methods to infer pair interaction
potentials, we do not rely here on an equilibrium assumption.

\section{Inhomogeneous diffusion}
\label{sec:inhomogeneous}

We have so far assumed that the diffusion tensor does not depend on
the state of the system. While this is a natural first approximation,
it is rarely strictly the case: for instance, the mobility of colloids
depends on their distance to walls and other colloids due to
hydrodynamic interactions~\cite{lau_state-dependent_2007}. In order to
mathematically describe Brownian dynamics in the presence of an
inhomogeneous diffusion tensor $D_{\mu\nu}(\mathbf{x})$,
\Eq{eq:Langevin} should be modified into
\begin{equation}
  \dot{x}_\mu = \Phi_\mu(\mathbf{x})  +  \sqrt{2D(\mathbf{x})}_{\mu\nu} \xi_\nu,
  \label{eq:Langevin_inhomogenous}
\end{equation}
written in the It\^o convention, \emph{i.e.} evaluating
$\mathbf{D}(\mathbf{x})$ at the start of the step. Here $\Phi_\mu$ is
the drift, which relates to the physical force through
\begin{equation}
  \label{eq:drift}
  \Phi_\mu(\mathbf{x}) = F_\mu(\mathbf{x}) + \partial_\nu
  D_{\mu\nu}(\mathbf{x}).
\end{equation}
The additional term $\partial_\nu D_{\mu\nu}$, sometimes called
``spurious force'', combines with the noise term to ensure that the
dynamics does not induce currents and probability gradients in the
absence of forces~\cite{lau_state-dependent_2007}. To our knowledge,
the only way to infer the physical force is to infer both terms in
\Eq{eq:drift} independently, and involves taking gradients of the
inferred diffusion. Here we show how to infer both the diffusion field
and the drift field, following the same idea as in \Sec{sec:SFI}.

We propose to approximate $D_{\mu\nu}(\mathbf{x})$ by its projection
as a linear combination of known functions,
$D_{\mu\nu}(\mathbf{x}) \approx D_{\mu\nu\alpha} c_\alpha(\mathbf{x})$
with
$ D_{\mu\nu\alpha} = \int D_{\mu\nu}(\mathbf{x}) c_\alpha(\mathbf{x})
\dP$.  As before, we can estimate the projectors $\hat{c}_\alpha$
using trajectory averages; the only missing ingredient is a local
estimate $\hat{d}_{\mu\nu}(t_i)$ for the diffusion tensor
$D_{\mu\nu}(\mathbf{x}(t_i))$. Such an estimator can be constructed as
$\hat{d}_{\mu\nu}(t_i) = \Delta x_\mu(t_i) \Delta x_\nu(t_i) / 2\Delta t$, so that our
estimator for $D_{\mu\nu\alpha}$ reads
\begin{equation}
  \hat{D}_{\mu\nu\alpha} = \frac{1}{\tau} \sum_i \hat{d}_{\mu\nu}(t_i) \hat{c}_\alpha (\mathbf{x}(t_{i})) \Delta t  
  \label{eq:D}
\end{equation}
The relative error on these projection coefficients is of order
$\sqrt{N_b \Delta t/\tau}$~\SI{sec:D}. Similarly to \Eq{eq:F_moments_empirical} for the force field, \Eq{eq:D} corresponds to a linear regression of $\hat{d}_{\mu\nu}(t_i)$, and was previously suggested for one-dimensional systems in~\cite{comte_penalized_2007}. We test this estimator using
two minimal models: a one dimensional ratchet process with sinusoidal
force and diffusion coefficient, inspired by the B\"uttiker-Landauer
model~\cite{buttiker_transport_1987,landauer_motion_1988}
(\Fig{fig:diffusion}A-D); and a two-dimensional process in a harmonic
trap with a constant diffusion gradient (\Fig{fig:diffusion}E-H). We
quantitatively recover the diffusion coefficient as a function of
position (\Fig{fig:diffusion}B,F) and confirm that the error vanishes
in the limit of long trajectories
(\Fig{fig:diffusion}D,H). Importantly, the estimator introduced in
\Eq{eq:D} is biased in the presence of noise on the measured
$\mathbf{x}$, and becomes effectively useless if this noise is larger
than the typical $\Delta \mathbf{x}$. Inspired by the estimator
proposed by Vestergaard \emph{et al.}~\cite{vestergaard_optimal_2014}
for homogeneous, isotropic diffusion, we define a bias-corrected local
estimator
\begin{equation}
 \hat{\mathbf{d}}(t_i) = \frac{(\Delta \mathbf{x}(t_{i-1}) +\Delta
\mathbf{x}(t_i))^2}{ 4\Delta t} +  \frac{\Delta \mathbf{x}(t_i) \Delta \mathbf{x}(t_{i-1})}{ 2\Delta t}
\label{eq:Vestergaard}
\end{equation}
where tensor products are implied.  Modifying \Eq{eq:D} accordingly
thus corrects measurement noise bias (\Fig{fig:diffusion}H), at the
price of an increased relative error for short
trajectories~\SI{sec:D}.

We also approximate the drift as a linear combination of functions,
$\Phi_\mu(\mathbf{x}) = \Phi_{\mu\alpha}
c_\alpha(\mathbf{x})$. Equation~\ref{eq:F_moments_empirical} provides
an estimator for the projection coefficients $\Phi_{\mu\alpha}$ in
terms of an It\^o integral. This estimator is however impractical for
experimental data, as even moderate measurement noise induces large
errors in these coefficients. As in \Eq{eq:F_corrected_estimator}, we
exploit the It\^o-to-Stratonovich conversion to obtain an estimator
that is not biased by measurement noise:
\begin{equation}
  \label{eq:phi_strato}
    \hat{\Phi}_{\mu\alpha} = \hat{v}_{\mu\alpha} - \frac{1}{\tau} \sum_i  \hat{d}_{\mu\nu}(t_i)  \partial_\nu 
\hat{c}_\alpha(\mathbf{x}(t_i)) \Delta t
\end{equation}
where $\hat{v}_{\mu\alpha}$ is the velocity projection coefficient
(\Eq{eq:v}), and $\hat{d}_{\mu\nu}(t_i)$ can either be the
local biased-corrected estimator (\Eq{eq:Vestergaard}) or another
estimator of $D_{\mu\nu}(\mathbf{x}_i)$. The convergence properties of
$\hat{\Phi}_{\mu\alpha}$ to its asymptotic value are similar to those
of \Eq{eq:F_moments_empirical}.

We can now combine our diffusion (\Eq{eq:D}) and drift
(\Eq{eq:phi_strato}) projection estimators to reconstruct the force
field,
\begin{equation}
  \label{eq:F_D}
    \hat{F}_{\mu}(\mathbf{x}) = \hat{\Phi}_{\mu\alpha}c_\alpha(\mathbf{x}) - \hat{D}_{\mu\nu\alpha} \partial_\nu \hat{c}_\alpha(\mathbf{x})
\end{equation}
using \Eq{eq:drift}. This estimator allows for quantitative inference
of the force provided that the divergence of the diffusion coefficient
is well approximated. We demonstrate this (\Fig{fig:diffusion}C,D,G)
for the simple processes presented in \Fig{fig:diffusion}A,E using an
adapted basis to fit the diffusion coefficient.

\section{Discussion}

In this article, we have introduced Stochastic Force Inference, a
method to reconstruct force and diffusion fields and measure entropy
production from Brownian trajectories. Based on the communication
theory notion of capacity, we have shown that such trajectories
contain a limited amount of information. With finite data, force
inference is thus limited by the information available per degree of
freedom to infer. SFI uses this information to fit the force field
with a linear combination of known functions. We have demonstrated its
utility on a variety of model systems and benchmarked its accuracy
using data comparable to current experiments.

\begin{figure}[ht]
  \centering
  \includegraphics[width=\columnwidth]{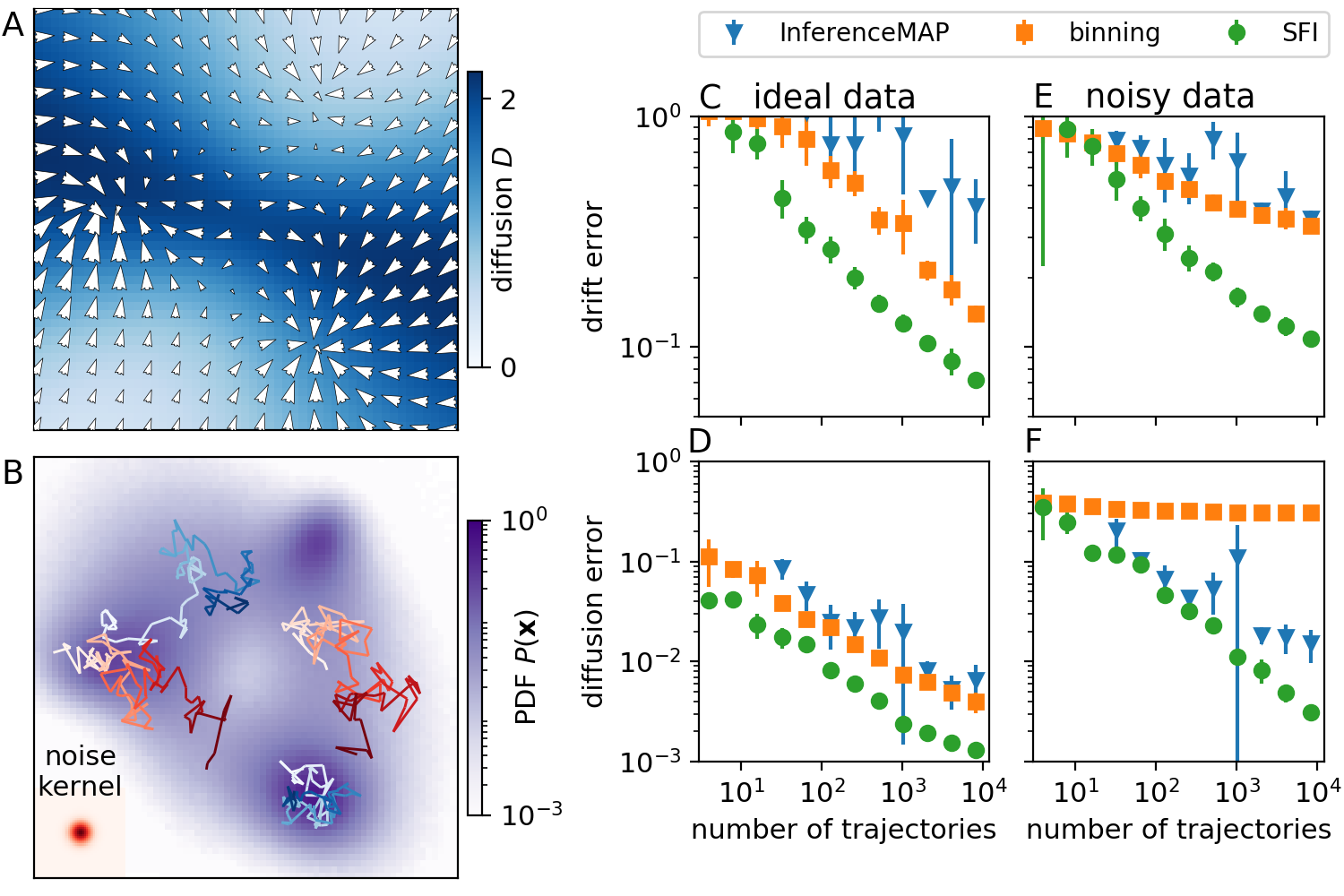}
  \caption{Quantitative comparison of SFI with other methods, on a
    simulated system mimicking 2D single molecule trajectories in a
    complex cellular environment with multiple potential wells,
    out-of-equilibrium circulation, and space-dependent isotropic
    diffusion. \textbf{A.} The diffusion field (blue gradient) and
    drift field (white arrows, scaled as $|\Phi|^{1/2}$ for better
    legibility). \textbf{B.} The steady-state probability distribution
    function of the process. The blue traces show two representative
    trajectories with $n=100$ time steps. The red traces show
    trajectories blurred by moderate Gaussian measurement error (with
    amplitude shown as a red kernel). \textbf{C-F.} Comparison of the
    performance of SFI with adaptive Fourier basis (green circles) and
    two widely used inference methods:
    InferenceMAP~\cite{beheiry_inferencemap:_2015}, a Bayesian method
    for single molecule inference (blue triangles), and grid-based
    binning with maximum-likelihood
    estimation~\cite{hoze_heterogeneity_2012,friedrich_approaching_2011}
    (\Eq{eq:F_moments_empirical}) and an adaptive mesh size (orange
    squares). We evaluate the performance of these methods on the
    approximation of the drift field (C,E) and diffusion field (D,F),
    as a function of the number $N$ of single-molecule trajectories
    (similar to those in B) used, with ideal data (C,D) and in the
    presence of measurement noise (E,F). The performance is evaluated
    as the average mean-squared error on the reconstructed field along
    trajectories. SFI outperforms both other methods in all cases; for
    noisy data, SFI is the only one that provides an unbiased
    estimation of the drift. Details and parameters in
    Appendix~\ref{sec:multitraps}. }
  \label{fig:comparison}
\end{figure}

We now briefly compare SFI to other existing methods to infer forces
from Brownian trajectories.  SFI combines the ability to infer
arbitrary force fields, for non-equilibrium processes, in high
dimensions and in the presence of measurement noise. In contrast, many
previous methods essentially rely on a specific
linear~\cite{penland_prediction_1993} or
parametric~\cite{bishwal_parameter_2008} form for the force, or are
specific to one-dimensional
systems~\cite{papaspiliopoulos_nonparametric_2012,kutoyants_statistical_2004,comte_penalized_2007}. Other approaches include spectral
methods~\cite{crommelin_diffusion_2011,gobet_nonparametric_2004},
Bayesian
methods~\cite{turkcan_bayesian_2012,ruttor_approximate_2013,yildiz_learning_2018,beheiry_inferencemap:_2015}
or methods that rely on coarse graining through
constant-by-parts~\cite{hoze_heterogeneity_2012,friedrich_approaching_2011,hoffmann_adaptive_1999}
or linear-by-parts~\cite{garcia_high-performance_2018}
approximations. However, these techniques become inefficient as the
system's dimensionality increases. Furthermore, none offers a generic
unbiased estimator in the presence of measurement noise. Few of these
general methods are being used on experimental data in soft matter and
biological systems. We quantitatively compare SFI to two of the most
popular such
methods~\cite{hoze_heterogeneity_2012,friedrich_approaching_2011,beheiry_inferencemap:_2015}
that rely on spatial binning (\Fig{fig:comparison}). Our method
significantly outperforms them for a two-dimensional process
simulating single molecule dynamics in a complex cellular environment,
in particular in the presence of realistic measurement noise.

An important by-product of SFI is the ability to quantify the
irreversibility of a system by measuring the entropy production
associated to its currents. Alternative methods to estimate entropy
production also exist, either by coarse-graining trajectories to
estimate currents~\cite{battle_broken_2016,seara_entropy_2018}, by
measuring cycling
frequencies~\cite{gladrow_broken_2016,mura_nonequilibrium_2018}, by
using non-Markovian signatures of irreversibility in hidden
variables~\cite{roldan_arrow_2018}, or by using thermodynamic bounds
on the fluctuations of dissipative
currents~\cite{li_quantifying_2019}. These methods are however
inherently limited to relatively low-dimensional systems with
homogeneous diffusion, and even then require large amounts of
well-resolved data; SFI, in contrast, performs well in high dimensions
-- even with trajectories too short to resolve the steady-state
density -- and in the presence of measurement noise and inhomogeneous
diffusion.

We have limited our scope here to systems whose dynamics is described
by \Eq{eq:Langevin} or \ref{eq:Langevin_inhomogenous}, with a
time-independent force field and white-in-time noise.  When the force
field varies in time, for instance due to the dynamics of unobserved
variables, SFI captures the average projection of the force onto the
observed variables~\SI{sec:time}. Furthermore, SFI could be extended
to capture an explicit time-dependence of the force by using a
time-dependent basis. Finally, force inference is notably complicated
by non-Markovian terms in the dynamics~\cite{daldrop_butane_2018},
such as colored noise; however, in such cases, our projection approach
to estimate phase-space velocities (\Eq{eq:v}) remains useful and
valid.

Our approach, all in all, proposes a solution to the inverse problem
of Brownian dynamics: inferring the force and diffusion fields from
trajectories. This method consists in a few intelligible equations,
and provides a powerful data analysis framework that could be used on
a broad class of stochastic systems where inferring effective forces
and currents from limited noisy data is of interest.  Our work thus
applies to microscopic systems where thermal noise is relevant, such
as single molecules~\cite{hoze_heterogeneity_2012}, active
colloids~\cite{palacci_living_2013,bricard_emergence_2013} and
cytoskeletal filaments~\cite{gladrow_broken_2016,seara_entropy_2018}.
Beyond thermal systems, for stochastic dynamical systems that can be
effectively modeled by Brownian dynamics, applications of our
framework range from the behavior of
cells~\cite{celani_bacterial_2010,li_dicty_2011,bruckner_stochastic_2019}
and animals~\cite{stephens_dimensionality_2008}, to modeling of
climate
dynamics~\cite{hasselmann_stochastic_1976,wheeler_all-season_2004,penland_prediction_1993}
and trend finding in financial data~\cite{oksendal_stochastic_2003}.
Our method could be combined with sparsity-promoting techniques, as
used to infer dynamical equations in deterministic
systems~\cite{brunton_discovering_2016}, to go from force fitting to
identifying the simple rules governing the dynamics.

\subsection*{Material and methods} 

All formulas presented in this article are derived in Appendix,
together with the details of each simulated system.
  
\textbf{Code availability.} A readily usable Python package to perform
Stochastic Force Inference is available at
\href{https://github.com/ronceray/StochasticForceInference}{https://github.com/ronceray/StochasticForceInference}. It
includes minimal
examples. 

\subsection*{Acknowledgments}

The authors warmly thank Eldad Afik, Theo Drivas, Kamesh
Krishnamurthy, David Lacoste, Martin Lenz, Ben Machta, Andreas Mayer,
Fr\'ed\'eric Van Wijland and especially Chase Broedersz and his group,
for multiple conversations and useful comments. This work was
initiated at the Aspen Center for Physics, which is supported by
National Science Foundation grant PHY-1607611. The authors acknowledge
support from the Princeton Center for Theoretical Science. PR is
supported by a Center for the Physics of Biological Function
fellowship.

\bibliography{SFI} 


\onecolumngrid
\vspace{2cm}

\appendix

\begin{center} {\centering \bf \Large{ Appendix}}
\end{center}

\section{Gaussian channel interpretation of Brownian dynamics}
\label{sec:capacity}

In this Section, we address the question of quantifying the rate at which
information can be read out, or is encoded in a trajectory. We
assume that the system follows the overdamped Langevin equation,
\begin{align}
  \label{eq:langevin}
  \dot{x}_{\mu}  = F_{\mu}(\mathbf{x}) + \sqrt{2 D }_{\mu\nu} \xi_{\nu} && \av{\xi_{\mu}(t)\xi_{\nu}(t')}= \delta(t-t').
\end{align}
Here and in the main text, what we refer to as a "force" is in fact
the physical force multiplied by the mobility matrix $M$, which has
the dimension of a mobility. So, in terms of our $F$, the system is
out-of-equilibrium if $D^{-1} F$ does not derive from a
potential. Indeed, a system in equilibrium has a physical force that
is derived from a potential, and a mobility matrix which is
proportional to the diffusion coefficient: $D = M T$ where $T$ is the
temperature. Our approach thus does not distinguish out-of-equilibrium
systems due to difference in temperature between components, such as
popular bead-spring models, from systems driven by non-reciprocal
force fields. We assume through most of this article that this
diffusion matrix is known and space-independent (although it can be
anisotropic); the case of a spatially variable diffusion matrix, and
how to infer it from data, is treated it \Sec{sec:D}. We also assume
that a steady state exists and that the system is ergodic, \emph{i.e.}
that time averages converge to phase space averages. Note however that
the discussion below can be readily extended to averages over an
ensemble of trajectories instead of time averages over a single long
trajectory.

The complete force field is characterized by an infinite number of
degrees of freedom, and thus in principle contains an infinite amount
of information (the value of the force components at each location in
phase space). It is therefore pertinent to ask if there is a bound to
the rate at which this information can be read off from the trajectory. We consider an infinite length trajectory, from which, in principle, all information about the force field can be recovered.  We
argue that indeed there is such a maximal rate, given by the capacity
(in natural information units, or nats)
\begin{equation}
C = \frac{1}{4} D^{-1}_{\mu\nu}  \int  F_{\mu}(\mathbf{x}) F_{\nu}(\mathbf{x})  \dP
 \label{eq:capacitySI}
\end{equation}

To explain this formula, let us first focus on a one dimensional
system. A trajectory which satisfies the dynamics given by \Eq{eq:langevin}
encodes the information about the force field in the form of a
continuous time signal $F(x(t))$ corresponding to the values of the
force field at the points $x(t)$ that the trajectory visits. However,
what can actually be read out from the trajectory is $\dot{x}$, \emph{i.e.}
the signal $F(x(t))$ with noise $\xi$ added to it (\Fig{fig:schematic}). 
Thus, we can think
of the dynamics \Eq{eq:langevin} as a noisy communication channel, with Gaussian
white correlated noise, where the information about the force is
transmitted in the form of a codeword $F(x(t))$ which satisfies
$\lim_{\tau\to \infty}1/\tau\int_0^\tau F^2 dt =\int F^2(x)  P(x)\  \dd x$.
In communication theory, such a channel is called an infinite
bandwidth Gaussian channel~\cite{cover_elements_2006}. 
It has a well
defined capacity, \emph{i.e.} a maximal rate of information transmission: for
codewords of duration $\tau$ that satisfy the so-called ``power constraint''
$1/\tau\int_0^\tau dt F^2(t) \leq {\cal P}$, and a white noise with
amplitude $2D$ the capacity is given by ${\cal P}/(4D)$ nats per
second.  Information cannot be transmitted through the channel at a
faster rate. Stated differently, the capacity quantifies the (exponential) rate with which the maximal number of distinguishable signals grows with the amount of time the channel is used for, in particular as $\tau \to \infty$. In our case, the capacity is related to the distinguishability of different force fields with the same power constraint. The maximal rate is obtained for a signal which saturates the power constraint, so that the relevant constraint to consider is ${\cal P}=\lim_{\tau\to \infty}1/\tau\int_0^\tau F^2 dt$. 
Thus, our trajectory which has
$\lim_{\tau\to \infty}1/\tau\int_0^\tau F^2 dt =\int F^2 P(x)\ \dd x$ cannot
produce information about the force field at a rate faster than the
capacity as defined in \Eq{eq:capacitySI}. Note that in contrast to the
usual communication theory setting, we do not control the codeword
through which the force field is encoded, only the decoding
scheme---the code word is determined by the dynamics, the force field
being sampled according to the probability density function (pdf) $P(x)$.  To go from the capacity
for a one dimensional process to that of a $d$ dimensional process, \Eq{eq:capacitySI}, we
have decomposed the channel into $d$ parallel channels and added up
their capacities. Indeed, let us first go into the basis where the noise is diagonal and normalize its amplitude to two, such that all components of the new force $D^{-1/2}_{\mu\nu}F_\nu$ have the same units ($t^{-1/2}$). 
The components of the noise become independent, and the $d$
components in that basis become parallel channels, with signals measured in the same units, whose capacities sum
up to \Eq{eq:capacitySI}.

\paragraph*{The Shannon-Hartley formula and infinite bandwidth channels.}
The infinite-bandwidth capacity of Brownian dynamics, as presented in
\Eq{eq:capacitySI}, corresponds to that of the continuous dynamics. It
can also be seen as the $\Delta t \to 0$ limit of a discrete signal
(\emph{i.e.} a finite bandwidth signal) such as can be acquired in
practice. The capacity of such a discrete Gaussian channel is given by
the Shannon-Hartley formula~\cite{cover_elements_2006}
\begin{equation}
C= \frac{1}{2 \Delta t} \log\left(1+\frac{{\cal P}\Delta t}{{\cal N}}\right)
\end{equation}
where we consider as before power-limited signals, where
${\cal P}\Delta t/{\cal N}$ is the signal-to-noise ratio: ${\cal P}$
is the signal power (note that it is not the power of the system in
the energetic sense, only in the signal theory sense), and
${\cal N}/\Delta t$ the noise power.  When the bandwidth is taken to
infinity, \emph{i.e.} $\Delta t \to 0$, we get
\begin{equation}
C_0= \frac{{\cal P}}{2{\cal N}} \log_2e \text{        bits per second}
\end{equation}
which corresponds to \Eq{eq:capacitySI}.  For a finite but small
$\Delta t$ the expression for the capacity becomes
\begin{equation}
C= \frac{{\cal P}}{2{\cal N}}  - \frac{{\cal P}^2\Delta t}{4{\cal N}^2}+...\approx C_0\left(1-C_0 \Delta t\right)  
\end{equation} 
The first correction to the continuous-time capacity due to finite
rate of sampling is thus of relative order $C_0 \Delta t$, \emph{i.e.}
the information per sample: the loss of information when monitoring
Brownian dynamics at a finite rate is thus negligible provided that
the information per sample remains small. This has an important
practical consequence for experimental applications, where there is
often a trade-off between acquisition rate and duration of the
experiment (for instance due to photobleaching of fluorescent
proteins): when the information per sample becomes small, very little
can be learned about the force field by increasing the acquisition
frequency.

\section{Information at the trajectory level}
\label{sec:trajectories}

In this Appendix, we relate the notion of capacity to trajectory-level
quantities, and relate it to other stochastic thermodynamics
quantities: the entropy production and the inflow rate. While
Appendix~\ref{sec:capacity} was restricted to the case of
constant-diffusion Brownian dynamics, here we consider the general
case with not only a state-dependent force, but also a state-dependent
diffusion tensor. In that case, the noise is no longer additive: it
has a multiplicative component, and care must be taken to specify the
convention within which the Langevin equation is written. We use the
It\^o convention here, writing:
\begin{equation}
\dot{x}_\mu = \Phi_\mu(\vec{x}) + \sqrt{2  D(\vec{x})}_{\mu\nu}  \xi_\nu
\label{eq:Langevin_D_cont}
\end{equation}
where
$\Phi_\mu =F_\mu(\vec{x}(t_i)) +\partial_\nu D_{\mu\nu}
(\vec{x}(t_i))$ is the drift term~\cite{lau_state-dependent_2007}, and
$F_\mu(\vec{x}(t_i))$ equals the mobility matrix times the physical
force.

\subsection{The capacity as a Kullback-Leibler divergence rate}
\label{sec:KL}

To relate the capacity to path-dependent quantities, we consider a
trajectory
${\cal C}^N =(\mathbf{x}(0),\mathbf{x}(\Delta t),..\mathbf{x}(N\Delta
t))$, with $t_i = i \Delta t$, and where we have defined the discrete
difference $\Delta x_{\mu}(t_i) = x_{\mu}(t_i+\Delta t)-x_{\mu}(t_i)$
and $\tau = N \Delta t$.  The path integral formula for the
probability density ${\cal P}({\cal C}^N|F)$ of a trajectory
${\cal C}^N$ in the force field $F$, written in the It\^o convention,
reads~\cite{risken_fokker-planck_1996}:
\begin{align}
&{\cal P}({\cal C}^N|F) =  \frac{P_0(\mathbf{x}(0))}{(4\pi)^{dN/2} }\prod_{i=0}^{N-1} \frac{1}{(\det D(\vec{x}(t_i))\Delta t)^{1/2}} \\
&\times\exp\left[-\frac{1}4 \Delta t\left(\frac{\Delta x_{\mu}(t_i)}{\Delta t} - F_{\mu}(\mathbf{x}(t_i))-\partial_\rho D_{\mu\rho}(\mathbf{x}(t_i))\right)D^{-1}_{\mu\nu} (\vec{x}(t_i))\left(\frac{\Delta x_{\mu}(t_i)}{\Delta t} - F_{\nu}(\mathbf{x}(t_i))-\partial_\sigma D_{\nu\sigma}(\mathbf{x}(t_i))\right)\right]
\label{eq:traj_probability_D_ito}
\end{align}
Note that in the limit of long trajectories, the initial point
probability becomes unimportant. We show here that the capacity of the system relates to
the Kullback-Leibler divergence rate between ${\cal P}({\cal C}^N|F)$ and
the probability density at zero force (but with the same diffusion
field), ${\cal P}({\cal C}^N|0) \equiv {\cal P}({\cal
  C}^N|F=0)$:
\begin{equation}
  C= \lim_{\tau\to \infty}\frac{1}{\tau}\int {\cal{D}}{{\cal C}^{\tau}} \  {\cal P}({\cal C}^{\tau}|F)\log\frac{ {\cal P}({\cal C}^{\tau}|F)}{{\cal P}({\cal C}^{\tau}|0)} =\av{\frac{1}4 F_\mu(\vec{x}(t)) D_{\mu\nu}^{-1}(\vec{x}(t))F_\nu(\vec{x}(t))}
\label{eq:traj_capacity_D}
\end{equation}
Indeed, for a constant diffusion coefficient the right hand side of the above equation reduces to the capacity discussed in \Sec{eq:capacitySI}, \Eq{eq:capacitySI}. Note that for systems with multiplicative noise,
to the best of our knowledge a formula for
the channel capacity, as defined in transmission
theory, has yet to be derived. Moreover, the interpretation from the standpoint of transmission theory is further complicated as, from
physical considerations, we wish to infer $F_\mu$ rather than
$ \Phi_\mu$. However, one may use the trajectory based formula in \Eq{eq:traj_capacity_D} as a general definition of the capacity for Brownian dynamics. Then, the generalization of
\Eq{eq:capacitySI} to systems with inhomogeneous diffusion is seen to be: 
\begin{equation}
C = \frac{1}{4}   \int  D^{-1}_{\mu\nu}(\mathbf{x}) F_{\mu}(\mathbf{x}) F_{\nu}(\mathbf{x})  \dP 
 \label{eq:capacity_nonH_SI}
\end{equation}

Let us proceed to show \Eq{eq:traj_capacity_D},
\begin{align}
C &= \lim_{\tau\to \infty}\frac{1}{\tau}\int {\cal{D}}{{\cal C}^{\tau}} \  {\cal P}({\cal C}^{\tau}|F)\log\frac{ {\cal P}({\cal C}^{\tau}|F)}{{\cal P}({\cal C}^{\tau}|0)} \\
&= \lim_{\tau\to \infty}\frac{1}\tau   \av{\frac{1}2\int^{\text{It\^o}} dt \ \dot{x}_\mu D_{\mu\nu}^{-1}F_\nu(\vec{x}(t)) -\frac{1}2 \int_0^{\tau} dt (\partial_\rho D_{\rho\mu}) D_{\mu\nu}^{-1}F_\nu(\vec{x}(t)) -\frac{1}4 \int_0^{\tau} dtF_\mu D_{\mu\nu}^{-1}F_\nu(\vec{x}(t))} \\
& =\av{\frac{1}4 F_\mu(\vec{x}(t)) D_{\mu\nu}^{-1}(\vec{x}(t))F_\nu(\vec{x}(t))}
\end{align}
where we have used that $\av{\int^{\text{It\^o}} dt\dot{x}_\mu D_{\mu\nu}^{-1}F_\nu(\vec{x}(t))}=\av{\int_0^{\tau}dt(F_\mu+ \partial_\rho D_{\rho\mu})D_{\mu\nu}^{-1}F_\nu(\vec{x}(t))}$.
Note that passing between the first and second line in the above equation is equivalent to deriving the Girsanov formula for diffusions. 

\subsection{The inflow rate}
\label{sec:inflow}

In the main text, we connect the capacity to the inflow rate
$G = \int d\vec{x} P(\vec{x}) g_{\mu} D_{\mu\nu} g_{\nu} $ with
$g_{\mu}= \partial_{\mu}\log P$. This quantity was originally
introduced and studied by Baiesi and Falasco~\cite{baiesi_inflow_2015}
in the case of Brownian dynamics with homogeneous diffusion (and for
discrete Markov processes, not discussed here). We generalize it here
to systems with inhomogeneous diffusion and discuss its properties.

\paragraph*{ Relation between the inflow rate and an instantaneous entropy production rate. }
 Let us show that it corresponds to an instantaneous entropy production rate that would be present if the force was suddenly set to zero. Consider the entropy $S(t) = -\int d\vec{x} P(x,t)\log P(x,t)$, after the force is set to zero: $F_\mu=0$, denoting that instant by $t=0$. At that instant one has $\partial_t P =  \partial_{\mu} [ D_{\mu\nu}\partial_\nu P ]$. Then
\begin{equation}
\begin{split}
\partial_t S |_{t=0} = -\int d\vec{x}  \log P(\vec{x})\partial_{\mu} (D_{\mu\nu}\partial_\nu P(\vec{x}) )+\int d\vec{x} \partial_{\mu} (D_{\mu\nu}\partial_\nu P )=   \int d\vec{x}  \frac{\partial_{\mu} P(\vec{x})}P D_{\mu\nu}\partial_\nu P(\vec{x}) \\
 =\int d\vec{x} P(\vec{x}) \partial_{\mu} \log P(\vec{x}) D_{\mu\nu}(\vec{x}) \partial_{\nu} \log P(\vec{x})= G
\end{split}
\end{equation}
where we have used integration by parts, assuming boundary terms vanish. 
We can define $ v_\mu^\mathrm{Fick}= -D_{\mu\nu} g_\nu$, a Fick velocity related to the current $j_{\mu}^{\mathrm{Fick}}=-D_{\mu\nu}\partial_{\nu} P$, that would result from diffusion of particles with an initial density profile $P(\vec{x})$ in the absence of forces. Indeed, in these notations $G$ has a similar form to the entropy production rate
\begin{equation}
  \label{eq:Sdot}
G = \int v^\mathrm{Fick}_\mu v^\mathrm{Fick}_\nu D^{-1}_{\mu\nu}  P(\mathbf{x}) d\mathbf{x}
\end{equation}

However, the inflow rate is nonzero even at equilibrium.
It measures the heterogeneity of the steady-state probability
distribution. Indeed, for an equilibrium process $F^{\mu} = D_{\mu\nu} \partial_\mu \log P$ (and $G= C$ trivially).
In a sense, it is the amount of information that the
force field needs to continuously inject into the system in order to
maintain its spatial structure; while the entropy production can be
seen as the amount of information the force field injects into the
system to maintain its currents. 

\paragraph*{ The inflow rate as a phase space contraction rate. }
The relation $D_{\mu\nu} g_\mu= F_\mu-v_\mu $ (which holds for a
space-dependent diffusion tensor) can be used to rewrite the inflow
rate as
\begin{equation}
\begin{split}
G = \int d\vec{x} P(\vec{x}) g_{\mu} D_{\mu\nu} g_{\nu} = \int d\vec{x} P(\vec{x})( \partial_\mu \log P)  (F_\mu - v_\mu )  
\\ =\int d\vec{x}  (\partial_\mu P(\vec{x}))  F_\mu  +  \int d\vec{x}  \partial_\mu(v_\mu P(\vec{x}) ) \log P = -\int d\vec{x} P(\vec{x}) \partial_\mu F_\mu    
\end{split}
\end{equation}
where in the second line the steady state relation $\partial_\mu(v_\mu P(\vec{x}) ) =\partial_\mu j_\mu   =0$ was employed.
We have thus obtained an expression for the inflow rate as (minus) the average divergence of the force. In a deterministic dynamical system this is equal to the average sum of the Lyapunov exponents and is called the average phase space contraction rate.  It then corresponds to the mean rate of entropy production in the environment~\cite{chetrite_fluctuation_2008}. For non-deterministic systems it was  mentioned in~\cite{chetrite_fluctuation_2008} as a "natural entropy production". It is worth stressing the difference between the deterministic case and overdamped Brownian dynamics in this context. While for a deterministic system at equilibrium, \emph{i.e.} a Hamiltonian system, the divergence of the force is identically zero due to the symplectic structure (there is no entropy production), for an equilibrium overdamped system that divergence is nonzero. Indeed the inflow rate (which does not correspond to an actual entropy production in this case) is positive, as discussed above.

\paragraph*{ Trajectory based interpretation of the inflow rate. }
Here we prove that an equivalent expression for the inflow rate is
\begin{equation}
G = \lim_{\tau\to \infty}\frac{1}{\tau}\int {\cal{D}}{{\cal C}^{\tau}} \  {\cal P}({\cal C}^{\tau}|F)\log\frac{ {\cal P}({\cal C}^{\tau}|F)}{{\cal P}({-\cal C}^{\tau}|-F)} =  \lim_{\tau\to \infty}\frac{1}{\tau}\av{ \log\frac{ {\cal P}({\cal C}^{\tau}|F)}{{\cal P}(-{\cal C}^{\tau}|-F)}}_F 
\label{eq:inflow_trajectory}
\end{equation}

The simplest way to do that is to express the probability density of a trajectory (\Eq{eq:traj_probability_D_ito}) in an alternative form, as we now show. We begin with the expression for the probability of a transition to the point $\vec{x}$ from the point $\vec{x'}$ in an infinitesimal time $\Delta t$~\cite{risken_fokker-planck_1996}
\begin{equation}
\begin{split}
P(\vec{x},t+\Delta t|\vec{x'},t)=\frac{1}{\sqrt{(4\pi)^d \det D(\vec{x})\Delta t}}  \exp\left[\Delta t \left\{  -\partial_\mu \Phi_\mu(\mathbf{x}) +\partial_\mu\partial_\nu D_{\mu\nu}(\vec{x}) \right.\right.\\
\left.\left.-\frac{1}4 \left(\frac{x_\mu-x'_\mu}{\Delta t}- \Phi_\mu(\vec{x}) +2\partial_\rho D_{\mu\rho}(\vec{x})\right) D^{-1}_{\mu\nu}(\vec{x})\left(\frac{x_\nu-x'_\nu}{\Delta t}- \Phi_\nu(\vec{x}) +2\partial_\sigma D_{\nu\sigma}(\vec{x})\right)\right\}\right]
\end{split}
\end{equation}
Note that here the diffusion coefficient and $\Phi_\mu$ are both evaluated at the point $\vec{x}$ to which the system transitions. The probability of a trajectory is then simply given by a product of such transition probabilities, and the distribution of the initial point. Using that  $\Phi_\mu = F_\mu+\partial_\nu D_{\mu\nu}$ we then get
\begin{equation}
\begin{split}
{\cal P}({\cal C}^N|F) = \frac{P_0(\mathbf{x}(0))}{(4\pi)^{dN/2} }\prod_{i=0}^{N-1} \frac{1}{(\det D(\vec{x}(t_{i+1}))\Delta t)^{1/2}} \exp\left[  -\partial_\mu F_\mu(\mathbf{x}(t_{i+1}))\Delta t \right.\\
\left.-\frac{1}4 \Delta t\left(\frac{\Delta x_{\mu}(t_{i})}{\Delta t} - F_{\mu}(\mathbf{x}(t_{i+1}))+\partial_\rho D_{\mu\rho}(\mathbf{x}(t_{i+1}))\right)D^{-1}_{\mu\nu} (\vec{x}(t_{i+1}))\left(\frac{\Delta x_{\nu}(t_{i})}{\Delta t} - F_{\nu}(\mathbf{x}(t_{i+1}))+\partial_\sigma D_{\nu\sigma}(\mathbf{x}(t_{i+1}))\right)\right]
\end{split}
\label{eq:traj_probability_D_aito}
\end{equation}

It follows that the probability of the time reversed trajectory $-{\cal C}^N=\{\vec{x}(t_N),\vec{x}(t_{N-1})...,\vec{x}(t_0)\}$ can be written in the form
\begin{equation}
\begin{split}
{\cal P}(-{\cal C}^N|F) = \frac{P_0(\mathbf{x}(N\Delta t))}{(4\pi)^{dN/2} }\prod_{i=0}^{N-1} \frac{1}{(\det D(\vec{x}(t_{i}))\Delta t)^{1/2}} \exp\left[  -\partial_\mu F_\mu(\mathbf{x}(t_{i}))\Delta t \right.\\
\left.-\frac{1}4 \Delta t\left(\frac{-\Delta x_{\mu}(t_{i})}{\Delta t} - F_{\mu}(\mathbf{x}(t_{i}))+\partial_\rho D_{\mu\rho}(\mathbf{x}(t_{i}))\right)D^{-1}_{\mu\nu} (\vec{x}(t_{i}))\left(\frac{-\Delta x_{\nu}(t_{i})}{\Delta t} - F_{\nu}(\mathbf{x}(t_{i}))+\partial_\sigma D_{\nu\sigma}(\mathbf{x}(t_{i}))\right)\right]
\end{split}
\label{eq:-traj_probability_D_aito}
\end{equation}

Now, it becomes straightforward to evaluate \Eq{eq:inflow_trajectory}, dividing term by term in the product in \Eq{eq:traj_probability_D_ito} by the product in ${\cal P}(-{\cal C}^N|-F)$, using \Eq{eq:-traj_probability_D_aito} with the reversed sign for the force. Indeed, we notice that all terms cancel out except for the divergence of $F_\mu$, which yields (we ignore the terms related to the initial and final distributions whose contribution vanishes in the limit of $\tau\to \infty$)
\begin{equation}
\begin{split}
G= \lim_{\tau\to \infty}\frac{1}{\tau}\int {\cal{D}}{{\cal C}^{\tau}} \  {\cal P}({\cal C}^{\tau}|F)\log\frac{ {\cal P}({\cal C}^{\tau}|F)}{{\cal P}(-{\cal C}^{\tau}|-F)} =   -\lim_{\tau\to \infty} \int_0^\tau \frac{dt}\tau \av{\partial_\mu F_\mu(\vec{x}(t)) }
\end{split}
\end{equation}

\subsection{Different decompositions of the capacity and the relation to traffic}
\label{sec:traffic}

The trajectory-based expression for the capacity,
\Eq{eq:traj_capacity_D}, is related to the "dynamical entropy"
introduced in~\cite{maes_steady_2008}: it is equal to the dynamical
entropy per unit time in the limit $\tau\to \infty$, i.e to a rate of
dynamical entropy. In~\cite{maes_steady_2008} the dynamical entropy
was split into two contributions: a time anti-symmetric contribution,
equal to $\dot{S}/2$ and a time symmetric contribution $-{\cal T}$,
where ${\cal T}$ is called the traffic (and is related to the
so-called \emph{frenesy} in Markov jump processes).  The relations
between the capacity, the inflow rate we have defined, the entropy
production and the steady state traffic ${\cal T}$ are
\begin{align}
C= -{\cal T}+\frac{1}2 \dot{S}  && {\cal T} = ( \dot{S}-G)/4
\end{align}

The decomposition of the capacity that we have presented in the main text can also be presented as the sum of time symmetric and anti-symmetric parts, but  corresponding to a different trajectory-based expression for the capacity:
\begin{align}
4 C &=  \lim_{\tau\to \infty}\frac{1}{\tau}\int {\cal{D}}{{\cal C}^{\tau}} \  {\cal P}({\cal C}^{\tau}|F)\log\frac{ {\cal P}({\cal C}^{\tau}|F)}{{\cal P}({\cal C}^{\tau}|-F)} =  \lim_{\tau\to \infty}\frac{1}{\tau}\av{ \log\frac{ {\cal P}({\cal C}^{\tau}|F)}{{\cal P}({\cal C}^{\tau}|-F)}}_F\\
& =\lim_{\tau\to \infty}\frac{1}\tau   \av{\int^{\text{It\^o}} dt\ \dot{x}_\mu D_{\mu\nu}^{-1}F_\nu(\vec{x}(t)) - \int_0^{\tau} dt (\partial_\rho D_{\rho\mu}) D_{\mu\nu}^{-1}F_\nu(\vec{x}(t))}\\
& =  \lim_{\tau\to \infty}\frac{1}\tau\av{\int^{\text{Strat}}  \!\!\!  \!\!\! dt\ \dot{x}_\mu D_{\mu\nu}^{-1}F_\nu(\vec{x}(t)) } -  \lim_{\tau\to \infty}\frac{1}\tau\av{\int_0^\tau dt D_{\mu\rho} \partial_\rho (D_{\mu\nu}^{-1}F_\nu)(\vec{x}(t)) - \int_0^{\tau} dt \partial_\rho (D_{\rho\mu}) D_{\mu\nu}^{-1}F_\nu(\vec{x}(t))} \\
& = \lim_{\tau\to \infty}\frac{1}\tau\av{\underbrace{\int^{\text{Strat}} \!\!\! dt\ \dot{x}_\mu D_{\mu\nu}^{-1}F_\nu(\vec{x}(t)) }_{\text{time anti-symmetric}}} + \lim_{\tau\to \infty}\frac{1}\tau\av{\underbrace{-\int_0^{\tau} dt \partial_\mu F_\mu(\vec{x}(t))}_\text{ time symmetric}} -\lim_{\tau\to \infty}\frac{1}\tau \av{ \int_0^{\tau} dt F_\nu\underbrace{\partial_\rho (D_{\rho\mu} D_{\mu\nu}^{-1})}_0}
\label{eq:capacity_trajectory2}
\end{align}
Indeed, the first term in the last line is time anti-symmetric, and is equal to the entropy production rate, and the second term is time symmetric and is equal to the inflow rate.  

One can think of the decomposition of the capacity into $\dot{S}$ and $G$ as decomposing the influence of the force field into two types of ``orders'': ``\emph{go
  there!}'' -- corresponding to a dissipative, irreversible motion
quantified by $\dot{S}$ -- and ``\emph{stay there!}'' -- corresponding
to a nondissipative, reversible motion fighting thermal diffusion, and
quantified by $G$.

\section{Stochastic Force Inference: estimating $F_{\mu\alpha}$ and its error}
\label{sec:trajectory}

In this Section, we derive the core results of our article: how to
perform SFI in practice, and self-consistently estimate the error in
the inference. 

\subsection{The force as a trajectory average}

To be able to deduce the force from the trajectory one
first needs an expression for the force in terms of measurable
quantities along the trajectory. We have 
\begin{equation}
\mathbf{F}(\mathbf{x})= \lim_{\epsilon\to0}\av{\left.\frac{(\mathbf{x}(t+\epsilon)-\mathbf{x}(t))}{\epsilon}\right|\mathbf{x}(t) = \mathbf{x}} =\av{\dot{\mathbf{x}}^{+}|\mathbf{x}(t)} = \av{\delta(\mathbf{x}(t)-\mathbf{x})\dot{\mathbf{x}}^{+}}/P(x)
\label{eq:full_force_from_traj}
\end{equation}
where $\av{\left. \cdot \right|\mathbf{x}(t) = \mathbf{x}}$ means
averaging over realizations of the noise, conditioned on being at
position $\mathbf{x}$ at time $t$. We have defined here
$\dot{\mathbf{x}}^{+}$ as the right hand derivative, corresponding to
It\^o calculus (see Appendix A of \cite{chetrite_eulerian_2009}).  The
coefficients of the force field in its decomposition with respect to
the phase space projector $c_\alpha(\mathbf{x})$ are:
\begin{equation}
\begin{split}
F_{\mu\alpha} =  \int d\mathbf{x} P(\mathbf{x}) F_{\mu}(\mathbf{x}) c_\alpha(\mathbf{x})=\int d\mathbf{x}  \av{\delta(\mathbf{x}(t)-\mathbf{x})\dot{x}_{\mu}^{+}}c_\alpha(\mathbf{x}) \\
=  \av{\int d\mathbf{x} \delta(\mathbf{x}(t)-\mathbf{x})\dot{x}_{\mu}^{+}c_\alpha(\mathbf{x})} = \av{\dot{x}_{\mu}^{+}c_\alpha(\mathbf{x})}
\end{split} 
\label{eq:Fn}
\end{equation}
Because of this last expression, the force projection coefficient
$F_{\mu\alpha}$ can be expressed as an average quantity along an
infinitely long trajectory, which can thus be estimated by computing
it on a finite trajectory.

Note that, similarly to the force, the phase space velocity can also be defined through an average of $\mathbf{\dot{x}}$, where the time derivative is taken in the Stratonovich sense:
\begin{equation}
\mathbf{v}(\mathbf{x})= \lim_{\epsilon\to0}\av{\left.\frac{(\mathbf{x}(t+\epsilon)-\mathbf{x}(t-\epsilon))}{2\epsilon}\right|\mathbf{x}(t)=\mathbf{x}} =\left.\av{\frac{1}2(\dot{\mathbf{x}}^{+}+\dot{\mathbf{x}}^{-})\right|\mathbf{x}(t)=\mathbf{x}} = \av{\delta(\mathbf{x}(t)-\mathbf{x})\frac{1}2(\dot{\mathbf{x}}^{+}+\dot{\mathbf{x}}^{-})}/P(\mathbf{x})
\end{equation}
(see Appendix A of \cite{chetrite_eulerian_2009}).
The phase space velocity in its decomposition with respect to the phase space basis $c_\alpha(\mathbf{x})$ is, analogously to the force,:
\begin{equation}
\begin{split}
v_{\mu\alpha}  = \av{\frac{1}2(\dot{x}_{\mu}^{+}+\dot{x}_{\mu}^{-})c_\alpha(\mathbf{x})}
\end{split}
\label{eq:vn}
\end{equation}

\subsection{Projection on the empirical basis}
The second difficulty in evaluating Eq.2 of the main text in practice
is that the phase space measure $P(\mathbf{x})$ is unknown in
practice. As a consequence, the phase space basis,
$c_\alpha(\mathbf{x})$ is not known either, as it is the
orthonormalized basis derived from $b$ using $P$ as the measure.  Our
approach consists in approximating $P(\mathbf{x})$ by the empirical
measure
\begin{equation}
\hat{P}_{\tau}(\mathbf{x}) = \frac{1}{\tau}\int_0^{\tau} \delta(\mathbf{x} - \mathbf{x}(t))dt
\end{equation}
corresponding to a time average along the trajectory.

We then define the empirical projector $\hat{c}_\alpha$ with respect
to this measure, as in the main text:
\begin{equation} 
\hat{c}_\alpha(\mathbf{x}) = \hat{B}^{-1/2}_{\alpha\beta} b_\beta(\mathbf{x}) \qquad \mathrm{with} \qquad
\hat{B}_{\alpha\beta} = \int b_\alpha(\mathbf{x}) b_\beta(\mathbf{x})
\frac{\dd t}{\tau}.
\label{eq:c_empirical}
\end{equation}
In the long-trajectory limit, these ``empirical projectors''
$\hat{c}_\alpha(\mathbf{x})$ converge to the phase-space projectors
${c}_\alpha(\mathbf{x})$; more precisely, we expect that for typical
trajectories
$\hat{c}_\alpha(\mathbf{x}) = {c}_\alpha(\mathbf{x}) +
O(\sqrt{\tau_0/\tau})$, where $\tau$ is the duration of the trajectory
and $\tau_0$ is a relaxation time of the system.  In the case of the
polynomial basis for instance, the convergence of the basis at order
$n$ is related to the convergence of the $n$-th cumulant of the
probability distribution function. We do not seek to make this
statement more mathematically precise here.

As an intermediate variable for this calculation, we define the
projection coefficients $F^{\tau}_{\mu\alpha}$ of the (exact) force
onto these empirical projectors. These coefficients are trajectory
dependent; however, $\hat{c}_\alpha$ are directly accessible from the trajectory, as is the empirical measure with respect to which they are projectors, so that obtaining the coefficients $F^{\tau}_{\mu\alpha}$ precisely, would result in an accurate approximation of the force field $F_\mu \approx F^{\tau}_{\mu\alpha}\hat{c}_\alpha$ along the trajectory. For this reason, we focus here on how the estimator
$\hat{F}_{\mu\alpha}$ as defined in \Eq{eq:F_moments} of the main text converges
to $F^{\tau}_{\mu\alpha}$.  The relative errors presented in the main text also refer to this convergence (rather than the convergence to the phase-space projection $F_{\mu\alpha}$). Recall that our estimator is given by
\begin{align}
\hat{F}_{\mu\alpha} & = \frac{1}{\tau}\int^{\text{It\^o}} \hat{c}_\alpha(\mathbf{x}) d\mathbf{x}_t^{\mu} \\
& =  \underbrace{\frac{1}{\tau}\int_0^{\tau} \hat{c}_\alpha(\mathbf{x}) F_{\mu}(\mathbf{x})dt}_{F^\tau_{\mu\alpha}} +\underbrace{\frac{1}{\tau}\int^{\text{It\^o}}\hat{c}_\alpha(\mathbf{x}) \sqrt{2} D^{1/2}_{\mu\nu}d\xi_t^{\nu}}_{Z_{\mu\alpha}} 
\label{eq:Fn_estimator}
\end{align}
using the Langevin equation (\ref{eq:langevin}).  Since
$F^{\tau}_{\mu\alpha}$ is what we wish to infer, we propose to study
now the statistics of
$Z_{\mu\alpha} = \hat{F}_{\mu\alpha} - F^{\tau}_{\mu\alpha}$,
\emph{i.e.} its mean and variance.

\subsection{Statistics of the error in the inference of the projection coefficients}
\label{sec:stats_Z}

We thus study the first and second moment of the random tensor
$Z_{\mu\alpha}$, \emph{i.e.} respectively the systematic bias and the
typical error of $\hat{F}_{\mu\alpha}$ as an estimator of
$F^{\tau}_{\mu\alpha}$. To make the norm of these moments meaningful,
it is necessary here to go to dimensionless coordinates: indeed,
different phase space coordinates can have different dimensions (such
as, for instance, a phase space comprising both distances and angles,
as in \Fig{fig:particles} of the main text), and thus different
coordinates of $Z_{\mu\alpha}$ cannot be compared or summed. To this
end, we define $W_{\mu\alpha} = D^{-1/2}_{\mu\nu} Z_{\nu\alpha}$, all
the coordinates of which have the dimension of $t^{-1/2}$.

First recall that we defined both phase-space and empirical
projectors as a linear combination of the basis functions $b$,
$c_\alpha=B_{\alpha\beta}^{-1/2} b_{\beta} $ and
$\hat{c}_\alpha =\hat{B}_{\alpha\beta}^{-1/2} b_{\beta} $, where
\begin{align} \label{eq:B}
B_{\alpha\beta} = \int d\mathbf{x} P(\mathbf{x}) b_{\beta}(\mathbf{x}) b_{\alpha}(\mathbf{x}) && \hat{B}_{\alpha\beta} = \int_0^{\tau}\frac{dt}\tau b_{\beta}(\mathbf{x}(t)) b_{\alpha}(\mathbf{x}(t))
\end{align}
Thus we have
$\lim_{\tau\to \infty} \hat{B}_{\alpha\beta}^{-1/2} =
B_{\alpha\beta}^{-1/2}$ and $\av{\hat{B}_{\alpha\beta}}=B_{\alpha\beta}$.
Let us denote
$\Delta_{\alpha\beta} = B_{\alpha\gamma}^{1/2}\hat{B}_{\gamma\beta}^{-1/2}-\delta_{\alpha\beta} 
$
the dimensionless error on the orthonormalization matrix (indeed, the
basis functions $b_\alpha$ can in principle have a dimension). We have
$\lim_{\tau\to \infty}\Delta_{\alpha\beta} =0$; typically, we'll have
more precisely $\Delta_{\alpha\beta} = O(1/\sqrt{\tau})$,
corresponding to the convergence of trajectory integrals to
phase-space integrals in \Eq{eq:B}. We then have
\begin{equation} \label{eq:Z_decomp}
Z_{\mu\alpha} \equiv \frac{1}{\tau}\int^{\text{It\^o}}\hat{c}_\alpha(\mathbf{x}) \sqrt{2} D^{1/2}_{\mu\nu}d\xi_t^{\nu} = B^{-1/2}_{\alpha\beta} \sqrt{2} D^{1/2}_{\mu\nu} \frac{1}{\tau}\int^{\text{It\^o}} b_\beta(\mathbf{x})d\xi_t^{\nu}+B^{-1/2}_{\alpha\beta}\Delta_{\beta\gamma} \sqrt{2} D^{1/2}_{\mu\nu} \frac{1}{\tau}\int^{\text{It\^o}} b_\gamma(\mathbf{x})d\xi_t^{\nu}.
\end{equation}
For the remainder of this Section we will denote the It\^o integral by a regular integration: $\int^{\text{It\^o}} d\xi_t^{\nu}=\int_0^{\tau} d\xi_t^{\nu}$.
We now put an upper bound on the first moment of $Z_{\mu\alpha}$,
\emph{i.e.} on the systematic bias.  Note that the first term in
\Eq{eq:Z_decomp} has zero average, as it is linear in the noise. In
contrast, due to possible correlations between the noise and the
random variable $\Delta_{\alpha\beta}$, the second term may not
average to zero.  Going to dimensionless coordinates, we use the
Cauchy-Schwarz inequality to bound the norm of this bias:
\begin{equation}
\begin{split}
\|\av{W_{\mu\alpha} }\|^2 =& \left\|\av{ B^{-1/2}_{\alpha\beta}\Delta_{\beta\gamma} \frac{1}{\tau}\int_0^{\tau}  b_\gamma(\mathbf{x}) D^{-1/2}_{\mu\nu}\sqrt{2}D^{1/2}_{\nu\rho}  d\xi_t^{\rho} }\right\|^2  \leq 2 B^{-1}_{\beta\delta}\av{\Delta_{\beta\rho}\Delta_{\rho\delta}}\av{\frac{1}{\tau^2}  \int_0^{\tau}  b_\gamma(\mathbf{x}) d\xi_t^{\mu} \int_0^{\tau}  b_\gamma(\mathbf{x})d\xi_{t'}^{\mu}} 
\end{split}
\end{equation}
We can then use the It\^o isometry
relation~\cite{gardiner_stochastic_2009} to prove that
\begin{equation}
  \av{\int_0^{\tau}  b_\alpha(\mathbf{x})d\xi_t^{\mu} \int_0^{\tau}  b_\beta(\mathbf{x})d\xi_{t'}^{\mu}} =  \av{ \int_0^{\tau}  b_\alpha(\mathbf{x}(t))   b_\beta(\mathbf{x}(t))dt} = \av{\hat{B}_{\alpha\beta} }
\label{eq:isometry}
\end{equation}
which implies that 
\begin{equation}
  \|\av{W_{\mu\alpha} }\|^2  \leq \frac{2}{\tau} B^{-1}_{\beta\delta}\av{\Delta_{\beta\rho}\Delta_{\rho\delta}}\av{\hat{B}_{\gamma\gamma} }
\end{equation}
Since $\Delta_{\alpha\beta} = O(\tau^{-1/2})$, we thus have
$\av{W_{\mu\alpha} } = O(1/\tau)$, which corresponds to a fast
convergence of the bias towards zero: the bias is negligible compared
to the fluctuating part of inference error, which goes as
$O(\tau^{-1/2})$.

Indeed, let us now compute the second moment of $W_{\mu\alpha}$.  We have
\begin{equation}
\av{W_{\mu\alpha}W_{\nu\beta} } = \frac{2}{\tau^2}   \av{ \hat{B}^{-1/2}_{\alpha\gamma} \hat{B}^{-1/2}_{\beta\delta} \int_0^{\tau} \int_0^{\tau} d\xi_t^{\mu} d\xi_{t'}^{\nu} b_\gamma(\mathbf{x}(t))  b_\delta(\mathbf{x}(t')) } 
\end{equation}
As $\hat{B}^{-1/2}_{\alpha\gamma}$ depends on all values of $t$, it is
not \emph{adapted} to the Wiener process $d\xi_t^{\mu}$, and thus we
cannot apply the It\^o isometry. However, we have
$\hat{B}^{-1/2}_{\alpha\gamma} =
{B}^{-1/2}_{\alpha\beta}(\delta_{\beta\gamma}+\Delta_{\beta\gamma})$.
Applying the It\^o isometry
(\Eq{eq:isometry}) yields:
\begin{align} \label{eq:varW}
\av{W_{\mu\alpha}W_{\nu\beta} } & = \frac{1}{\tau^2} \delta_{\mu\nu}  {B}^{-1/2}_{\alpha\gamma} {B}^{-1/2}_{\beta\delta}  2 \tau \av{\hat{B}_{\gamma\delta} } + R_{\mu\alpha\nu\beta}    \\
& = \frac{2}{\tau} \delta_{\mu\nu}\delta_{\alpha\beta} + R_{\mu\alpha\nu\beta} 
\end{align}
where we have defined the remainder 
\begin{equation}
R_{\mu\alpha\nu\beta} = \frac{2}{\tau^2}   \av{\left({B}^{-1/2}_{\alpha\gamma}{B}^{-1/2}_{\beta\lambda}\Delta_{\lambda\delta}+ {B}^{-1/2}_{\alpha\lambda}\Delta_{\lambda\gamma} \hat{B}^{-1/2}_{\beta\delta} \right)\int_0^{\tau} \int_0^{\tau} d\xi_t^{\mu} d\xi_{t'}^{\nu} b_\gamma(\mathbf{x}(t))  b_\delta(\mathbf{x}(t')) } 
\end{equation}
which is, as we show now, subleading in \Eq{eq:varW}. We now wish to bound the amplitude of the remainder $|\av{W_{\mu\alpha}W_{\mu\alpha} }-\frac{2}{\tau} N_b| =|R_{\mu\alpha\mu\alpha}|$. Since for typical trajectories $\Delta_{\alpha\beta} = O(\tau^{-1/2})$, we can bound every element of the matrix $|{B}^{-1/2}_{\alpha\gamma}{B}^{-1/2}_{\alpha\lambda}\Delta_{\lambda\delta}+{B}^{-1/2}_{\alpha\lambda}\Delta_{\lambda\gamma} \hat{B}^{-1/2}_{\alpha\delta}| \leq R\cdot O_{\gamma\delta}$ for such trajectories, where $R = O(1/\sqrt{\tau})$ is a (non-fluctuating) number and $O_{\gamma\delta}$ is the matrix with ones at all places. We get
\begin{equation}
\begin{split}
|\av{W_{\mu\alpha}W_{\mu\alpha} }-\frac{2}{\tau} N_b| = \frac{2}{\tau^2}  \left|\av{ \left({B}^{-1/2}_{\alpha\gamma}{B}^{-1/2}_{\alpha\lambda}\Delta_{\lambda\delta}+{B}^{-1/2}_{\alpha\lambda}\Delta_{\lambda\gamma} \hat{B}^{-1/2}_{\alpha\delta}\right) \int_0^{\tau} \int_0^{\tau}   d\xi_t^{\mu} d\xi_{t'}^{\mu} b_\gamma(\mathbf{x}(t))  b_\delta(\mathbf{x}(t')) } \right|  \\
\leq  \frac{2}{\tau^2}  \av{ \left|\left({B}^{-1/2}_{\alpha\gamma}{B}^{-1/2}_{\alpha\lambda}\Delta_{\lambda\delta}+{B}^{-1/2}_{\alpha\lambda}\Delta_{\lambda\gamma} \hat{B}^{-1/2}_{\alpha\delta}\right)\right|\left| \int_0^{\tau} \int_0^{\tau}d\xi_t^{\mu} d\xi_{t'}^{\mu}  b_\gamma(\mathbf{x}(t))  b_\delta(\mathbf{x}(t'))\right| }
 \\
\leq  \frac{2}{\tau^2} R\cdot O_{\gamma\delta}\av{\left| \int_0^{\tau} \int_0^{\tau} d\xi_t^{\mu} d\xi_{t'}^{\mu}  b_\gamma(\mathbf{x}(t))  b_\delta(\mathbf{x}(t'))\right|} \\
\leq \frac{2}{\tau^2} R\cdot O_{\gamma\gamma}\av{ \left|\int_0^{\tau} \int_0^{\tau} d\xi_t^{\mu} d\xi_{t'}^{\mu}  b_\delta(\mathbf{x}(t))  b_\delta(\mathbf{x}(t'))\right|}
\\
= \frac{2}{\tau^2} R\cdot O_{\gamma\gamma}\av{  \int_0^{\tau}   d\xi_t^{\mu}  b_\delta(\mathbf{x}(t))  \int_0^{\tau}  d\xi_t^{\mu}  b_\delta(\mathbf{x}(t)) } = \frac{1}{\tau^2} R\cdot O_{\gamma\gamma} 2\tau \av{\hat{B}_{\delta\delta}} = O(1/\tau^{3/2}).
\end{split} 
\end{equation}
In the fourth line we have used that for two semi-definite matrices $M_{\alpha\beta}$ and $N_{\alpha\beta}$, $M_{\alpha\beta}N_{\beta\alpha} \leq\sqrt{ M^2_{\alpha\alpha}N^2_{\beta\beta}}\leq M_{\alpha\alpha}N_{\beta\beta}$, an identity based on the Cauchy-Schwarz inequality. In the fifth line we employed the the It\^o isometry
(\Eq{eq:isometry}).
Again, this subleading term originates from the convergence of the
empirical projected basis to its long-trajectory limit.

\subsection{Self-consistent estimate of the error on the projected
  force}
\label{app:information_error}

The previous error estimates are rigorous, but require knowledge of
the exact force field to assess their amplitude. The goal of this
section is to provide approximate estimates of the typical error that
can be obtained using only the inferred force field, and are thus
useful in practical situations. Now that we know the statistical
properties of the dimensionless error term $W_{\mu\alpha}$, we can
write the covariance of the inferred force projection coefficients
explicitly:
\begin{equation}
\av{\left(\hat{F}_{\mu\alpha}-F^{\tau}_{\mu\alpha}\right)\left(\hat{F}_{\nu\alpha}-F^{\tau}_{\nu\alpha}\right)} = \frac{2 D_{\mu\nu}}{\tau}\delta_{\alpha\beta}(1+O(1/\sqrt{\tau}))
\label{eq:force_cov}
\end{equation}

Now, let us define the information along the trajectory by
\begin{equation}
I^{\tau}_b = \frac{1}{4}\tau F^{\tau}_{\mu\alpha}D^{-1}_{\mu\nu}F^{\tau}_{\nu\alpha}.
\end{equation}
In the long time limit, the rate of information $I^{\tau}_b /\tau$ converges to the capacity we had discussed previously. Similarly, we define the empirical estimate of the information along the trajectory,
\begin{equation}
\hat{I}_b=  \frac{\tau}{4}\hat{F}_{\mu\alpha}D^{-1}_{\mu\nu}\hat{F}_{\nu\alpha} =I^\tau_b + \frac{1}2\tau F^{\tau}_{\mu\alpha}D^{-1}_{\mu\nu}Z_{\nu\alpha}+\frac{1}4\tau Z_{\mu\alpha}D^{-1}_{\mu\nu}Z_{\nu\alpha}=I^\tau_b+\frac{1}{2}\tau\hat{F}_{\mu\alpha}D^{-1}_{\mu\nu}Z_{\nu\alpha} -\frac{1}4\tau Z_{\mu\alpha}D^{-1}_{\mu\nu}Z_{\nu\alpha}.
\end{equation}
so that
\begin{equation}
I^{\tau} _b= \hat{I}_b -\frac{1}2\tau\hat{F}_{\mu\alpha}D^{-1}_{\mu\nu}Z_{\nu\alpha}+\frac{1}{4} \tau Z_{\mu\alpha}D^{-1}_{\mu\nu}Z_{\nu\alpha}
\end{equation}

We can also relate the average of the empirical information to the
trajectory information:
\begin{equation} 
\av{\hat{I}_b}-I_b^{\tau} = \frac{1}{2} N_b
\end{equation}
at leading order. The estimator $\hat{I}_b$ is thus biased, with bias
$\frac{1}2 N_b$.  The variance of this estimator is well approximated
by $\av{(I^{\tau}_b - \hat{I}_b)^2} \approx 2 \av{\hat{I}_b} + N_b^2/4$.

In practice, the ``true'' force field is not known -- inferring it is
the goal here. It is therefore important to provide an estimate of the
inference error using only the inferred quantities.  \Eq{eq:force_cov}
allows us to propose such a self-consistent estimate of the
error. Indeed, it can be interpreted as the (squared) typical error on
the force projection coefficients, its right-hand-side can be
estimated using only trajectory-dependent quantities (again, we assume
that the diffusion matrix is known). We can also combine these
quantities in a single number quantifying the relative inference
error, as
\begin{equation}
  \label{eq:self_consistant_error}
\frac{(F^{\tau}_{\mu\alpha}-\hat{F}_{\mu\alpha}) D^{-1}_{\mu\nu} (F^{\tau}_{\nu\alpha}-\hat{F}_{\nu\alpha})}{\hat{F}_{\mu\alpha} D^{-1}_{\mu\nu}\hat{F}_{\nu\alpha}} \sim  N_b/2\hat{I}.
\end{equation}
Thus $N_b/2\hat{I}$ provides a self-consistent estimate of the
relative error. Note that in the absence of forces,
$\av{\hat{I}} = N_b/2$, corresponding to an inferred error of $1$,
which is consistent. Similarly, based on our estimate of the variance
of $\hat{I}_b$, we define a self-consistent confidence interval around
this inferred information as $\delta\hat{I}_b^2 = 2 \hat{I}_b + N_b^2/4$.

\subsection{The force estimator and maximum likelihood}
Here we show that the estimator we propose in \Eq{eq:Fn_estimator} is also the maximum log-likelihood estimator for $F_{\mu\alpha}$. Indeed, given a measured trajectory $C^{\tau}$, we use the expression for the probability of a trajectory, \Eq{eq:traj_probability_D_ito}, to calculate
\begin{equation}
0=\frac{\partial\log {\cal P}({\cal C}^{\tau}|F)}{\partial F^{\tau}_{\mu\alpha} }= \int d\vec{x}\frac{\partial\log {\cal P}({\cal C}^{\tau}|F)}{\partial F_\nu(\vec{x})}\frac{\partial F_\nu(\vec{x})}{\partial F^{\tau}_{\mu\alpha}}.
\end{equation}
We have
\begin{equation}
\frac{\partial\log {\cal P}({\cal C}^{\tau}|F)}{\partial F_\nu(\vec{x})} = \frac{1}2\int_0^\tau dt D^{-1}_{\nu\mu} (\dot{{x}}_\mu(t) -F_\mu(\vec{x}(t))) \delta(\vec{x} -\vec{x}(t))
\end{equation}
Next, the empirical projectors $\hat{c}_\alpha$, corresponding to the trajectory, give the decomposition of the force as
\begin{equation}
F_\nu(\vec{x}) = F^{\tau}_{\nu\alpha} \hat{c}_\alpha(\vec{x})+ F^{\perp}_\nu
\end{equation}
so that
\begin{equation}
\frac{\partial F_\nu(\vec{x})}{\partial F^{\tau}_{\mu\alpha}} = \hat{c}_\alpha(\vec{x})\delta_{\mu\nu}
\end{equation}
and 
\begin{equation}
0=\int d\vec{x}\frac{\partial\log {\cal P}({\cal C}^{\tau}|F)}{\partial F_\nu(\vec{x})}\frac{\partial F_\nu(\vec{x})}{\partial F^{\tau}_{\mu\alpha}}=\int d\mathbf{x}  \hat{c}_\alpha(\vec{x})\int_0^\tau dt ({\vec{x}}_\nu (t)-F_\nu(\vec{x}(t))) \delta(\vec{x} -\vec{x}(t))
\end{equation}
resulting in
\begin{equation}
 \int_0^\tau dt \dot{x}_\nu (t) \hat{c}_\alpha(\vec{x}(t))\underbrace{\int d\vec{x} \delta(\vec{x} -\vec{x}(t))}_1= \int d\vec{x}  \hat{c}_\alpha(\vec{x})F_\nu(\vec{x}) \underbrace{\int_0^\tau dt\delta(\vec{x} -\vec{x}(t)) }_{\tau\hat{P}(\vec{x})} = \tau F^\tau_{\nu\alpha}
\end{equation}
which is solved by our estimator in \Eq{eq:Fn_estimator}. 
This estimator indeed maximizes the log-likelihood, since $  \hat{c}_\alpha(\vec{x})$ is independent of $F_{\mu\alpha}^\tau$ so that 
\begin{equation}
\begin{split}
\frac{\partial\log {\cal P}({\cal C}^{\tau}|F)}{\partial F^{\tau}_{\mu\alpha}\partial F^{\tau}_{\nu\beta} } = \frac{\partial}{\partial F^{\tau}_{\nu\beta} }  \int d\vec{x}\frac{1}2\int_0^\tau dt D^{-1}_{\mu\rho} ({\vec{x}}_\rho(t) -F_\rho(\vec{x}(t))) \delta(\vec{x} -\vec{x}(t))\hat{c}_\alpha(\vec{x})\\
 =- \int d\vec{x}\frac{1}2\int_0^\tau dt D^{-1}_{\mu\nu} \delta(\vec{x} -\vec{x}(t))\hat{c}_\alpha(\vec{x}(t)) \hat{c}_\beta(\vec{x}(t)) = -\frac{\tau}2 \delta_{\alpha\beta}D^{-1}_{\mu\nu}
\end{split}
\end{equation}
which is a negative definite matrix.

\section{Inference of velocities and entropy production}
\label{sec:entropy_production}

In this section, we show how our approach allows the inference of
entropy production, and phase space currents (or more specifically
phase space velocities). We start by some phase-space reminders about
the entropy production, then discuss how to infer the entropy produced
from a given trajectory.

\paragraph*{Phase space entropy production.}

The steady state entropy production rate is defined
via~\cite{seifert_stochastic_2012}
\begin{equation}
\Sdot = \int \dd \mathbf{x} P(\mathbf{x}) v_{\mu}(\mathbf{x})D^{-1}_{\mu\nu}v_{\nu}(\mathbf{x}) = \int \dd \mathbf{x} P(\mathbf{x}) v_{\mu}(\mathbf{x})D^{-1}_{\mu\nu}F_{\nu}(\mathbf{x})
\label{eq:phase_space_S} 
\end{equation}
where $v_{\nu}(\mathbf{x}) = j_{\nu}(\mathbf{x}) /P(x)$ is the phase space velocity, explicitly given by
\begin{equation}
v_{\mu} = F_{\mu}-D_{\mu\nu} \partial_{\nu}\log P(\mathbf{x}) 
\end{equation}
and $j_{\nu}$ is the phase space current. The equality between the two expressions for the entropy production arises from the steady state condition: $\partial_{\mu} j_{\mu} =0$, implying that $g_{\mu} =  \partial_{\nu}\log P(\mathbf{x}) $ is orthogonal to $v_{\mu} $ with respect to the phase space measure. 

The quantity
$\int \dd \mathbf{x} P(\mathbf{x}) v_{\mu}(\mathbf{x})D^{-1}_{\mu\nu}F_{\nu}(\mathbf{x})$ is
the entropy production related to the heat produced in the
bath. Indeed, if the Einstein relation between the mobility and diffusion matrix holds, then this
term corresponds to the average work performed by the force divided by
the temperature. As the system is overdamped, any work performed is
dissipated into heat. Note however that even if we do not assume the Einstein relation holds (\emph{i.e.} that the origin of the white noise is a heat bath), this quantity is related to time irreversibility. 

\paragraph*{Entropy production along a trajectory.}

One can define the entropy production along the trajectory, or equivalently the dissipated heat divided by temperature corresponding to the work performed by the force, as~\cite{seifert_stochastic_2012}
\begin{equation} \label{eq:heat}
\Delta \heat^{\tau} = \int^{\mathrm{Strat}} \mathbf{ D^{-1} F}\cdot \dot{\mathbf{x}} \ \dd t =  \int_0^{\tau}D^{-1}_{\mu\nu}F_{\nu}(\mathbf{x}(t))\circ  dx^t_{\mu}
\end{equation}
where the integral is to be understood in the Stratonovich sense
(which following usual notations we denote as $\circ
dx^t_{\mu}$).
This entropy production is often referred to as the entropy produced
in the medium, and one can also define what is called the total
entropy production along the trajectory
(medium+system)~\cite{seifert_stochastic_2012}. Assuming the initial point is drawn from the steady state pdf, the total entropy production is given by
\begin{equation}
\Delta S^{\tau} = \int_0^{\tau}D^{-1}_{\mu\nu}v^{\nu}(\mathbf{x}(t))\circ  dx^t_{\mu}
\end{equation}

In the limit $\tau\to \infty$, when divided by $\tau$, the two definitions for the entropy production converge to the same limit, equal to the entropy production rate in the system: $\Sdot =\int D^{-1}_{\mu\nu}v_{\nu}(\mathbf{x})v_{\mu}(\mathbf{x})  \dP = \int D^{-1}_{\mu\nu}F_{\nu}(\mathbf{x})v_{\mu}(\mathbf{x}) \dP$.

\paragraph*{Velocity and entropy production inference.}
The probability density $P(\mathbf{x})$ is generally not accessible, so that the phase space velocity cannot be directly computed. However, we have already discussed the empirical density $\hat{P}(\mathbf{x}) $ and we can also define the empirical current (see for example~\cite{touchette_introduction_2018}):
\begin{equation}
\hat{j}_{\mu}(\mathbf{x}) = \frac{1}{\tau}\int_0^{\tau} \delta(\mathbf{x}(t)-\mathbf{x}) \circ dx^t_{\mu}= F_{\mu}(\mathbf{x})\hat{P}(\mathbf{x})-D_{\mu\nu}\partial_{\nu}\hat{P}(\mathbf{x})+\frac{1}{\tau}\int^{\text{It\^o}} \delta(\mathbf{x}(t)-\mathbf{x})  d\xi_{\mu}^t
\end{equation}
using in the last line that $\mathbf{x}(t)$ satisfies the Langevin equation and the relation between It\^o and Stratonovich integrals.
This motivates the definition for the empirical phase space velocity
\begin{equation}\label{eq:empirical_v}
\hat{v}_\mu = \hat{j}_{\mu}(\mathbf{x}) /\hat{P}(\mathbf{x})
\end{equation}
and allows to write
\begin{equation}
\Delta \heat^{\tau}=\tau  \int F_{\mu}(\mathbf{x})  D^{-1}_{\mu\nu} \hat{j}_\nu(\mathbf{x})  d\mathbf{x}=\tau  \int F_{\mu}(\mathbf{x})  D^{-1}_{\mu\nu} \hat{v}_\nu(\mathbf{x}) \hat{P}(\mathbf{x})  d\mathbf{x}
\end{equation}
Note that in this last equation, the force is the exact force, but the
velocity (and probability measure) is the empirical one, defined in \Eq{eq:empirical_v}, so that
we obtain the trajectory-wise entropy production related to the heat, as in \Eq{eq:heat}. If we now
insert into this relation the projection onto the empirical basis of
the force and phase space velocity we get the entropy production
corresponding to that basis:
\begin{equation}
\Delta \heat^{\tau}_b=\tau F^{\tau}_{\mu\alpha}  D^{-1}_{\mu\nu} \hat{v}_{\nu\alpha}
\end{equation}
where 
\begin{equation}
\hat{v}_{\mu\alpha} = \int^{\text{Strat}} \dot{x}_{\mu} \hat{c}_{\alpha}(\mathbf{x}) \frac{dt}{\tau} =  F^{\tau}_{\mu\alpha}+D_{\mu\nu}\int \partial_{\nu}\hat{c}_{\alpha}(\mathbf{x})\frac{dt}{\tau}  + \frac{1}{\tau}\int^{\text{It\^o}} \hat{c}_{\alpha}(\mathbf{x})  \sqrt{2} D^{1/2}_{\mu\nu}d\xi_{\nu}^t = \hat{F}_{\mu\alpha}+D_{\mu\nu}\int \partial_{\nu}\hat{c}_{\alpha}(\mathbf{x})\frac{dt}{\tau}
\end{equation}
using integration by parts.
The estimator for the entropy production related to the basis is
\begin{equation}
\Delta \hat{\heat}_b=\tau \hat{F}_{\mu\alpha}  D^{-1}_{\mu\nu} \hat{v}_{\nu\alpha}= \Delta \heat^{\tau}_b +\tau Z_{\mu\alpha}D^{-1}_{\mu\nu}\hat{v}_{\nu\alpha}.
\end{equation}
It is important to note that the projected entropy production corresponding to the heat is not positive definite, unless we are able to resolve the entire force. Therefore, it does not give a bound on the entropy produced. Furthermore, recall that as was the case for the projection onto the phase space basis, the projected total entropy and that related to heat generically differ.

On the other hand, the projection of the total entropy production is positive definite, and therefore does give a lower bound on the entropy production. The expression $\hat{v}_{\mu\alpha}D^{-1}_{\mu\nu}\hat{v}_{\nu\alpha}$ may be viewed as an estimator of the projection of the total entropy production $\Delta S^{\tau}_b/\tau$, or the total entropy production rate in the steady state $\Sdot_b = v_{\mu\alpha} D^{-1}_{\mu\nu}v_{\mu\alpha}$, however some caution is required. Indeed, consider
\begin{equation}
\Delta S^{\tau}/\tau=  \int v_{\mu}(\mathbf{x})  D^{-1}_{\mu\nu} \hat{j}_\nu(\mathbf{x})  d\mathbf{x}=  \int v_{\mu}(\mathbf{x})  D^{-1}_{\mu\nu} \hat{v}_\nu(\mathbf{x})  \hat{P}(\mathbf{x})d\mathbf{x}.
\end{equation}
(note that one velocity is empirical and the other is exact in this
equation). then
\begin{equation}
\Delta S^{\tau}_b/\tau = v_{\mu\alpha}^{\tau} D^{-1}_{\mu\nu} \hat{v}_{\nu\alpha}
\end{equation}
and we can define the estimator
\begin{equation}
\hat{\Sdot}_b = \hat{v}_{\mu\alpha}D^{-1}_{\mu\nu} \hat{v}_{\nu\alpha}
\end{equation}

This estimator is less controlled than the estimator we have for $\dot{\heat}_b^{\tau}$. Indeed, the estimator $\hat{v}$ has two sources of error as an estimator of $v$. Defining
\begin{equation}
\tilde{v}_{\mu} =F_{\mu} - D_{\mu\nu}\partial_{\nu}\log \hat{P}(\mathbf{x})
\end{equation} 
with the empirical pdf rather than the actual one, we have $\hat{v} = \tilde{v}_{\mu}+\frac{1}{\tau}\int^{\text{It\^o}} \delta(\mathbf{x}(t)-\mathbf{x})  \sqrt{2} D^{1/2}_{\mu\nu}d\xi^{\nu}_t /  \hat{P}(\mathbf{x})$. In particular, for the projection onto the empirical basis:
\begin{equation}
\hat{v}_{\mu\alpha} = \tilde{v}_{\mu\alpha}^{\tau} +Z_{\mu\alpha}
\end{equation}
where $\tilde{v}_{\mu\alpha}^{\tau} -v_{\mu\alpha}^{\tau}= \delta v_{\mu\alpha}^{\tau}\neq 0$, $v_{\mu\alpha}^{\tau}$ being the projection of the actual phase space velocity onto the empirical basis. This is in contrast to the force, where our estimator includes the projection of the actual force.

We write
\begin{equation}
\hat{\Sdot}=\Delta\hat{ S}_b/\tau=\hat{v}_{\mu\alpha}  D^{-1}_{\mu\nu} \hat{v}_{\nu\alpha}= \Delta S^{\tau}_b/\tau +Z_{\mu\alpha}D^{-1}_{\mu\nu}\hat{v}_{\nu\alpha}+ \delta v_{\mu\alpha}^{\tau}D^{-1}_{\mu\nu}\hat{v}_{\nu\alpha}.
\end{equation}
This is a biased estimator, since $\av{Z_{\mu\alpha}D^{-1}_{\mu\nu}\hat{v}_{\nu\alpha}}\approx\av{Z_{\mu\alpha}D^{-1}_{\mu\nu}Z_{\nu\alpha}} = \frac{2N_b}{\tau}$ . We do not have a formal estimate for the last term, $\delta v_{\mu\alpha}^{\tau} D^{-1}_{\mu\nu}\hat{v}_{\nu\alpha}$, but we expect $\delta v^{\tau}_{\mu\alpha} \sim O(1/\sqrt{\tau})$ so that a reasonable estimate seems to be:
\begin{equation}
\hat{\Sdot}_b = \frac{\Delta S^{\tau}_b}\tau +\frac{2N_b}{\tau} +O\left(\sqrt{\frac{2\hat{v}_{\mu\alpha}D^{-1}_{\mu\nu}\hat{v}_{\nu\alpha}}{\tau}+\left(\frac{2N_b}{\tau}\right)^2}\right).
\end{equation}
Here, for the estimate of the fluctuating part (the error term) we
have estimated
$\av{Z_{\mu\alpha}D^{-1}_{\mu\nu}Z_{\nu\alpha}Z_{\rho\beta}D^{-1}_{\rho\sigma}Z_{\sigma\beta}}\sim
O((2N_b/\tau)^2)$,
and the contribution in the square root is the dominant term when
$v^{\tau}_{\mu\alpha}$ is non-zero, i.e there is signal.  We focus on
the long time limit, $\tau\to \infty$,
$\Sdot^{\tau}\to  \Sdot$ so that naturally an
estimator of $\Delta S^{\tau}_b/\tau$ becomes also an estimator of
$\Sdot_b$, with deviations which are again of order
$O(1/\sqrt{\tau})$.  Thus, we may finally estimate
\begin{equation}
\hat{\Sdot}_b  =  \Sdot_b  +\frac{2N_b}{\tau} +O\left(\sqrt{\frac{2\hat{\Sdot}_b}\tau+\left(\frac{2N_b}{\tau}\right)^2}\right).
\end{equation}
Where $\Sdot_b = v_{\mu\alpha} D^{-1}_{\mu\nu}v_{\mu\alpha}$. Note
that this is an order-of-magnitude error estimate, not a fully
rigorous one.

\section{Incomplete observations and time-dependent forces}
\label{sec:time}

In this article, we make strong assumptions on the dynamics of the
system we observe: that it obeys a Langevin dynamics for the observed
degrees of freedom $\mathbf{x}$, and that the force field in phase
space is time-independent. These two assumptions are linked. Indeed,
consider the very relevant case of systems which obey a Langevin
dynamics, but for which not all degrees of freedom are observable. In
that case, the force on the observed degrees of freedom depends on the
state of the hidden variables, therefore apparently violating the
assumptions of our formalism. It is interesting to note however that
this violation is only superficial. Indeed, ``hiding'' some degrees of
freedom of the system is completely equivalent to using a projection
basis where these degrees of freedom do not appear explicitly
(\emph{i.e.} functions that are constant with respect to these degrees
of freedom). Therefore, provided that the system as a whole obeys a
constant-force Langevin equation, SFI will capture the projection of
the dynamics onto the observed degrees of freedom, effectively
averaging over the hidden ones.  Indeed, assume that the force takes the form $F(\mathbf{x},\mathbf{y})$ where only $\mathbf{x}$ can be measured. We thus project the force field onto a set of function $b_\alpha(\mathbf{x})$ that depends only on $\mathbf{x}$. Hence 
\begin{equation}
F_{\mu\alpha}= \int d\mathbf{x} d\mathbf{y} P(\mathbf{x},\mathbf{y}) c_\alpha(\mathbf{x}) F_\mu(\mathbf{x},\mathbf{y})  = \int d\mathbf{x} dy P(\mathbf{y}|\mathbf{x}) F_\mu(\mathbf{x},\mathbf{y}) P(\mathbf{x}) c_\alpha(\mathbf{x}) = \int d\mathbf{x} \bar{F}(x)P(x) c_\alpha(\mathbf{x}) 
\end{equation}
where
$\bar{F}(\mathbf{x}) = \int dy P(\mathbf{y}|\mathbf{x})
F_\mu(\mathbf{x},\mathbf{y})$ is the force at $\mathbf{x}$ averaged
over $\mathbf{y}$. A similar formula applies to the phase space
velocity, as well as when replacing the phase space integral by a time
integral--- in which case one replaces the phase space measure with
the empirical measure.  As a consequence, our formulas for the
projected entropy production and capacity remain valid, and provide
lower bounds to total entropy production and capacity of the system:
\begin{align}
\Sdot_\mathrm{tot} = \int d\mathbf{x} d\mathbf{y} P(\mathbf{x},\mathbf{y}) v_{\mu}(\mathbf{x},\mathbf{y})D^{-1}_{\mu\nu}v_{\nu}(\mathbf{x},\mathbf{y})= \int d\mathbf{x} P(\mathbf{x}) D^{-1}_{\mu\nu}\overline{v_{\mu}v_{\nu}}(\mathbf{x})\geq \int d\mathbf{x} P(\mathbf{x}) D^{-1}_{\mu\nu}\bar{v}_{\mu}(\mathbf{x})\bar{v}_{\mu}(\mathbf{x})\geq D^{-1}_{\mu\nu}v_{\mu\alpha}v_{\mu\alpha} = \Sdot_b
\end{align} 
where we have applied Jensen's inequality twice.

\section{Inference with imperfect data: measurement noise and time discretization}
\label{sec:noise}

Our inference method relies heavily on computing $\dot{\mathbf{x}}$,
\emph{i.e.} the first time derivative of the signal, and on being able
to resolve the difference between It\^o and Stratonovich time
derivatives for (the white noise part of) the signal. One expects that
measurement noise would then swamp the signal and make the distinction
between the two, and thus our inference method, impractical. It turns
out, however, that even in the presence of measurement noise we can
suggest estimators $\hat{v}_{\mu\alpha}$ and $\hat{F}_{\mu\alpha}$
which are unbiased by the measurement noise and accurately capture the
currents and forces, respectively. 

Indeed, let us consider a noisy measure $\mathbf{y}$ of the system's
state $\mathbf{x}$ at discrete times $t_i = i\Delta t$, defined as
\begin{align}
y_\mu(t_i) = x_\mu(t_i) +\eta_\mu^i && \av{\eta_\mu^i \eta_\nu^j} = \Lambda_{\mu\nu} \delta_{i,j}
\end{align}
where $\mathbf{x}$ obeys the dynamics (\ref{eq:langevin}) and $\eta$
is the measurement noise (which we assume to be of zero average,
without loss of generality). We assume this noise to be uncorrelated
between different (discrete) time points. Consider first the estimator
$\hat{F}_{\mu\alpha}^\mathrm{(noisy)}$ for the force projection
coefficient in the presence of noise (we define, as before,
$\Delta y_\mu(t_i) = y_\mu(t_{i+1})-y_\mu(t_i)$ and
$\Delta x_\mu(t_i) = x_\mu(t_{i+1})-x_\mu(t_i)$):
\begin{align} \label{eq:estimator_F_noisy}
  \hat{F}_{\mu\alpha}^\mathrm{(noisy)} & = \frac{1}\tau\sum_i \Delta y_\mu(t_i) c_{\alpha}\left(\mathbf{y}(t_i)\right) \\ 
& = \frac{1}\tau\sum_i \Delta x_\mu(t_i) c_{\alpha}\left(\mathbf{y}(t_i)\right)+\frac{1}\tau\sum_i \Delta t \ c_{\alpha}\left(\mathbf{y}(t_i)\right)  \frac{\eta_\mu^{i+1}-\eta_\mu^i}{\Delta t} 
\end{align}
There are two parts to the error due to measurement noise, one
stemming from the noise in the position and the other from the noise
in the velocity. We assume here that the former is relatively small,
\emph{i.e.} that we can write
\begin{align}
  \label{eq:measurement_noise_ito}
   c_{\alpha}\left(\mathbf{y}(t)\right) \approx  c_{\alpha}\left(\mathbf{x}(t)\right) + \eta_\mu(t) \partial_\mu c_{\alpha}\left(\mathbf{x}(t)\right) + \frac{\eta_\mu\eta_\nu}{2} \partial_{\mu\nu}^2 c_{\alpha}\left(\mathbf{x}(t)\right) + \dots
\end{align}
Then the average (over measurement noise) of the estimator for the
force projection reads
\begin{align}
  \av{\hat{F}_{\mu\alpha}^\mathrm{(noisy)}} = \hat{F}_{\mu\alpha} -  \frac{\av{\eta_\mu \eta_\nu}}{\Delta t} \int \partial_\nu c_{\alpha}(\mathbf{x}(t)) \frac{\dd t}{\tau} + \dots
\end{align}
This second term is a ``dangerous'' bias, as it diverges with
$\Delta t \to 0$, which is symptomatic of the influence of measurement
noise on force inference. \Eq{eq:estimator_F_noisy} is thus
impractical in this case.

In contrast, it is interesting to notice than when doing the same
expansion with the velocity projection coefficients, we have
\begin{align}
  \hat{v}_{\mu\alpha}^{\mathrm{(noisy)}} & = \frac{1}\tau\sum_i \Delta y_\mu(t_i) c_{\alpha}\left(\frac{\mathbf{y}(t_i)+\mathbf{y}(t_{i+1})}{2}\right) \\ 
& = \frac{1}\tau\sum_i \Delta x_\mu(t_i) c_{\alpha}\left(\frac{\mathbf{y}(t_i)+\mathbf{y}(t_{i+1})}{2}\right)+\frac{1}\tau\sum_i \Delta t \ \frac{\eta_\mu^{i+1}-\eta_\mu^i}{\Delta t} c_{\alpha}\left(\frac{\mathbf{y}(t_i)+\mathbf{y}(t_{i+1})}{2}\right)
\end{align}
and
\begin{align}
  \label{eq:measurement_noise_strato}
   c_{\alpha}\left(\frac{\mathbf{y}(t_i)+\mathbf{y}(t_{i+1})}{2}\right) \approx & c_{\alpha}\left(\frac{\mathbf{x}(t_i)+\mathbf{x}(t_{i+1})}{2}\right) + \frac{(\eta_\mu^{i+1}+\eta_\mu^i)}2 \partial_\mu c_{\alpha}\left(\frac{\mathbf{x}(t_i)+\mathbf{x}(t_{i+1})}{2}\right) + \dots
\end{align}
Now all the dangerous terms in $1/\Delta t$ have zero average. Indeed, averaging over the measurement noise,
\begin{eqnarray}
  \av{\frac{(\eta_\mu^{i+1}-\eta_\mu^i)}{\Delta t}   \frac{(\eta_\mu^{i+1}+\eta_\mu^i)}{2} \partial_\nu c_{\alpha}\left(\frac{\mathbf{x}(t_i)+\mathbf{x}(t_{i+1})}{2}\right)} = \frac{ \av{\eta_\mu^{i+1}\eta_\nu^{i+1}-\eta_\mu^i\eta_\nu^i}}{2\Delta t}  \partial_\nu c_{\alpha}\left(\frac{\mathbf{x}(t_i)+\mathbf{x}(t_{i+1})}{2}\right)  = 0
\end{eqnarray}
The reason for these useful cancellations is that by construction, the
velocity projection coefficient is odd under time-reversal of the
trajectory; in contrast, all moments of the measurement noise are even
under time reversal, as it is assumed to be time-uncorrelated. Note
that there remains a fluctuating term which is of the order
$O(\sqrt{\Lambda/\tau\Delta t})$, where $\Lambda$ is the magnitude of the measurement
noise variance. Up to this zero-mean error term, our estimator for the
velocity projection coefficients is thus unaffected by measurement
noise on time derivatives.

To obtain an unbiased estimator for the force, we may use the relation
between It\^o and Stratonovich integration for a variable $x$ which
satisfies the stochastic differential equation (\Eq{eq:langevin}):
\begin{equation}
\begin{split}
\frac{1}\tau\sum_i\Delta x_\mu(t_i) c_{\alpha}\left(\mathbf{x}(t_i)\right)= \frac{1}\tau\sum_i\Delta x_\mu(t_i) c_{\alpha}\left(\frac{x(t_{i+1})+x(t_i)}{2}\right) - D_{\mu\nu}\frac{1}\tau\sum_i \partial_{\nu}c_{\alpha}\left(\frac{x(t_i)+x(t_{i+1})}2\right)\Delta t.
\end{split}
\end{equation}
We can therefore use for the force estimator
\begin{equation} \label{eq:modified_estimator}
\hat{F}_{\mu\alpha} = \hat{v}_{\mu\alpha} -D_{\mu\nu}\frac{1}\tau\sum_i \partial_{\nu}c_{\alpha}\left(\frac{y(t_i)+y(t_{i+1})}2\right)\Delta t
\end{equation}
where we have seen that $ \hat{v}_{\mu\alpha} $ is unbiased by the
noise, and the last term does not include a time derivative of the
measurement and so is also under control.

Note that both the empirical information $\hat{I}_b$ and the estimated
entropy production $\hat{\Sdot}_b$ are now biased by the measurement
noise, the bias being of order $O(1/(\tau\Delta t))$. Thus our
treatment of the measurement noise remains incomplete, and if no other
method is used to take care of the measurement noise, requires
sufficiently large $\tau$ as well as not too small time steps
$\Delta t$. In addition, if the amplitude of the noise is not small
compared to the typical spatial variation of the trajectory then there
are additional biases coming from evaluating the projectors at the
wrong points.
 
Finally, in order to resolve the force correctly, the time step
$\Delta t$ must not be too large: indeed, force variations during
the time step result in a blurring of the inferred force
field. Specifically, the force variation over a time step is, on
average, $\av{\Delta F_\mu} \sim \Delta t \ F_\nu \partial_\nu
F_\mu$. This results in a discretization bias
$\delta \hat{F}_{\mu\alpha}$ in the force estimator
(\Eq{eq:estimator_F_noisy}), the magnitude $\epsilon_\mathrm{discretization}$ of which can be
self-consistently estimated as:
\begin{equation}
  \label{eq:discretization_bias}
   \epsilon^2_\mathrm{discretization} = \frac{\delta \hat{F}_{\nu\alpha} D^{-1}_{\mu\nu} \delta \hat{F}_{\mu\alpha}}{\hat{F}_{\nu\alpha} D^{-1}_{\mu\nu} \hat{F}_{\mu\alpha}} \sim \frac{\Delta t^2}{4\hat{C}} \av{ (\hat{F}_\rho \partial_\rho
\hat{F}_\mu ) D^{-1}_{\mu\nu} (  \hat{F}_\sigma \partial_\sigma
\hat{F}_\mu ) }
\end{equation}
where
$\hat{F}_\mu(\mathbf{x}) = \hat{F}_{\mu\alpha}
\hat{c}_\alpha(\mathbf{x})$ is the inferred force field,
$\hat{C}=\hat{F_{\nu\alpha}} D^{-1}_{\mu\nu} \hat{F_{\mu\alpha}}/4$ is
the inferred capacity, and $\av{\ \cdot \ }$ denotes average over the
trajectory. Note however that when using, as we suggest for ``real''
data, \Eq{eq:modified_estimator} as an estimator for the force
projections, the discretization error is only for the dissipative part
of the force field, \emph{i.e.} only on $\hat{v}$. Indeed, the second
term in \Eq{eq:modified_estimator} does not involve the time ordering
of the data, and is therefore independent of $\Delta t$. Furthermore,
the use of a Stratonovich average for the estimate of
$\hat{v}_{\mu\alpha}$ reduces the squared error in
\Eq{eq:discretization_bias} by a factor $4$.

Comparing the discretization error estimate
(\Eq{eq:discretization_bias}) with the error stemming from the limited
amount of information, \Eq{eq:self_consistant_error}, allows to
self-consistently determine whether the limiting factor to force
inference is the total trajectory length or the frame rate. This is
particularly important for the optimization of the acquisition
protocol in applications such as tracking of fluorescently labeled
biological objects, where photobleaching limits the total number of
frames that can be captured.

\section{Inference in the presence of an inhomogeneous diffusion coefficient}
\label{sec:D} 

We now provide proofs of the results presented in
\Sec{sec:inhomogeneous} of the main text, regarding the inference of
diffusion and drift in the presence of a state-dependent diffusion tensor.  Our
method of inference for the diffusion coefficient follows a similar
logic to that of the inference of the force.  We start with the local
expression
\begin{equation}
D_{\mu\nu}(\mathbf{x}) =\frac{1}2 \lim_{\Delta t\to0}\av{\left.\frac{(\mathbf{x}(t+\Delta t)-\mathbf{x}(t))_\mu(\mathbf{x}(t+\Delta t)-\mathbf{x}(t))_\nu}{\Delta t}\right|\mathbf{x}(t) = \mathbf{x}} 
\end{equation}
and define the projections
\begin{equation}
\begin{split}
D_{\mu\nu\alpha} =  \frac{1}2\int d\mathbf{x} P(\mathbf{x}) D_{\mu\nu}(\mathbf{x}) c_\alpha(\mathbf{x})=\int d\mathbf{x}  \lim_{\Delta t\to0}\av{\delta(\mathbf{x}(t)-\mathbf{x})\frac{(\mathbf{x}(t+\Delta t)-\mathbf{x}(t))_\mu(\mathbf{x}(t+\Delta t)-\mathbf{x}(t))_\nu}{\Delta t}}c_\alpha(\mathbf{x}) \\
= \frac{1}2\av{\int d\mathbf{x} \delta(\mathbf{x}(t)-\mathbf{x})\lim_{\Delta t\to0}\frac{(\mathbf{x}(t+\Delta t)-\mathbf{x}(t))_\mu(\mathbf{x}(t+\Delta t)-\mathbf{x}(t))_\nu}{\Delta t}c_\alpha(\mathbf{x})} \\
=\frac{1}2\lim_{\Delta t\to0}\av{\frac{(\mathbf{x}(t+\Delta t)-\mathbf{x}(t))_\mu(\mathbf{x}(t+\Delta t)-\mathbf{x}(t))_\nu}{\Delta t}c_\alpha(\mathbf{x})}
\end{split} 
\end{equation}
from which we get our estimator
\begin{equation}
\hat{D}_{\mu\nu\alpha}= \frac{1}{\tau} \sum_{i=0}^N \Delta t \ \hat{d}_{\mu\nu}(t_i) \hat{c}_\alpha(x(t_i)) 
\label{eq:D_estimator}
\end{equation}
where we have defined the local diffusion estimator,
\begin{equation}
 \hat{d}_{\mu\nu}(t_i) = \frac{\Delta x_\mu (t_i)\Delta x_\nu (t_i)}{2\Delta t}.
  \label{eq:Dloc_MSD}
\end{equation}

\subsection{Estimate of the error on the projected diffusion coefficient}
\label{sec:D_error}

We now compute the typical error between the estimator
$\hat{D}_{\mu\nu\alpha}$ and the exact projection coefficient
$D_{\mu\nu\alpha}$. We work with the discrete version of the
overdamped Langevin equation (\Eq{eq:Langevin_inhomogenous}), written
using the It\^o convention:
\begin{equation}
  \label{eq:langevin_inh_discrete}
  \Delta x_\mu (t_i) = x_\mu(t_{i+1})-x_\mu(t_{i})  = \Phi_\mu(\vec{x}(t_i))\Delta t + \sqrt{2  D(\vec{x}(t_i))}_{\mu\nu} \Delta \xi_\nu^{t_i}
\end{equation}
where $\Delta \xi^{t_i}_\nu$ is a centered Gaussian
variable with variance
$\av{ \Delta\xi_\nu^{t_i} \Delta\xi_\mu^{t_i}} =\Delta t
\delta_{\mu\nu}\delta_{ij}$.  For error calculations we only consider
the leading order terms in $\Delta t$, so that we can replace
$\Delta x_\mu (t_i)\Delta x_\nu (t_i)$ by
$2 D^{1/2}_{\mu\rho}D^{1/2}_{\nu\sigma} \Delta \xi_\rho^{t_i} \Delta
\xi_\sigma^{t_i}$. Hence:
\begin{equation}
\hat{D}_{\mu\nu\alpha}-D_{\mu\nu\alpha}= \frac{1}{N} \sum_{i=0}^N  D^{1/2}_{\mu\rho}(\vec{x}(t_i))D^{1/2}_{\nu\sigma} (\vec{x}(t_i)) \left(\frac{\Delta \xi_\rho^{t_i}  \Delta \xi_\sigma^{t_i}}{\Delta t} -\delta_{\rho\sigma}\right)\hat{c}_\alpha(\vec{x}(t_i)).
\end{equation} 
We define the normalized (dimensionless) error
\begin{equation}
E_\alpha = \bar{D}^{-1}_{\mu\nu} (\hat{D}_{\mu\nu\alpha}-D_{\mu\nu\alpha}) =  \frac{1}{N} \sum_{i=0}^N \tilde{D}_{\rho\sigma}(\vec{x}(t_i))\zeta_{\rho\sigma}^{t_i} \hat{c}_\alpha(\vec{x}(t_i))
\end{equation}
where $\bar{D}_{\mu\nu} $ is a reference constant diffusion matrix used for the normalization, which could be taken as the average diffusion tensor: $\bar{D}_{\mu\nu} =  \int D_{\mu\nu} (\vec{x}) P(\vec{x})d\vec{x}  $. We have also denoted $\tilde{D}_{\rho\sigma} = D^{1/2}_{\rho\mu}\bar{D}^{-1}_{\mu\nu}D^{1/2}_{\nu\sigma}$ and $\zeta_{\rho\sigma}^{t_i} = \Delta \xi_\rho^{t_i} \Delta \xi_\sigma^{t_i}/\Delta t -\delta_{\rho\sigma}$. Note that $\av{\zeta_{\rho\sigma}^{t_i} }=0$ and
$\av{\zeta_{\rho\sigma}^{t_i}\zeta_{\mu\nu}^{t_j} }=\delta_{ij}(\delta_{\rho\mu}\delta_{\sigma\nu}+\delta_{\rho\nu}\delta_{\sigma\mu})$:
\begin{equation}
\av{\zeta_{\rho\sigma}^{t_i}\zeta_{\mu\nu}^{t_i} } = \av{\left(\frac{\Delta \xi_\rho^{t_i}  \Delta \xi_\sigma^{t_i}}{\Delta t} -\delta_{\rho\sigma}\right)\left(\frac{\Delta \xi_\mu^{t_i}  \Delta \xi_\nu^{t_i}}{\Delta t} -\delta_{\mu\nu}\right)} = \frac{\av{\Delta \xi_\rho^{t_i}  \Delta \xi_\sigma^{t_i}\Delta \xi_\mu^{t_i}  \Delta \xi_\nu^{t_i}}}{\Delta t^2}-\delta_{\rho\sigma}\delta_{\mu\nu} =\delta_{\rho\mu}\delta_{\sigma\nu}+\delta_{\rho\nu}\delta_{\sigma\mu}
\end{equation}
using Wick's theorem in the last equality.
The normalized squared error is then given by:
\begin{equation}
\av{E_\alpha E_\alpha} =\frac{1}{N^2}\sum_{i=0}^N \sum_{j=0}^N \av{\zeta_{\mu\nu}^{t_j} \zeta_{\rho\sigma}^{t_i}  \tilde{D}_{\rho\sigma}(\vec{x}(t_i))\tilde{D}_{\mu\nu}(\vec{x}(t_j))\hat{c}_\alpha(\vec{x}(t_i))\hat{c}_\alpha(\vec{x}(t_j))} 
\end{equation}
We compute the leading order of this error, replacing $\hat{c}_\alpha(\vec{x}(t_j))$ by $c_\alpha(\vec{x}(t_j))$:
\begin{flalign}
\nonumber &\frac{1}{N^2}\sum_{i=0}^N \sum_{j=0}^N \av{\zeta_{\mu\nu}^{t_j} \zeta_{\rho\sigma}^{t_i}  \tilde{D}_{\rho\sigma}(\vec{x}(t_i))\tilde{D}_{\mu\nu}(\vec{x}(t_j))c_\alpha(\vec{x}(t_i))c_\alpha(\vec{x}(t_j))} \\ \nonumber
&= \frac{1}{N^2}\sum_{i=0}^N \av{\zeta_{\mu\nu}^{t_i} \zeta_{\rho\sigma}^{t_i}}\av{  \tilde{D}_{\rho\sigma}(\vec{x}(t_i))\tilde{D}_{\mu\nu}(\vec{x}(t_i))c_\alpha(\vec{x}(t_i))c_\alpha(\vec{x}(t_i))}  =  \frac{1}{N^2}\sum_{i=0}^N\av{  \tilde{D}_{\nu\mu}(\vec{x}(t_i))\tilde{D}_{\mu\nu}(\vec{x}(t_i))c_\alpha(\vec{x}(t_i))c_\alpha(\vec{x}(t_i))}\\ \nonumber
&= \frac{1}N \av{ \int_0^\tau\frac{dt}{\tau}\tilde{D}_{\nu\mu}(\vec{x}(t_i))\tilde{D}_{\mu\nu}(\vec{x}(t_i))c_\alpha(\vec{x}(t_i))c_\alpha(\vec{x})} \\ 
&\leq  \frac{d (D_{\text{max}})^2}N \av{ \int_0^\tau\frac{dt}{\tau}c_\alpha(\vec{x}(t_i))c_\alpha(\vec{x}(t_i))} =\frac{(D_{\text{max}})^2N_b}N = \frac{(D_{\text{max}})^2N_b \Delta t}\tau
\end{flalign}
in the equality in the second line we have used that $\zeta_{\rho\sigma}^{t_i}$ is white in time correlated and centered: \emph{i.e.} that it is uncorrelated with $\vec{x}(t_j)$ for $j\leq i$ and that $\av{\zeta_{\rho\sigma}^{t_i}}=0$, which gives an It\^o isometry type of result for the double sum. In the line before last we have passed to the continuous limit of the sum, using $\tau = N \Delta t$. In the last line we have assumed that $\tilde{D}_{\nu\mu}(\vec{x}(t_i))$ is bounded from above in the domain.  We denote by $D_{\text{max}}$ the maximum eigenvalue of $\tilde{D}_{\nu\mu}(\vec{x}(t_i))$ in the domain, and bound $\tilde{D}_{\nu\mu}(\vec{x}(t_i))\tilde{D}_{\mu\nu}(\vec{x}(t_i))\leq d (\tilde{D}_{\text{max}})^2$.

Let us comment that the correction to the above result, due to the difference between $\hat{c}_\alpha(\vec{x}(t_j))$ and $c_\alpha(\vec{x}(t_j))$ can be bounded in a similar fashion as was done in Section \ref{sec:stats_Z}, if one again uses the assumption that $\tilde{D}_{\nu\mu}(\vec{x}(t_i))$ is bounded in the domain. This correction should result in a term of order $O(\tau^{-3/2})$, which is sub-leading.

To summarize, we have the error estimate
\begin{equation}
\av{\left|\left|\bar{D}^{-1}_{\mu\nu}(\hat{D}_{\mu\nu\alpha}-D_{\mu\nu\alpha})\right|\right|^2} \leq  \frac{(D_{\text{max}})^2N_b \Delta t}\tau
\end{equation}
with $\tilde{D}_{\text{max}}$ the maximum eigenvalue of $\tilde{D}_{\rho\sigma} = D^{1/2}_{\rho\mu}\bar{D}^{-1}_{\mu\nu}D^{1/2}_{\nu\sigma}$ in the domain. Here the choice of normalization $\bar{D}$ is arbitrary, and it may be chosen as a diagonal matrix with the maximal diffusion coefficients in the domain on the diagonal, in a dimensionally consistent way (i.e if there are directions in phase space with different units each has its own maximal diffusion). In that case $D_{\text{max}}$ becomes of order unity.

\subsection{Inference of the diffusion coefficient with measurement noise}
As in \Sec{sec:noise}, we now consider the case where the exact trajectory is not known, but only a noisy approximation of it, due to imperfections of the measurement device.   To correct for such measurement noise, we suggest using the modified estimator 
\begin{equation}
\hat{D}^{\text{(noisy)}}_{\mu\nu\alpha}= \frac{1}{\tau} \sum_{i=0}^N  \Delta t \ \hat{d}_{\mu\nu}^{\mathrm{(noisy)}}(t_i) \ \hat{c}_\alpha(y(t_i))
\label{eq:D_estimator_noise}
\end{equation}
where as in \Eq{eq:Vestergaard} of the main text,
\begin{equation}
  \hat{d}_{\mu\nu}^{\mathrm{(noisy)}}(t_i) = \frac{1}{4\Delta t}\left[\Delta y_\mu (t_i)\Delta y_\nu (t_i) + \Delta y_\mu (t_{i-1})\Delta y_\nu (t_{i-1}) + 2\Delta y_\mu (t_{i-1})\Delta y_\nu (t_i) +2\Delta y_\mu (t_i)\Delta y_\nu (t_{i-1})\right]
  \label{eq:D_local_noisy}
\end{equation}
is the bias-corrected estimator proposed by Vestergaard \emph{et al.}
for homogeneous diffusion inference in the presence of measurement
noise~\cite{vestergaard_optimal_2014}. Indeed, the measurement
noise-induced terms compensate in \Eq{eq:D_local_noisy}, thanks to the
additional cross-terms $\Delta y_\mu (t_{i-1})\Delta y_\nu (t_i)$.

Let us compare the squared error for the corrected estimator
(\Eq{eq:D_estimator_noise}) with that for the estimator
(\Eq{eq:D_estimator}): on the one hand the squared error for
\Eq{eq:D_estimator} has a non-vanishing bias of order
$\bar{D}^{-2}\Lambda^2/\Delta t^2$ due to measurement noise, while
\Eq{eq:D_estimator_noise} only has a contribution of order
$\bar{D}^{-2}\Lambda^2/(\tau \Delta t)$, which vanishes for long
trajectories. On the other hand, the squared error for the corrected
estimator (\Eq{eq:D_estimator_noise}) has an additional contribution
coming from the signal, due to the contributions to
$\av{\zeta_{\rho\sigma}^{t_i}\zeta_{\mu\nu}^{t_i} } $ from
$\Delta y_\mu (t_{i-1})\Delta y_\nu (t_i)/\Delta t +\Delta y_\mu
(t_i)\Delta y_\nu (t_{i-1})/\Delta t$ when squared:
\begin{equation}
\begin{split}
\av{\frac{\Delta \xi_\mu ^{t_{i-1}}\Delta \xi_\rho ^{t_{i-1}}}{\Delta t}} \av{\frac{\Delta \xi_\nu^{t_{i}}\Delta \xi_\sigma^{t_{i}}}{\Delta t}}+\av{\frac{\Delta \xi_\mu ^{t_{i}}\Delta \xi_\rho ^{t_{i}}}{\Delta t}} \av{\frac{\Delta \xi_\nu^{t_{i-1}}\Delta \xi_\sigma^{t_{i-1}}}{\Delta t}}
+\av{\frac{\Delta \xi_\nu ^{t_{i-1}}\Delta \xi_\rho ^{t_{i-1}}}{\Delta t}} \av{\frac{\Delta \xi_\mu^{t_{i}}\Delta \xi_\sigma^{t_{i}}}{\Delta t}}\\
+\av{\frac{\Delta \xi_\nu ^{t_{i}}\Delta \xi_\rho ^{t_{i}}}{\Delta t}} \av{\frac{\Delta \xi_\mu^{t_{i-1}}\Delta \xi_\sigma^{t_{i-1}}}{\Delta t}}= 2\delta_{\mu\rho}\delta_{\nu\sigma}+2\delta_{\nu\rho}\delta_{\mu\sigma}
\end{split}
\end{equation}
giving a squared error that is four times larger than that of the
biased estimator in \Eq{eq:D_estimator}. There is therefore a
trade-off where for short trajectories with sufficiently small
measurement noise the estimator \eqref{eq:D_estimator} may outperform
the corrected estimator, but the (squared) error on it would saturate
at $\bar{D}^{-2}\Lambda^2/\Delta t^2$ for sufficiently long
trajectories, for which the error on the corrected estimator would
continue decreasing. This behavior is demonstrated in
\Fig{fig:diffusion}D in the main text.

\subsection{Drift inference for an inhomogeneous diffusion coefficient}

We now turn to the inference of the It\^o drift (\Eq{eq:drift}).  As
discussed in the main text, in the presence of inhomogeneous diffusion
the force estimator we have used before (\eqref{eq:Fn_estimator})
becomes an estimator for the drift:
\begin{align}
\hat{\Phi}_{\mu\alpha}  = \frac{1}{\tau}\int^{\text{It\^o}} \hat{c}_\alpha(\mathbf{x}) d\mathbf{x}_t^{\mu}  =  \underbrace{\frac{1}{\tau}\int_0^{\tau} \hat{c}_\alpha(\mathbf{x}) \Phi_{\mu}(\mathbf{x})dt}_{\Phi^\tau_{\mu\alpha}} +\underbrace{\frac{1}{\tau}\int^{\text{It\^o}}\hat{c}_\alpha(\mathbf{x}) \sqrt{2} D^{1/2}_{\mu\nu}(\mathbf{x})d\xi_t^{\nu}}_{Z_{\mu\alpha}} 
\label{eq:Phi_estimator}
\end{align}
where as in \Sec{sec:trajectory} we define $\Phi^\tau_{\mu\alpha}$ as
the projection of the exact drift onto
$\hat{c}_\alpha(\mathbf{x})$. This estimator is however biased by
measurement noise, as discussed in \Sec{sec:noise}. To circumvent this
limitation and make our estimators applicable to real, noisy data, we
use again the relation between It\^o and Stratonovich integrals. As in
\Sec{sec:noise}, we thus relate $\hat{\Phi}_{\mu\alpha}$ to
$\hat{v}_{\mu\alpha}$, which can be inferred as before (it is
unaffected by inhomogeneous diffusion) and is unbiased by measurement
noise. We have:
\begin{equation}
\frac{1}{\tau}\int^{\text{It\^o}} \hat{c}_\alpha(\mathbf{x}) d\mathbf{x}_t^{\mu} = \frac{1}{\tau}\int^{\text{Strat}} \hat{c}_\alpha(\mathbf{x}) d\mathbf{x}_t^{\mu} - \frac{1}{\tau}\int D_{\mu\nu} (\mathbf{x})\partial_\nu \hat{c}_\alpha(\mathbf{x})dt = \hat{v}_{\mu\alpha}- \frac{1}{\tau}\int D_{\mu\nu} (\mathbf{x})\partial_\nu \hat{c}_\alpha(\mathbf{x})dt
\label{eq:Ito_Strat_relation}
\end{equation}
To make this a practical estimator, however, one needs to substitute
the unknown $D_{\mu\nu} (\mathbf{x})$ with an accessible value. Using
the standard diffusion estimator (\Eq{eq:Dloc_MSD}) results in an
expression that is mathematically equivalent to \Eq{eq:Phi_estimator}:
it is correct with ideal data, but flawed in the presence of
measurement noise. With ideal data, we thus recommend the use of
\Eq{eq:Phi_estimator}, which is significantly less complex
computationally. In the presence of measurement noise, using the
modified diffusion estimator $\hat{d}_{\mu\nu}^{\mathrm{(noisy)}}(t)$
(\Eq{eq:D_local_noisy}) corrects for the bias induced by measurement
noise. This yields our drift projection estimator adapted to systems
with measurement noise, \Eq{eq:phi_strato} of the main text:
\begin{align}
  \hat{\Phi}_{\mu\alpha} = \hat{v}_{\mu\alpha} - \frac{1}{\tau} \sum_i  \hat{d}_{\mu\nu}^{\mathrm{(noisy)}}(t_i) \  \partial_\nu  \hat{c}_\alpha(\mathbf{x}(t_i)) \ \Delta t
\label{eq:Phi_estimator_strato}
\end{align}
Indeed,
$\av{\hat{d}_{\mu\nu}^{\mathrm{(noisy)}}(t) \ \partial_\nu
  \hat{c}_\alpha(\mathbf{x}(t))} = \av{D_{\mu\nu}(\mathbf{x}(t)) \
  \partial_\nu \hat{c}_\alpha(\mathbf{x}(t))}$: to first order, the
use of the modified local diffusion estimator does not result in a
bias in \Eq{eq:Phi_estimator_strato}.

\subsection{Estimate of the error on the projected drift}

Here we estimate the error on the inference of
$\hat{\Phi}_{\mu\alpha}$. To this end, we employ the It\^o version of
the estimator, \Eq{eq:Phi_estimator}. The error on
\Eq{eq:Phi_estimator_strato} has a similar form, but is analytically
less tractable.

We thus want to estimate the relative magnitude of the error term
$Z_{\mu\alpha}$ in \Eq{eq:Phi_estimator}. The statistics of
$Z_{\mu\alpha}$ can be derived following the derivation in
\Sec{sec:stats_Z}, except that now the diffusion coefficient depends
on $\vec{x}$. Thus, the normalized error $W_{\mu\alpha}$ is defined
using the average diffusion coefficient $\bar{D}_{\mu\nu}$ and the
calculations go through resulting in the same asymptotic
behavior. However, now the variance of the error reads
\begin{equation}
\av{Z_{\mu\alpha}Z_{\nu\beta}}=\frac{2}{\tau}\av{D_{\mu\nu}c_\alpha c_\beta}(1+O(1/\sqrt{\tau}))
\end{equation}
where the space dependence of $D_{\mu\nu}$ prevents us from using the
orthonormality of $c_\alpha$. We thus have
\begin{equation}
\av{\left(\hat{\Phi}_{\mu\alpha}-\Phi^{\tau}_{\mu\alpha}\right)\left(\hat{\Phi}_{\nu\beta}-\Phi^{\tau}_{\nu\beta}\right)}=\frac{2}{\tau}\av{D_{\mu\nu}c_\alpha c_\beta}(1+O(1/\sqrt{\tau}))
\end{equation}
Finally, we can normalize by the average diffusion tensor
$\bar{D}_{\mu\nu}$ to obtain the estimate: 
\begin{equation}
\av{\left(\hat{\Phi}_{\mu\alpha}-\Phi^{\tau}_{\mu\alpha}\right) \bar{D}_{\mu\nu}^{-1} \left(\hat{\Phi}_{\nu\beta}-\Phi^{\tau}_{\nu\beta}\right)} \leq \frac{2 n_b }{\tau}D_{max}
\end{equation}
where we have defined $D_{max}$ as the maximal eigenvalue of the matrix $\bar{D}_{\mu\rho}^{-1} D_{\rho\nu}(\mathbf{x})$ in the domain.

Finally, we note that in our method, the inferred physical force
$\hat{F}_\mu(\mathbf{x})$ is obtained in \Eq{eq:F_D} by combining the
drift with the divergence of the inferred diffusion tensor. As there
is no control of the error on this latter term -- the error on the
gradient is \emph{a priori} independent of the error on the function
estimate, in the absence of regularity assumptions -- we cannot
provide an error estimate for the inferred physical force.

\section{Model details and simulation parameters for numerical results}
\label{sec:simus}

\subsection{Overdamped Langevin simulations}

To benchmark our Stochastic Force Inference method, we test it on
several simple models of Brownian dynamics. We discretize the
overdamped Langevin equation, $\dot{x}_\mu = F_\mu + \xi_\mu$, into
\begin{equation}
  \label{eq:overdamped}
  \mathbf{x}(t+\dd t) = \mathbf{x}(t) +\dd t \ \mathbf{F}(\mathbf{x}(t)) + \sqrt{2 \mathbf{D} \dd t}\ \zeta
\end{equation}
or, in the case of a state-dependent diffusion tensor inducing multiplicative noise, 
\begin{equation}
  \label{eq:overdamped_mult}
  \mathbf{x}(t+\dd t) = \mathbf{x}(t) +\dd t\ \mathbf{F}(\mathbf{x}(t)) + \sqrt{2 \mathbf{D}(\mathbf{x}(t)) \dd t}\ \zeta + \dd t\ \nabla \cdot \mathbf{D}(\mathbf{x}(t))
\end{equation}
Here $\zeta$ is a vector of independent normal random variables with
zero mean and unit variance.  Again, the force here includes the
mobility matrix: the system is out-of-equilibrium if
$\mathbf{D}^{-1} \mathbf{F}(\mathbf{x})$ does not derive from a
potential, regardless of whether this comes from violations of
fluctuation-dissipation relations (such as interacting components at
different temperatures), non-reciprocal interactions or the presence
of curl in the external force fields.  Note that in order to ensure
numerical stability of this equation, the interval $\dd t$ must be
sufficiently small, while SFI can accommodate a moderately large value
of $\dd t$ (see \Sec{sec:capacity}). We therefore run the simulations
at a higher rate than the input for SFI; the value of $\Delta t$
indicated in the parameters is that of the SFI input, while the
elementary time step used to generate the trajectories is denoted
$\dd t$. All simulations presented here have an initial state
pre-equlibrated.

In the simulations presented in this article, the diffusion matrix is
assumed to be known, except in \Fig{fig:diffusion} where
inferring it is part of the object of the simulations. In all other
figures, it could however be inferred using our method (but fitting it
only with a constant). In general, in the strong-noise cases
considered in this article, inferring the diffusion coefficient is
significantly less demanding than force inference, and results in very
little additional error.

\subsection{2D Ornstein-Uhlenbeck processes (Figure~\ref{fig:2dOU})}

The first model we benchmark our method on is a 2D process in a linear
trap, also known as an Ornstein-Uhlenbeck process. We consider here an
anisotropic equilibrium process with isotropic diffusion; we set the
diffusion to unity, $D_{\mu\nu} = \delta_{\mu\nu}$. The force field is
$F_\mu = -\Omega_{\mu\nu} (x_\nu-x^0_\mu) $ (black arrows in Fig. 1F), where we
choose
\begin{equation}
\mathbf{x}^0 =   \left( \begin{matrix}
    0 \\
    0
  \end{matrix} \right)
\qquad
  \Omega = \left(
  \begin{matrix}
    1 & 0.5 \\
    0.5 & 1 
  \end{matrix} \right)
\end{equation}
We use a simulation timestep $\dd t = 0.005$ and $\Delta t = 0.01$.
The trajectory presented in Fig. 1C and analyzed in Fig 1G of the main
text has a length $N_\mathrm{samples} = 4000$. It is analyzed by SFI
with basis $b=\{1,x_1,x_2\}$. The inferred projected force field on
this basis (blue arrow in Fig. 1F) has the form
$\hat{F}_\mu(\mathbf{x}) = -\hat{\Omega}_{\mu\nu}( x_\nu - \hat{x}_0)
$ where the $N_b=6$ inferred parameters are
\begin{equation} \label{eq:OU-inf}
\hat{x}_0 =   \left( \begin{matrix}
    0.27 \\
    0.13 
  \end{matrix} \right)
\qquad
  \hat{\Omega} = \left(
  \begin{matrix}
    1.15 & 0.27 \\
    0.42 & 0.76 
  \end{matrix} \right)
\end{equation}
Quantitatively, as mentioned in the main text, this results in a
(squared) relative error on the inferred projection coefficient
$[(\hat{F}_{\mu\alpha}-F_{\mu\alpha})D^{-1}_{\mu\nu}(\hat{F}_{\nu\alpha}-F_{\nu\alpha})]/[\hat{F}_{\mu\alpha}D^{-1}_{\mu\nu}\hat{F}_{\nu\alpha}]
= 0.15$.
The inferred information along this trajectory is
$\hat{I}_b = \hat{F}_{\mu\alpha}D^{-1}_{\mu\nu}\hat{F}_{\nu\alpha} =
19.1$
(\emph{i.e.} $27.6$ bits with the $1/log(2)$ nat-to-bit conversion
factor). The self-consistent confidence interval for this error is
$N_b/2\hat{I}_b = 0.16$: the actual error is thus within the
confidence interval.

It is interesting to note that the inferred matrix $\hat{\Omega}$
(\Eq{eq:OU-inf}) is not symmetric, meaning that the inferred model is
out-of-equilibrium (it exhibits phase space cycling). This does not,
however, result in significant entropy production. Indeed, the
inferred entropy produced is $\hat{\Delta S} = 0.5 k_B$.

In Fig 1G of the main text, we study the statistics of the relative error,
obtained over 64 realizations of trajectories of the same model, with
varying length $N_\mathrm{samples}=2^4,2^5,...,2^{17},2^{18}$.  We
present the average (and standard deviation, blue symbols and error
bars) of the squared relative error
$[(\hat{F}_{\mu\alpha}-F_{\mu\alpha})D^{-1}_{\mu\nu}(\hat{F}_{\nu\alpha}-F_{\nu\alpha})]/[\hat{F}_{\mu\alpha}D^{-1}_{\mu\nu}\hat{F}_{\nu\alpha}]$;
the average self-consistent estimate of this error $N_b/2\hat{I}_b$
(orange solid curve), and the asymptotic convergence to
$N_b/2\tau C_b$, \emph{i.e.} the actual information per degree of
freedom (black dashed line). These quantities match quantitatively in
the long trajectory limit, as predicted from our analytical reasoning
(\Sec{sec:trajectory}). Interestingly, in the regime where there is
little information available in the trajectory, our self-consistent
formula reliably predicts a relative error of order $1$, consistent
with the fact that there is no signal.

\subsection{6D circulating Ornstein-Uhlenbeck processes (Figure~\ref{fig:6D})}

The next example we use to test SFI is another Ornstein-Uhlenbeck
process with force $F_\mu = -\Omega_{\mu\nu} (x_\nu-x^0_\mu) $, but this
time with several complications: it is high-dimensional ($d=6$), with
anisotropic diffusion and trapping, and such that we exert a torque in
a given plane. We challenge our method by applying it to the short
trajectories displayed in Fig. 1D in the main text, and even further
in Fig. 1E in the presence of strong measurement noise.

The diffusion and harmonic trapping matrices are obtained as random
matrices constructed to have a moderate degree of anisotropy. The
diffusion matrix is symmetric, while the confinement is not and
induces circulation. Specifically we choose:
\begin{equation}
\mathbf{\Omega} = \left(
  \begin{matrix}
    1.34 & -0.25 & -0.   &  0.73 &  0.38 &  0.23\\
    -0.07 &  1.77 & -0.45 &  1.92 &  0.88 & -0.09\\
    0.24 &  0.52 &  0.81 & -0.63 &  0.05 &  0.97\\
    -0.24 & -1.14 &  0.52 &  0.93 & -0.32 & -0.69\\
    0.16 & -0.01 &  0.07 &  0.66 &  0.92 & -0.02\\
    0.51 &  0.52 &  0.27 &  0.79 &  0.61 &  2.45
  \end{matrix} \right) \quad 
\mathbf{D} = \left(
  \begin{matrix}
    1.92 &  1.27 &  0.29 & -0.18 &  0.2  & -0.02\\
    1.27 &  1.87 &  0.26 & -0.1  &  0.11 & -0.25\\
    0.29 &  0.26 &  0.98 & -0.45 &  0.06 &  0.09\\
    -0.18 & -0.1  & -0.45 &  1.03 & -0.17 & -0.15\\
    0.2  &  0.11 &  0.06 & -0.17 &  0.84 &  0.09\\
    -0.02 & -0.25 &  0.09 & -0.15 &  0.09 &  0.81
  \end{matrix} \right) 
\end{equation}
and $\mathbf{x}_0=0$. Our simulation parameters are $\Delta t = 0.05$
and $\dd t = 0.01$. The trajectory presented in Fig. 1E has
$N_\mathrm{samples} = 400$ points, and the three plots correspond to
three projections of the same trajectory, respectively (from left to
right) along directions $(x_1,x_2)$, $(x_3,x_4)$ and $(x_5,x_6)$.

In Fig. 1H, we present the results of SFI at linear order
($b=\{1,x_\mu\}$) for the specific trajectory displayed in 1E. The inferred parameters are:
\begin{equation}
\hat{x}_0 =   \left( \begin{matrix}
    -0.86\\
    -0.64\\
    -0.29\\
    -0.46\\
    -0.25\\
    0.25
  \end{matrix} \right)
\qquad
  \hat{\Omega} = \left(
    \begin{matrix}
       2.38& -1.24&  0.47&  0.4 &  0.19&  0.29\\
        0.96&  1.06& -1.01&  0.92&  1.59& -0.91\\
       -0.16&  0.44&  1.09& -1.13&  0.58&  0.96\\
        0.18& -1.36&  1.07&  1.27& -0.91& -0.87\\
        0.61& -0.28& -0.  &  0.36&  1.01&  0.22\\
        0.25&  0.29&  0.86&  0.91&  0.29&  3.  
  \end{matrix} \right)
\end{equation}
with a squared relative error of $0.24$, consistent with the
self-consistent estimate $N_b/2\hat{I}_b = 0.22$. 

We show in Fig. 2H in the main text a 2D-slice of the inferred force
field (blue) and the exact force field (black). This slice is chosen
as the plane of maximal inferred circulation. To determine this plane,
we consider the  non-dimensionalized velocity projection coefficients,
$R_{\alpha\beta} = C^{-1/2}_{\alpha\mu}\hat{v}_{\mu\beta}$, with
$\mathbf{C}$ the covariance matrix of the data. With this choice of
normalization, the rows and columns of $\mathbf{R}$ are normalized in
the same way, and it thus makes sense to consider its antisymmetric
part to quantify circulation. The eigenvalues of
$\frac{1}{2}(R_{\alpha\beta} - R_{\beta\alpha})$ are imaginary and
come in conjugate pairs. We define the inferred principal circulation
plane as the real-space plane $(\mathbf{u},\mathbf{v})$ spanned by
$(C^{1/2}_{\mu\alpha} r^1_\alpha,C^{1/2}_{\mu\alpha} r^2_\alpha)$,
where $(r^1_\alpha,r^2_\alpha)$ is the pair of eigenvectors of
$\mathbf{R}$ associated to the eigenvalue of largest norm. We compare
this inferred plane to the exact plane of maximal circulation
$(\mathbf{u}^0,\mathbf{v}^0)$, obtained through the same procedure but
with an asymptotically long trajectory ($N_\mathrm{steps}=2.10^6$).
In Fig 1J, we present the statistics of the angular error in this
cycle detection. This angular error is defined as
$\delta = \|\mathbf{u} -
(\mathbf{u^0}.\mathbf{u})\mathbf{u^0}-(\mathbf{v^0}.\mathbf{u})\mathbf{v^0}\|^2
+ \|\mathbf{v} -
(\mathbf{u^0}.\mathbf{v})\mathbf{u^0}-(\mathbf{v^0}.\mathbf{v})\mathbf{v^0}\|^2$,
where $(\mathbf{u},\mathbf{v})$ and $(\mathbf{u}^0,\mathbf{v}^0)$ are
the pairs of orthogonal unit vectors defining the inferred and exact
maximal circulation planes, respectively. This error is equal to
$0.12$ for the trajectory presented in Fig 1D, and decays to zero as
$\delta\sim\tau^{-1}$ with increasing trajectory length, as the
inferred matrix $\hat{\Omega}$ converges to $\Omega$. Fig 1K shows the
statistics of the de-biased entropy production,
$\hat{\Sdot}-2N_b/\tau$.

\paragraph*{Measurement noise.}

In Fig 1E, we present the same trajectories as in Fig 1D, with an
added challenge to force detection: a strong ``measurement noise'',
\emph{i.e.} a time-uncorrelated error on the input data $x_\mu$. We
model such noise by adding Gaussian white noise to each coordinate of
$x_\mu$, with standard deviation equal to $0.5$ (half the standard
deviation of the data). In the presence of such time-uncorrelated
noise, the estimate of $\dot{x}$ becomes strongly noisy, and we have
to used the modified estimator for $\hat{F}_{\mu\alpha}$,
\Eq{eq:modified_estimator}. With this estimator, we infer:
\begin{equation}
\hat{x}_0 =   \left( \begin{matrix}
-0.83\\
-0.64\\
-0.07\\
-0.51\\
-0.24\\
0.12
\end{matrix} \right)
\qquad
  \hat{\Omega} = \left(
  \begin{matrix}
 1.92& -1.04&  0.26& -0.09&  0.18& -0.26\\
 1.07&  0.71& -1.12&  0.28&  1.01& -1.31\\
 -0.19&  0.48&  0.88& -1.19&  0.21&  0.83\\
 0.03& -0.66&  0.78&  0.92& -0.2 & -1.12\\
 0.27&  0.06& -0.04& -0.1 &  1.01& -0.17\\
 0.21&  0.23&  0.45&  0.67&  0.32&  1.91
  \end{matrix} \right)
\end{equation}
with a squared relative error of $0.6$ on $\hat{F}_{\mu\alpha}$ and an
angular error on cycle detection of $0.156$.  

\subsection{Nonlinear obstacle process (Figure~\ref{fig:nonlinear}A, C, E, G, I)}

In Figure 2A, we study the case of a 2D stochastic process with
circulation in a nonlinear force field, using Stochastic Force
Inference with a polynomial basis at different orders. The force field
we use is:
\begin{equation}
  \label{eq:obstacle}
  F_\mu(\mathbf{x}) = -\Omega_{\mu\nu} x_\nu + \alpha e^{-x^2/2\sigma^2} x_\mu \qquad \mathrm{with}\qquad \alpha = 10\quad , \quad  
\Omega = \left(
  \begin{matrix}
    2 & 2 \\
   -2 & 2 
  \end{matrix} \right)
\end{equation}
which is a non-polynomial force field, \emph{i.e.}  it cannot be
captured exactly in our choice of basis. We use isotropic diffusion
with $D=1$. We simulate this process with $\Delta t = 0.01$ and
$\dd t = 0.001$; the trajectory in Fig 2A has
$N_\mathrm{samples}=4096$. We perform SFI on the trajectory with a polynomial basis at orders
$n=1,3,5$ in Figs.2C,E,G; note that as the force field is odd under
reversal $\mathbf{x}\to-\mathbf{x}$, the even orders in the polynomial
expansion do not contribute to it (as apparent in the $n$-dependency
of the capacity in Fig 1I). The bootstrapped trajectories presented on
the right column of Fig 2C,E,G are obtained using the inferred
projected force field,
$\hat{F}_{\mu\alpha}\hat{c}_\alpha(\mathbf{x})$, to simulate new
trajectories with the same starting point, $\tau$, $\dd t$ and
$\Delta t$ as the original trajectory.

In Fig 2I, we present the capacity $C_b$ and entropy production
$\Sdot_b$ captured by the projection of a long trajectory with
$N_\mathrm{samples}=2^{18}$ onto three different bases:
\begin{itemize}
\item Polynomials of order $n=0...7$.
\item Fourier modes of order $n=0...7$; specifically, we use all
  functions of the form
  $\cos\left( 2\pi \sum_\mu k_\mu (x_\mu - \av{x_\mu} )/R_\mu \right)$
  and
  $\sin\left( 2\pi \sum_\mu k_\mu (x_\mu - \av{x_\mu} )/R_\mu \right)$
  with non-negative integers $k_\mu$ such that
  $\sum_\mu k_\mu \leq n$. Here we choose $R_\mu$ to be $1.05$ times
  the diameter of the trajectory in direction $\mu$.
\item A constant-by-part grid coarse-graining with $n=2...7$ grid
  cells in each direction, centered on $\av{x_\mu}$ and with width
  $R_\mu$.
\end{itemize}

\subsection{Lorenz process (Figure~\ref{fig:nonlinear}B, D, F, H, J)}

Our second nonlinear process is a stochastic variant of a popular
model for dynamical systems, the Lorenz
system~\cite{allawala_statistics_2016}. Its 3D Brownian dynamics is
described by the force field
\begin{equation}
  \label{eq:Lorenz}
  F_x = s(y-x) \qquad ; \qquad F_y = rx-y-zx \qquad ; \qquad F_z = xy - bz
\end{equation}
In our simulations, we employ the parameters $r=10$, $s=3$ and
$b=1$. Diffusion is isotropic with $D=1$. We use
$\Delta t = \dd t = 0.02$, and the trajectory in Fig 2B has
$N_\mathrm{samples}=2^{12}$. All images of trajectories are in the
$(xz)$ plane. It should be noted that this force field is polynomial
of order 2, implying that it can be fully captured by the order $n=2$
of our polynomial expansion. Indeed, with polynomial SFI at orders 2
and 3 (Fig 2F,H) we capture precisely the force field, and
bootstrapped trajectories are very similar to the original data. As
apparent in Fig 2J, the order $n=1$ polynomial approximation only
captures a fraction of the capacity and entropy
production. Interestingly, the order $n=2$ polynomial approximation
captures the whole capacity, but \emph{not} the full
entropy production, as there are nonzero exchange terms with higher
order moments (corresponding to the fact that the logarithm of the pdf
is not itself a polynomial).

\subsection{Active Brownian Particles simulations (Figure~\ref{fig:particles})}
\label{sec:ABP}
 
The next system studied in this article corresponds to a model of
self-propelled Brownian particles, mimicking in a somewhat realistic
manner experimental systems such as studied in
Refs.~\cite{palacci_living_2013}. Specifically,
we simulate $N_\mathrm{particles}=25$ self-propelled 2D particles,
each characterized by its coordinates $\mathbf{x}$ and orientation
$\theta$. These particles interact through soft repulsive pair
interactions $f(r)$ between particles at distance $r$,
are self-propelled towards the direction $\theta$ at velocity $v$, and
are harmonically confined with strength $\omega$: the force exerted on
particle $i$ is thus
\begin{equation}
  \label{eq:ABP}
  \mathbf{F}_i = -\omega \mathbf{x}_i + v  \left( \begin{matrix}
    \cos \theta_i \\
    \sin \theta_i
  \end{matrix} \right)
 - \sum_{j \neq i} f(r_{ij}) \frac{\mathbf{r}_{ij}}{{r}_{ij}}
\end{equation}
where $\mathbf{r}_{ij} = \mathbf{x}_j-\mathbf{x}_j$. The angle
$\theta$ is freely diffusing (note that we could include alignment
interactions in this model). In our simulations we use
$f(r) = 1/(r^2 + 1)$, $\omega = 0.2$, $v = 1$, isotropic diffusion
with $D=1$ in spatial coordinates and angular diffusion with
$D_\theta=0.1$. We use a large sampling time step $\Delta t = 1$,
while the simulation step is $\dd t =0.01$. The number of frames for
our study is very limited, $N_\mathrm{frames}=25$, with significant
positional and angular measurement noise (on both $x$, $y$ and
$\theta$ with standard deviation $0.4$). These limitations are chosen
to mimic those of experimental data. Note that we assume that the
identity of the particles can be tracked along the trajectory.

\paragraph*{Symmetrization of the forces.}

Each of the $25$ particles being characterized by three degrees of
freedom, the phase-space of this system is $75$-dimensional, making
any ``brute-force'' approximation of the force field in phase space
hopeless: even a simple form such a linear polynomial (which would be
a terrible approximation of $\Eq{eq:ABP}$) would have $5700$
variables. Here we propose to use a more subtle projection basis,
making use of the invariance of the force field when exchanging two
particles. More precisely, instead of using a projection basis
$b_\alpha(\{\mathbf{x}_i\}_{i=1..N_\mathrm{particles}})$ that depends
on each phase space coordinate in an explicit way, we will project on
symmetrized functions
$b_\alpha(\mathbf{x}_i,\{\mathbf{x}_j\}_{j\neq i})$ that consider the
interaction between one particle $i$ and all others, regardless of the
identity of $i$. The projected force field thus consists in an
approximation of the force on any particle $i$ as
\begin{equation}
  \label{eq:projection_particles}
  F_{i,\mu} \approx F_{\mu\alpha} c_{\alpha}\left(\mathbf{x}_i,\{\mathbf{x}_j\}_{j\neq i}\right)
\end{equation}
where, crucially, the projection coefficient $F_{\mu\alpha} $ and the
projector $c_{\alpha}$ are independent of the identity of $i$. This
drastically reduces the number of degrees of freedom of our
approximation: now the data on each particle contributes to the
inference of the same coefficients $F_{\mu\alpha} $, and thus a large
number of particles actually facilitates force inference. These
additional symmetry constraints on the projection do not fit strictly
speaking in the framework developed in the rest of this
article. Specifically, the orthonormalization of the projector is now
performed with an additional average over all particles:
\begin{equation}
  \label{eq:particles_orthonormal} \hat{c}_\alpha = \hat{B}_{\alpha\beta} b_\beta \qquad \mathrm{with}\qquad
  \hat{B}_{\alpha\beta} = \frac{1}{\tau N_\mathrm{particles}} \sum_i \int \dd t \ b_\alpha\left(\mathbf{x}_i(t),\{\mathbf{x}_j(t)\}_{j\neq i}\right) b_\beta\left(\mathbf{x}_i(t),\{\mathbf{x}_j(t)\}_{j\neq i}\right)
\end{equation}
and all integrals are adapted accordingly; for instance, the It\^o
integral for the force projection now reads
\begin{equation}
  \label{eq:Fmualpha_particles}
  \hat{F}_{\mu\alpha} =  \frac{1}{\tau N_\mathrm{particles}} \sum_i \sum_t \left(x_{i,\mu}(t+\Delta t) - x_{i,\mu}(t) \right) \hat{c}_\alpha\left(\mathbf{x}_i(t),\{\mathbf{x}_j(t)\}_{j\neq i}\right) 
\end{equation}
with $\Delta t$ the time step.

\paragraph*{Choice of the basis.}
So far, we have only use the indiscernibility of the particles,
without any assumption on the nature of their interactions:
\Eq{eq:projection_particles} is completely generic, and could in
principle approximate any type of interactions -- provided that the
choice of projection basis is adapted. For instance, a natural choice
would be to expand the interaction in single-particle terms
(\emph{i.e.} external fields), pair interactions, and possibly higher
orders, as 
\begin{equation}
  \label{eq:projection_particles_pairs}
  F_{i,\mu} \approx F^{(1)}_{\mu\alpha} c^{(1)}_{\alpha}\left(\mathbf{x}_i\right) +  F^{(2)}_{\mu\beta} \sum_{j\neq i}c^{(2)}_{\beta}\left(\mathbf{x}_i,\mathbf{x}_j\right) +  F^{(3)}_{\mu\gamma} \sum_{j,k\neq i}c^{(3)}_{\gamma}\left(\mathbf{x}_i,\mathbf{x}_j,\mathbf{x}_k\right) + \dots
\end{equation}
where $c^{(1)},c^{(2)},c^{(3)}\dots$ are the respective projectors
onto the space spanned by the 1-, 2- and 3-body interaction terms in
the basis. It is important to note that these projectors should be
orthonormalized as a whole, either hierarchically (through the
Gram-Schmidt process, for instance by orthonormalizing the 1-body
term, then the 2-body term with respect to itself and the 1-body term,
etc.) or in a single step as in \Eq{eq:particles_orthonormal}, but
with the index $\alpha$ now understood as comprising all terms in the
expansion.

Let us also note that while polynomials constitute a natural
``default'' basis for generic processes in an unstructured phase
space, no such natural choice exist for the interaction
terms. Symmetries can serve as a guide: for instance, for
radially/spherically symmetric particles the magnitude of the pair
interaction should depend on the distance $r_{ij}$ between
particles. The use of such symmetries warrants some caution: indeed,
the choice of projection basis should be compatible with these
symmetries. For instance, for radial symmetry, the basis
$b = \{(\mathbf{x}_i,\mathbf{x}_j) \mapsto r_{ij}^n \}_{n=0,1,2...}$,
\emph{i.e.} polynomials in the distance between particles, is
\emph{not} adapted. Indeed, a force written as a linear combination of
these functions would transform as a scalar under rotations, not as a
vector. Instead,
$b = \{(\mathbf{x}_i,\mathbf{x}_j) \mapsto r_{ij,\mu} r_{ij}^{n-1}
\}_{\mu = 1..d, n=0,1,2...}$
would be adapted. This does not constrain the force to be invariant
under rotation, but allows it. Finally, let us note that while this
choice is fine, it is not great: indeed, polynomials in $r$ put most
of their weight in the far-field, \emph{i.e.} in interaction between
far-away particles: SFI will thus put most weight on capturing the
tail of the interaction. In most cases, interactions decay with
distance, and it is more interesting to capture the details of the
interaction forces between nearby particles. For this reason, decaying
functions of $r$, such as inverse power-laws or exponentials, are
better adapted. We finally note that non-power-law functions typically
have a characteristic scale, or shape parameters. These parameters are
not optimized upon by SFI, which only fits the signal as a linear
combination of the basis functions: the outcome will thus depend on
the choice of parameter. While such shape parameters could in
principle be optimized upon (for instance to maximize the inferred
information captured by SFI), we find that in practice it is simpler,
both computationally and analytically, to improve the precision of SFI
by expanding the basis than by performing such shape parameter
optimization. We leave this possibility open for future work.

Motivated by these considerations, in practice, our choice of basis for
Figure 3 of the main text is
\begin{equation}
  \label{eq:Fig3_basis}
  b^{(1)} = \{ x_\mu,\cos \theta,\sin\theta\} \qquad b^{(2)} = \{ r_{ij,\mu} r_{ij}^{k-1} \exp(-r_{ij} / r_0 )\}_{k=0..5} 
\end{equation}
where we choose $r_0 = 2$, corresponding to half the first peak in the
radial distribution function. The outcome of SFI is not significantly
affected by small changes in the number of functions or their shape.

\subsection{One-dimensional ratchet process (Figure~\ref{fig:diffusion}A-D)}
\label{sec:Buttiker}

Figure 8 of the main text deals with the case of Brownian dynamics
with multiplicative noise, \emph{i.e.} with a space-dependent
diffusion tensor. Panels A-D treat a minimal example of it: a 1D
ratchet process, where an out-of-equilibrium current is driven by the
combination of a periodic space-dependent diffusion coefficient and a
periodic force, such that the fluctuation-dissipation relation is not
satisfied for a unique temperature. This model falls within the class
described by Buttiker~\cite{buttiker_transport_1987} and
Landauer~\cite{landauer_motion_1988}. Specifically, we consider a
process on the segment $[0,1]$, with periodic boundary conditions. The dynamics is
described by \Eq{eq:overdamped_mult}, with
\begin{equation}
  \label{eq:BL}
  F(x) = F_0 \cos(2\pi x) \qquad \mathrm{and} \qquad   D(x) = D_0 + a \cos(2\pi x)
\end{equation}
where we choose $F_0 = -2$, $D_0 = 1$, $a = 0.5$, and the
discretization step is $\Delta t = 0.005$. The trajectory presented in
Fig. 8A has 10,000 steps.

In Fig 8B-C we perform SFI on the trajectory in A, using an adapted
basis, with $b = \{1, cos(2\pi x), \sin(2\pi x)\}$, for both the
diffusion and the force. In Fig 8D we present the convergence of the
inferred fields as a function of the trajectory duration for $n=32$
repeats.

\subsection{Minimal 2D model with diffusion gradient (Figure~\ref{fig:diffusion}E-H)}
\label{sec:Harmonic_gradient}
 
We next consider a minimal 2D equilibrium model with inhomogeneous
diffusion: an Ornstein-Uhlenbeck process with a constant gradient of
isotropic diffusion coefficient.  Specifically, we choose the
following form for the space-dependent diffusion tensor:
\begin{equation}
  \label{eq:4A-D}
  D_{\mu\nu}(\mathbf{x}) = (1 + a_\rho x_\rho) \delta_{\mu\nu} \qquad \mathrm{with} \qquad  \mathbf{a} = \left(
  \begin{matrix}
    0.25 \\
   0 
  \end{matrix} \right)
\end{equation}
and the following force field:
\begin{equation}
  \label{eq:4A-force}
  F_\mu (\mathbf{x}) = - D_{\mu\nu}(\mathbf{x}) x_\nu
\end{equation}
corresponding to a potential well with energy
$E(\mathbf{x}) = \mathbf{x}^2/2$ and a space-dependent mobility matrix
equal to the diffusion tensor $D_{\mu\nu}(\mathbf{x})$ (\emph{i.e.}
the system obeys the Einstein relation with $k_B T=1$). This choice
ensures that the probability distribution function of the process is
unaffected by the inhomogeneity of $\mathbf{D}(\mathbf{x})$. We
simulate this process using the discretized version of Eq. 8 of the
main text, with $\Delta t = dt = 0.02$. The trajectory showed in panel
4A and analyzed in panels B and C has length
$n_\mathrm{steps}=4096$. The blue symbols in panel 4D show the
convergence of the diffusion estimator with increasing trajectory
length $N_\mathrm{steps} = 2^4 \dots 2^{15}$. The green and orange
symbols correspond to the same data, with added measurement noise with
amplitude 0.075.

\subsection{Reconstruction of the drift and diffusion field for a complex 2D process (Figure~\ref{fig:comparison})}
\label{sec:multitraps}

In our last Figure, we present a comparison of SFI with two
pre-existing methods, grid binning and InferenceMAP. To this end, we
simulate a model designed to mimic the diffusion of single molecules
in a complex cellular environment. To allow for quantitative
comparison with the other methods, we consider here the inference of
the drift field, rather than the physical force, and an isotropic
space-dependent diffusion tensor. The diffusion coefficient is
constructed as the ratio of two second-order polynomials in the
coordinates, with randomly generated coefficients. The drift field is
chosen as the sum of an overall harmonic trap with constant torque,
three attractive Gaussian traps in a triangle, and a repulsive one at
the center. Typical scales are $\Phi\sim 1$, $D\sim 1$, the spatial
extent of the process is $\sim 4$, and we choose a time step
$\Delta t = 0.01$. We consider two types of input signal: exact data,
and noisy data where each coordinate is blurred by a Gaussian white
noise of amplitude $0.1$ (represented as a red dot in
\Fig{fig:comparison}B).

In single molecule contexts, the total duration of a trajectory is
typically limited by photobleaching: the exploration of a cellular
environment is only possible by accumulating many such tracks. To
reproduce this fact, we use $N=4\dots 10^4$ independently generated
finite-duration trajectories with $100$ time steps (four of which
are depicted in \Fig{fig:comparison}B), each starting at
steady-state. Each individual track contains, on average, an
information gain of $\av{I} = 2.8$ bits about the drift field. We
study the convergence of each method to the true drift and diffusion
fields as $N \to \infty$. The performance of drift and diffusion
inference are assessed as the mean-squared-error between exact and
inferred fields along the trajectory, normalized by the mean squared
inferred value. We now detail the parameters employed for each of the
three methods.

\subsubsection*{Stochastic Force Inference.}

We employ a Fourier basis over a window spanning $1.1\times$ the total
process extent for both diffusion and drift inference. The order
$n=1\dots 9$ of the basis is adapted to the number $N$ of
trajectories, as $n = \lfloor \log(N) \rfloor$. We employ noise-free
estimators for the exact signal, and noise-corrected estimators
(\Eq{eq:Vestergaard} and \Eq{eq:phi_strato}) for noisy data.

\subsubsection*{Maximum-likelihood grid binning.}

The principle of this method is simple: decomposing the phase space as
a regular grid, and inferring a constant drift vector and diffusion
coefficient in each bin using maximum-likelihood estimators. The
estimators are:
$$ \hat{\Phi}(\mathbf{x}) = \frac{1}{N(\mathbf{x})} \sum_{i,\ \mathbf{x}(t_i)\in \mathbf{x}} \frac{\mathbf{x}(t_{i+1})-\mathbf{x}(t_i)}{\Delta t} \qquad ; \qquad \hat{D}(\mathbf{x}) = \frac{1}{N(\mathbf{x})} \sum_{i,\ \mathbf{x}(t_i)\in \mathbf{x}} \frac{(\mathbf{x}(t_{i+1})-\mathbf{x}(t_i))^2}{2\Delta t}$$
where the sum runs over all $N(\mathbf{x})$ data points that are
inside the bin $\mathbf{x}$. We use an adaptive grid size with
$n = \sqrt{N_\mathrm{steps}}$ bins (width and height $\sqrt{n}$),
where $N_\mathrm{steps}$ is the total number of time points in all
trajectories in the data. This ensures that both the spatial
resolution and the accuracy of inference in each bin increase with the
amount of data.

This method, or slight variants of it, is used in a large number of
contexts~\cite{friedrich_approaching_2011,hoze_heterogeneity_2012},
and also often adapted to infer phase space
velocities~\cite{battle_broken_2016,gnesotto_broken_2018,seara_entropy_2018}. With
ideal data, we find that it performs reasonably well and converges to
exact values, although not as fast as SFI. With noisy data, it becomes
biased and does not converge.

\subsubsection*{InferenceMAP.}

\begin{figure}[H]
  \centering
  \includegraphics[width=\textwidth]{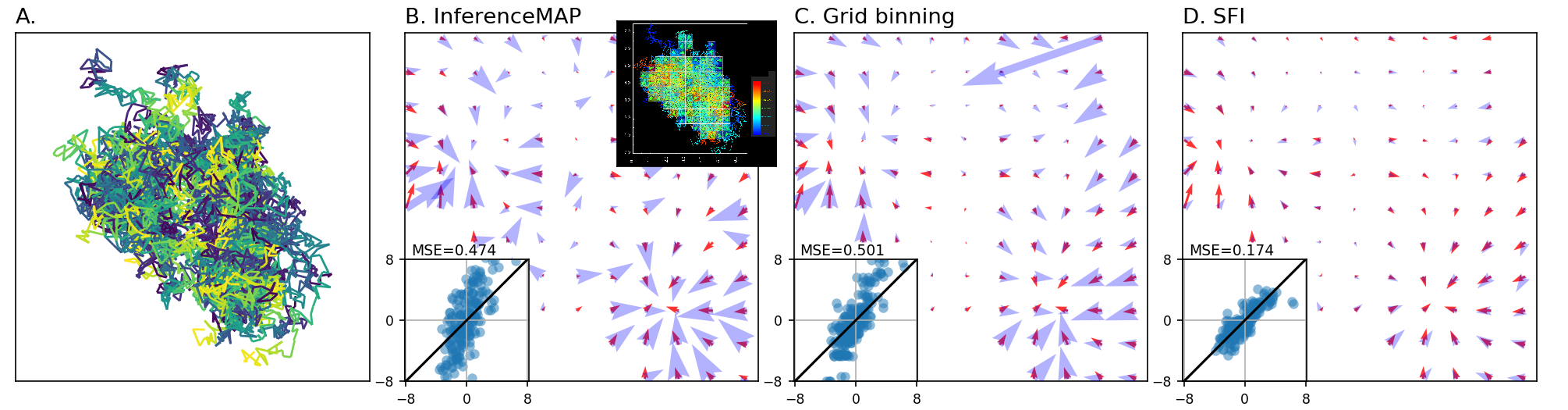}
  \caption{Comparison of the performance of the three methods on a set
    of $N=128$ noisy trajectories. \textbf{A.} The trajectories in the
    raw data set exploited by each method. \textbf{B-D.} The inferred
    drift field (thick light blue arrows) and the exact one used to
    generate the data (thin red arrows). The lower left insets show
    scatter plots of the inferred \emph{versus} exact drift
    components, with an indication of the normalized mean-squared
    error (MSE). Top right inset of B: screen capture of the
    InferenceMAP software used for this drift field. While methods B
    and C capture the qualitative shape of the drift field, they
    appear biased, and consistently overestimate the drift for noisy
    data. In contrast, our method shows quantitative agreement with
    the exact drift field. }
  \label{fig:inferencemap}
\end{figure}

The last method we compare to is InferenceMAP, a Bayesian method
relying on space discretization, introduced by Beheiry and
Masson~\cite{beheiry_inferencemap:_2015}. This method is commonly used
for the analysis of trajectories of single molecules inside
cells~\cite{knight_dynamics_2015,sungkaworn_single-molecule_2017,turkcan_bayesian_2012}. We
use the public implementation of this software. Upon trying many
different parameters, we find that the best results are obtained with
a square mesh, with maximum mesh size (the software adapts it to the
amount of data), and the (D,drift) inference option. We manually
provide the amplitude of the measurement noise (0 or 0.1). Typical
outcome of the method is presented on \Fig{fig:inferencemap}. The
performance on the inference of $D$ is slightly lower to that of SFI;
it significantly outperforms grid binning in the presence of
measurement noise. However, we find that the performance on drift
inference does not exceed that of grid binning, and our method
significantly outperforms InferenceMAP. This is demonstrated
quantitatively in \Fig{fig:comparison}, and on an example data set in
\Fig{fig:inferencemap}.

\end{document}